\documentclass{article}
\font\cero=cmss10 scaled 1728 
\usepackage{tensor}
\usepackage{verbatim}
\usepackage{amsfonts}
\usepackage{graphicx}
\usepackage{amssymb}
\usepackage{float}
\begin{document}
\begin{flushleft}
{\cero Quantum field theory of a hyper-complex scalar field  on a commutative ring}\\
\end{flushleft}
{\sf R. Cartas-Fuentevilla, and A.J.C.  Ju\'arez-Dom\'{\i}nguez }\\
{\it Instituto de F\'{\i}sica, Universidad Aut\'onoma de Puebla,
Apartado postal J-48 72570 Puebla Pue., M\'exico};

E-mail: rcartas@ifuap.buap.mx

ABSTRACT: Inspired by the structural unification of unitary groups (quantum field theory) with orthogonal groups (relativity) proposed recently through a non-division algebra, we construct a hypercomplex field theory with an internal symmetry that unifies the $U(1)$ compact gauge group with the $SO(1,1)$ noncompact gauge group, using the commutative ring of hypercomplex numbers. From the quantum field theory point of view, the hypercomplex field encodes two charged bosons with opposite charge, and corresponds thus to a neutral compound boson. Furthermore,
normal ordering of operators is not required for controling the vacuum divergences; in an analogy with SUSY, the theory under study contains $U(1)$ boson particles and their hyperbolic $SO(1,1)$ boson partners, whose contributions to the vacuum energy cancel out exactly to a zero value.
In fact the present scheme allows us to compare finite measuments of squeezed boson-number statistics obtained with and without normal ordering. Additionally we discuss on the potential applications of the squeezed boson states constructed on the commutative ring, in quantum teleportation and in related areas.\\


\section{Introduction and antecedents}
\label{introante}
Classical and quantum field theories with non-compact symmetries have appeared long time ago in different scenarios, from condensed matter systems such as disorded electron systems \cite{electron, electron1}, integrable sectors of QCD \cite{qcd,qcd1}, to dimensionally reduced quantum gravity \cite{qg,qg1}. The non-compact $\sigma$ models used in those contexts show basically a hyperbolic symmetry, and physical fields belonging to a non-compact space.
Furthermore, if the non-compact group is non-amenable, then surprising new features appear; for example spontaneous symmetry breakdown is possible at low dimensions, whereas it is forbidden in the case of compact internal symmetries according to the Mermin-Wagner theorem \cite{seiler}.

An original way of incorporating non-compact symmetries that we have explored recently, is through 
hypercomplex formalisms that contain various complex units; such schemes offer rich algebraic structures from the mathematical and field-theoretic point of views. For example,
the commutative ring of the hypercomplex numbers allows us to formulate gauge theories that contain the usual U(1) interactions together with their hyperbolic counterparts associated with the non-compact group $SO(1,1)$. Such theories have shown novel and interesting properties; the spontaneous symmetry breaking scenarios show running masses for vectorial and scalar fields, and running electromagnetic coupling that mimic the corresponding renormalization flow groups \cite{oscar,raa}. Furthermore, in the cosmological context of topological defects interacting with matter, the results reported in \cite{oscar} show that the Aharonov-Bohm type interaction is the only one, and not only the dominant interaction of cosmic strings with matter \cite{wil}.
Moreover the incorporation of the $SO(1,1)$ group as an additional internal symmetry has allowed us to bring up, from a new point of view, the so called gauge hierarchy, and certain dualities between weak and strong self-interaction regimes have been established \cite{raa}.

Other recent developments based on a hypercomplex formalism have appeared in the context of Calabi-Yau compactification, where a split-complex representation is the most natural way to parametrize the scalar fields of the universal multiplet \cite{emam1,emam5}. Similarly the description of D-instantons in terms of a supergravity formalism requires by consistency the use of split-complex algebra \cite{perry}.
Para-complex manifolds endowed with a para-K\"ahler geometry have allowed to find solutions in supergravity and M-theory \cite{cortes1, cortes2}.

\section{Motivations and results}
\label{mot}
Motivated with these antecedents,
this work will concern with the quantum field-theoretic formulation of a hypercomplex scalar field on a commutative ring; our results will have a direct impact in the following aspects of quantum field theory:

{\bf 1.- Strong connections with Quantions}:
There are 
strong connections between this work and the formulation of the so called quantions \cite{moldo,grgin}, an approach that attempts a structural unification of unitary groups (quantum mechanics) and orthogonal groups (relativity), and based on a non-division algebra; the quantal algebra constitutes also the natural framework in the formulation of the electroweak theory on a curved background.
In the present work, the full internal group of the model unifies 
 the unitary group $U(1)$ with the orthogonal group $SO(1,1)$;
the consequences of such an unification are the following.

{\bf 2.- Finite vacuum energy and an analogy with SUSY}: From the viewpoint of the quantum-field-theoretic formulations, 
it is desirable to construct theo\-ries or models in which the vacuum energy is finite; (unbroken) supersymmetry is the example by excellence: fermion and boson contributions to the vacuum energy cancel each other to an exact zero value \cite{weinberg}.
Roughly speaking, in the theory constructed in the present work, the incorporation of the non-compact hyperbolic symmetry will allow us to control the divergence of the $U(1)$-vacuum energy, which traditionally requires to invoke {\it normal ordering} of operators.

In analogy with SUSY, we can consider that the model under study will contain $U(1)$ (charged) boson particles and their hyperbolic $SO(1,1)$ boson partners, whose contributions to the vacuum energy cancel out exactly to a zero value. Hence, the conventional $U(1)$ scalar field theory will be embedded in an extended framework in which each $U(1)$ particle has a hyperbolic partner; since such partners are not observed in nature, then that extended symmetry must be broken, and the mentioned cancelation does not take place, and the vacuum energy is nonzero and large.

{\bf 3.-Normal ordering of operators}: {\it Normal ordering} corresponds to a formal subtraction of an infinite vacuum energy, by reordering the operators occuring in the Hamiltonian; however, such a procedure does not remove higher order contributions coming from the perturbative expansion.
The vacuum divergence and the remotion by normal ordering are considered as inherent properties of the standard quantum field theory formalism.

The justification of the normal ordering is found in the transition from classical to quantum description, since the ordering of the operators are not fixed from the beginning; it is a byproduct of the ambiguities of the quantization (for more details, see for example \cite{rugh}). However, in spite of such justifications, in the present quantization scheme it is no necessary to invoke {\it normal ordering} in order to obtain finite measurements; for example it will be fully unnecessary for controlling the vacuum divergences. Furthermore, we shall be able to compare finite measurements of the boson-number distributions in squeezed states, obtained with and without normal ordering; there will be a difference of orders of magnitude between them.

{\bf 4.- General Relativity and Quantum Field Theory}:
In the standard QFT formalism only the differences in vacuum energy have physical meaning; thus an infinite value may be circumvented by a redefinition of the energy scale. 
However, our model constructed on a ring, is sensitive to an absolute value of the vacuum energy, in consistency with a General Relativity point of view, which establishes that it is the total energy of a system that has physical meaning, and not just the energy differences \cite{ait}; there is no a {\it recalibration} of energy levels by an infinite constant in GR.
Hence our results may throw light on the problematic link between 
GR and QFT, specifically on the connection between quantum vacuum and the cosmological constant \cite{rugh}.

{\bf 5.- Squeezed states}: Although these states are of common use in quantum optics, they also appear in the following high energy physics scenarios:

Field configurations following the gravitational collapse to a black hole with subsequent eva\-po\-ra\-tion, have been described as squeezed states \cite{evaporation}; similarly the quantum fluctuations that are amplified due to the accelerated expansion of the universe during the inflation, are transformed under evolution into highly squeezed states \cite{evaporation,relic}.

In the present context, the model is constructed on the Minkowski spacetime; hence the vacuum state must be invariant under the action of the entire Poincar\'e group. Such a requirement leads to invoke nontrivial (bosonic) Bogoliubov transformations, which convert a vacuum state with no particles into a squeezed state with many particles, intimately related with condensation phenomena.

{\bf 6.- Quantum teleportation}: Squeezed states correspond to entangled states that are non-clasically correlated; using the traditional scheme with fictional experimentalists Alice and Bob in the famous quantum non-locality EPR paradox, such states are valuable resources in testing the quantum mechanics as a complete description of reality; currently such states are of commun use in quantum teleportation, and quantum computing. 
Therefore, our results on squeezed boson states constructed on a ring may be valuable resources in visualizing and understanding the physics of teleportation.

\section{Plan of the paper}
 In section \ref{hf} we outline basic aspects of the hypercomplex formalism, which are known in the literature; additionally we develop new results within this formalism. Especifically we develop an algorithm for determining the roots of Hermitian numbers, which generalize the concept of a real number of the conventional complex formalism. This result will allow us to determine in its turn the roots of polynomials with Hermitian coeffients; quartic characteristic polynomials will appear in the section \ref{vacuum}, in order to establish an isomorphism of the algebra of field commutators. In section \ref{intro} we introduce the classical field theory of a hypercomplex scalar field; such a field is encoding two charged scalar fields, and its norm contains a quadratic interaction term between them. The equations of motion, the solutions, and conserved quantities are ontained.
In section \ref{quantum} the classical functionals are promoted to quantum operators. In this point we realize that if the usual definition of the vacuum state is invoked, then one obtains a trivial quantum field theory; hence in section \ref{vacuum}, canonical transformations are required for defining a vacuum state that is appropriately annihilated by the quantum observables as the Hamiltonian, linear momentum, charge operators, etc; such canonical transformations allow us to eliminate the {\it normal ordering} of operators. The finite vacuum expectation values for observables are discussed in section \ref{finite}. The Fock space, the Hilbert space, and in particular the eigen-states for the number, and Hamiltonian  operators are discussed in section 
\ref{fock}; such eigen-states are required for describing the squeezing of the vacuum. In section \ref{squeezed} the single-mode squee\-zing operators that generate the canonical transformations are obtained; in order to determine the squeezing effect on the vacuum state the operators must be disentangled, and exclusive techniques on a ring are employed for this purpose; in this manner, the squeezed-boson states are constructed explicitly. The boson-number statistics is estudied in detail; then, finite measurements obtained with and without normal ordering  can be compared directly. Furthermore, along the same lines, multimode squezed boson states are obtained in section \ref{multi}.
In concluding remarks we discuss the perpectives, in particular future explorations on models involving quantum {\it friction}; additionally we shall give the basic ideas for future applications in quantum teleportation.
In the appendix we describe formal aspects on the disentangling
of exponential operators on a ring.

\section{The hypercomplex formalism}
\label{hf}
The commutative ring ${\cal H}$ of hypercomplex numbers $\psi $ is defined as \cite{U},
\begin{eqnarray}
     \psi =  u + iy +jv + ijw, \quad \overline{\psi}  =  u - iy -jv + ijw, \quad  u,y,v,w \in {\cal R} 
          \label{ring}
\end{eqnarray} where the hyperbolic unit $j$ has the properties $j^{2}=1$, and $\overline{j} = -j$, and, as usual, $i^{2}=-1$, and $\overline{i}=-i$, and with commuting complex units $ij=ji$. Hence, with respect to the conjugation involving both complex units, the square of the hypercomplex number is given by
\begin{equation}
     \psi \overline{\psi} = u^{2} + y^{2} -v^{2} - w^{2} +2ij (uw-yv),
     \label{square}
\end{equation} 
which is not a real number, instead it is in general a {\it Hermitian} number. 
The expression (\ref{square}) is invariant under the usual circular rotations $e^{i\theta}$ represented by the Lie group $U(1)$; similarly it is invariant under {\it hyperbolic} rotations 
\begin{equation}
e^{j\chi}=cosh\chi+j sinh \chi, \quad \chi \in R;
\label{rothyp}
\end{equation}
that can be represented by the connected component of the Lie group $SO(1,1)$ containing the group unit. Hence, the norm ({\ref{square}) is simultaneously invariant under the action of a compact and and non-compact group, $U(1)\times SO(1,1)$. The model that we analyse in this work will correspond to a hypercomplex version of the $U(1)$ field theory, which implies that the Lagrangian will be valued in the Hermitian numbers, instead of the conventional real numbers. The physical consequence of this extension is the incorporation of the non-compact group $SO(1,1)$ as a part of the internal symmetry of the model. Hence, an object that is invariant under the full action 
of the group $U(1)\times SO(1,1)$, necessarily requires to be valued in the Hermitian extension of the real numbers.

The commutating product of the complex units can be considered as a new complex unit with certain properties,
\begin{equation}
     k \equiv ij, \qquad \overline{k} = k, \qquad k^{2} = -1. \label{k}
\end{equation}
With this hybrid complex unit, one can define the pseudo-real or Hermitian numbers, which constitute a sub-set of ${\cal H}$ closed respect to the sum and the product,
\begin{eqnarray}
     \psi_{0} \!\! & = & \!\! u_{0} + k w_{0}, \qquad \overline{\psi}_{0} = \psi_{0}, \qquad \psi^{2}_{0} = u^{2}_{0} - w^{2}_{0} + 2ku_{0}w_{0}; \nonumber \\
     \psi_{0} + \psi'_{0} \!\! & = & \!\! (u_{0}+u'_{0}) + k(w_{0}+w'_{0}); \qquad \psi_{0} \psi'_{0} = u_{0}u'_{0} - w_{0}w'_{0} + k(w_{0}u'_{0} + w'_{0}u_{0});
     \label{subset}
\end{eqnarray}
the norm of a hypercomplex number (\ref{square}) belongs to this sub-set. Furthermore, as a consequence of the product in (\ref{subset}), any power of a Hermitian  number
is also Hermitian; the expression $(\psi_{0})^n = (u_{0} + k w_{0})^n$ can be developed using the binomial formula and the fact that $k^2=-1$. Note that the subset of Hermitian elements is an {\it integral domain}; under the condition of a vanishing product (\ref{subset}), then $\psi_{0}=0$, or $\psi'_{0}=0$.
In general the ring ${\cal H}$ is not an {\it integral domain}, due to the existence of zero-divisors as we shall see below.

Furthermore, for a number of the form (\ref{subset}), for which at least one of the real quantities $(u_{0}, w_{0})$ is different from zero, one can define its inverse,
\begin{equation}
     \psi_{0} \psi^{-1}_{0} = \psi^{-1}_{0} \psi_{0} = 1, \qquad \psi^{-1}_{0} = \frac{u_{0}-ijw_{0}}{u^{2}_{0}+w^{2}_{0}}; \label{inv}
\end{equation}
hence, the inverse belongs to the sub-set. Note that the real quadratic form in the denominator of $\psi^{-1}_{0}$ is not the norm of $\psi_{0}$ in (\ref{subset}); similarly the numerator is not the complex conjugate of $\psi_{0}$, since such a number is Hermitian, invariant under complex conjugation.

With the hybrid complex unit (\ref{k}) we can define a Hermitian exponential, $e^{k\theta}$, which can be expressed in the following form, considering the usual series expansion for the exponential,
\begin{equation}
     e^{k\theta} = \cos \theta +k\sin\theta ; \label{hybridexp}
\end{equation}
furthermore, with this hybrid exponential, we can construct a polar representation for a Hermitian number (\ref{subset}),
\begin{equation}
     \eta_{1} + k\eta_{2} = \rho e^{k\theta}, \qquad \eta_{1}, \eta_{2}; \rho , \theta \in R; \label{herpol}
\end{equation}
with
\begin{equation}
     \eta_{1} = \rho\cos\theta , \qquad \eta_{2}= \rho\sin\theta , \qquad \rho = \sqrt{\eta_{1}^{2}+\eta_{2}^{2}}, \qquad \theta = \arctan \frac{n_{2}}{n_{1}}. \label{herpol2}
\end{equation}
This polar expression allows us to find the roots of a Hermitian number; we develop first the square roots,
\begin{eqnarray}
     \sqrt{\eta_{1}+k\eta_{2}} \!\! & = & \!\! \sqrt{\rho} e^{k\frac{\theta}{2}} \nonumber \\
     \!\! & = & \!\! \pm \frac{1}{\sqrt{2}} \Big[ \sqrt{\eta_{1} + \sqrt{\eta_{1}^{2} +\eta_{2}^{2}}} + \frac{k\eta_{2}}{\sqrt{\eta_{1}+ \sqrt{\eta_{1}^{2}+\eta_{2}^{2}}}} \Big], \label{herroot}
\end{eqnarray}
where we have used the formula for the half-angle identities. Since the expression $\eta_{1}+\sqrt{\eta_{1}^{2}+\eta_{2}^{2}}$ under the square root, is definite positive for $\eta_{2}\neq 0$ and $\eta_{1} \in R$, then we conclude that the square roots of a  Hermitian number are Hermitian too. For $\eta_{2}=0$ one obtains the usual roots for the real (positive) number $\eta_{1}$. The figure (\ref{realhermitian}) shows that there is a continuous limit between the square roots of Hermitian numbers and of conventional (positive) real numbers. 
\begin{figure}[H]
  \begin{center}
    \includegraphics[width=.4\textwidth]{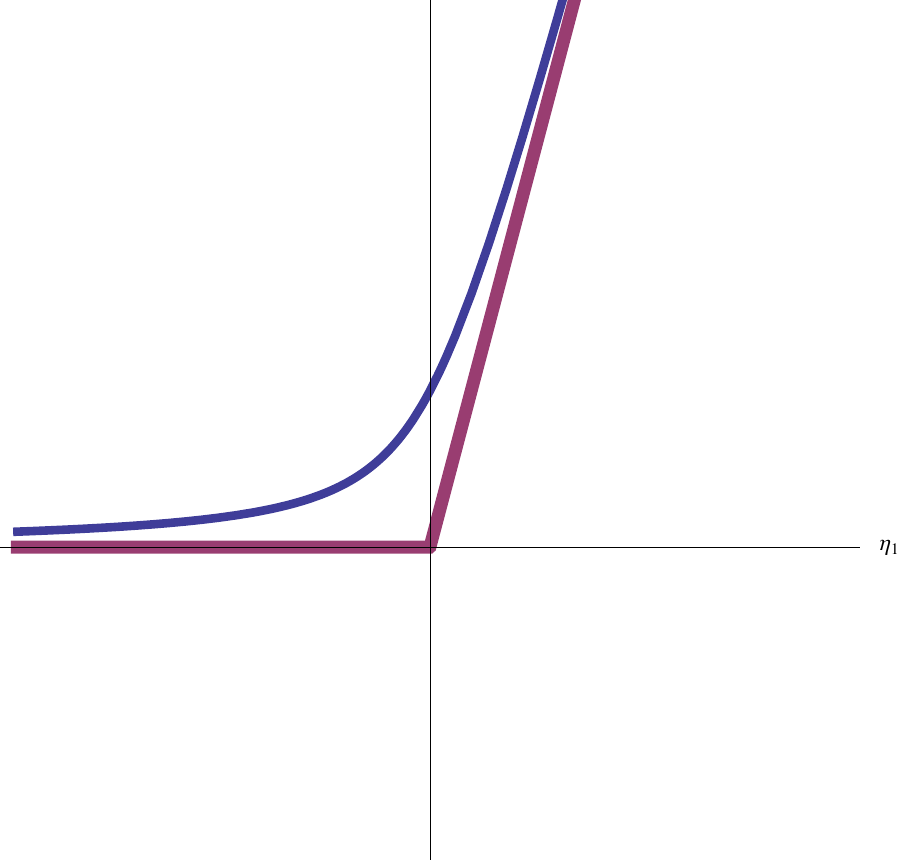}
  \caption{In blue the expression $\eta_{1}+\sqrt{\eta_{1}^{2}+\eta_{2}^{2}}$ as a function of $\eta_{1}$, and for a given $\eta_2$; for $\eta_{1}=0$ the ordinate is given by $|\eta_2|$. Note that $\lim_{_{\eta_{2}\rightarrow 0}}[\eta_{1}+\sqrt{\eta_{1}^{2}+\eta_{2}^{2}}]$ is a definite positive quantity, consistent with the fact that $\Big[\eta_{1}+\sqrt{\eta_{1}^{2}+\eta_{2}^{2}}\Big]_{\eta_{2}=0}=\eta_{1}+|\eta_1|$, which is shown by the red line.} 
   \label{realhermitian}
  \end{center}
\end{figure}
Similarly the cubic roots for the Hermitian number can be expressed as
\begin{equation}
     \sqrt[3]{\eta_{1}+k\eta_{2}} = \sqrt[3]{\rho} \big( \cos \frac{\theta}{3} + k\sin\frac{\theta}{3}\big) , \qquad \sqrt[3]{\rho} = \sqrt[6]{\eta^{2}_{1} + \eta^{2}_{2}}; \label{cubicroot}
\end{equation}
using the one-third angle identities we can find multiple roots,
\begin{eqnarray}
     \sqrt[3]{\rho} \sin\frac{\theta}{3} \!\! & = & \!\! \frac{1}{2} \Big[ \sqrt[3]{i\eta_{1}-\eta_{2}}+ \sqrt[3]{-i\eta_{1}-\eta_{2}}\Big], \nonumber \\
     \sqrt[3]{\rho} \cos\frac{\theta}{3} \!\! & = & \!\! \frac{1}{2} \Big[ \sqrt[3]{i\eta_{2}+\eta_{1}}+ \sqrt[3]{-i\eta_{2}+\eta_{1}}\Big];
     \label{cubicroot1}
\end{eqnarray}
and
\begin{eqnarray}
     \sqrt[3]{\rho} \sin\frac{\theta}{3} \!\! & = & \!\! -\frac{1}{4} \Big[ (1\mp i\sqrt{3}) \sqrt[3]{i\eta_{1}-\eta_{2}} + (1\pm i\sqrt{3}) \sqrt[3]{-i\eta_{1}-\eta_{2}} \Big], \nonumber \\
     \sqrt[3]{\rho} \cos\frac{\theta}{3} \!\! & = & \!\! -\frac{1}{4} \Big[ (1\mp i\sqrt{3}) \sqrt[3]{i\eta_{2}+\eta_{1}} + (1\pm i\sqrt{3}) \sqrt[3]{-i\eta_{2}+\eta_{1}} \Big].
     \label{cubicroot2}
\end{eqnarray}
Note that in spite of using (conventional) complex numbers under the cubic root, the expressions (\ref{cubicroot1}), and (\ref{cubicroot2}) are real numbers. Hence, the three roots in the Eq. (\ref{cubicroot}) are Hermitian again; a Hermitian number has three Hermitian cubic roots. As a conclusion the Hermitian 
numbers are closed respect to the sum, product, taking powers, and extracting square, and cubic roots.

 It can be of interest from the algebraic point of view to continue further with quartic roots, and beyond; however such a problem is not a trivial issue, since it will require to find the roots of polynomials of higher order, and it is well known that for every integer $n$ greater than four there can not be a formula for the roots for a degree-$n$ polynomial \cite{irving}. However, we conject that the subset of Hermitian numbers is closed respect to the extraction of roots of any order. Fortunately in this work we shall use only square and cubic  roots; especifically in the construction of the quantum field theory the roots of a quartic characteristic polynomial are required for establishing an isomorphism of the algebra of field commutators, and all these algebraic results will make sense.

In contrast to the conventional complex numbers with idempotents $0$, and $1$, the ring ${\cal H}$ contains additionally non-trivial idempotents,
\begin{eqnarray}
     J^{+} \!\! & \equiv & \!\! \frac{1}{2} (1+j), \qquad (J^{+})^{n} = J^{+}, \nonumber \\
     J^{-} \!\! & \equiv & \!\! \frac{1}{2} (1-j), \qquad (J^{-})^{n} = J^{-}, \qquad n=1,2,3 \ldots ;
     \label{nullbasis}
\end{eqnarray}
which are also {\it null},
\begin{equation}
     J^{+}J^{-} = 0; \label{ortho}
\end{equation}
this property can be understood as an orthogonality property between the idempotents, and indicates that the commutative ring is not an {\it integral domain} \cite{leclair}.
The hyperbolic rotation (\ref{rothyp}) can be expressed in terms of a combination of the idempotents $(J^{+},J^{-})$,
\begin{equation}
     e^{j\chi} = e^{\chi}J^{+} + e^{-\chi}J^{-}.\label{rothyp1}
\end{equation}
Such idempotents work as {\it projectors} in the ring ${\cal H}$, and will play a key role in the construction of the quantum formulation in a hypercomplex scheme. Additionally these idempotents have the property of {\it absorbing} the hyperbolic complex unit;
\begin{equation}
     j J^{+} = J^{+}, \qquad j J^{-}= -J^{-}. \label{absorb}
\end{equation}
Furthermore,
for any element of the ring $\rho \equiv \rho_{1}+i\rho_{2} +j\rho_{3} +k\rho_{4}$, we have
\begin{eqnarray}
     J^{+} \rho \!\! & = & \!\! J^{+} [\rho_{1} + \rho_{3} + i(\rho_{2}+\rho_{4})], \nonumber \\
     J^{-} \rho \!\! & = & \!\! J^{-} [\rho_{1} - \rho_{3} + i(\rho_{2}-\rho_{4})];
     \label{factorization}
\end{eqnarray}
hence, the idempotents $(J^{+},J^{-})$ have the effect of factoring out any element as the product of a purely hyperbolic number and a purely (conventional) complex number. Furthermore, if the number $\rho$ has a vanishing $J^-$-projection, then necessarily takes the form
\begin{eqnarray}
\rho_1=\rho_3, \quad \rho_2=\rho_4, \quad \rho=2J^+(\rho_1+i\rho_4), \quad \rho\overline{\rho}=0;
\label{projJ}
\end{eqnarray}
and conversely in the case when the $J^+$-projection vanishes; a vanishing norm in the above expression is due to the Eq. (\ref{ortho}).
From Eqs. (\ref{factorization}) it is evident that if a number has vanishing $J^-$, and $J^+$ projections, then the number vanishes,
\begin{eqnarray}
J^-\rho=0, \quad J^+\rho=0, \quad \rightarrow \rho=0.
\label{bothprojection}
\end{eqnarray}
In the applications below, we shall need to impose the
vanishing of some projection of certain hypercomplex coefficients, and in order to avoid its vanishing, one will require to maintain at least one projection as different from zero.
Consistently the direct sum of the projections (\ref{factorization}) leads to the complete number,
\begin{eqnarray}
J^-\rho+J^+\rho=\rho;
\label{directsum}
\end{eqnarray}
therefore any hypercomplex number can be decomposed in the {\it null} basis $(J^+, J^-)$, with conventional complex numbers as components. Finally, 
a hypercomplex number of real modulus, normalized to $\pm 1$ can be expressd as
\begin{eqnarray}
\rho\! & = & Ie^{i\theta}e^{j\chi}, \quad \rho\overline{\rho}=1; \quad I=\pm 1, \pm i;\nonumber\\
\rho\! & = &\! \pm je^{i\theta}e^{j\chi}, \quad \rho\overline{\rho}=-1; \quad \theta, \chi \in R.
\label{norm1}
\end{eqnarray}

\section{The Lagrangian for hypercomplex fields: The classical field theory}
\label{intro}
As already commented in the introduction, hypercomplex scalar field models with a $"\lambda \phi^4 "$ self-interaction term have been considered previously in spontaneous symmetry breaking scenarios (\cite{oscar,raa}); motivated by these previous results, by the quantionic perspective of a structural unification of compact and non-compact internal symmetries, and by the potential applications in quantum teleportation, we start now a second quantization program, formulating the simplest field theory on the commutative ring ${\cal H}$,
which will involve only a quadratic term in the hypercomplex field. As we shall see, such a model does not correspond to a free field theory, since the interactions will be present naturally in the simplest invariant objects constructed on the ring.

We describe first the physical content of a hypercomplex field;
the expression (\ref{ring}) can be rewritten as
\begin{eqnarray}
\psi=\psi_{1}+j\psi_{2}, \quad \psi_1\equiv u+iy, \quad \psi_2\equiv v+iw;
\label{physcon}
\end{eqnarray}
thus it is encoding two spin-zero charged fields. Furthermore, the norm (\ref{square}) can be rewritten as 
\begin{eqnarray}
\psi\overline{\psi}= \psi_1 \overline{\psi}_1 -\psi_2 \overline{\psi}_2+j
(\overline{\psi}_1\psi_2-\psi_1\overline{\psi}_2),
\label{physcon2}
\end{eqnarray}
with $\psi_1 \overline{\psi}_1=u^2+y^2$, $\psi_2 \overline{\psi}_2=v^2+w^2$, and $(\overline{\psi}_1\psi_2-\psi_1\overline{\psi}_2)=2i(uw-yv)$; thus, the hybrid term of the form $ij$ in (\ref{square}) can be understood as an {\it interacting} term between the two charged fields. Since the norm (\ref{physcon2}) is the simplest invariant quantity that one can construct in the ring $\cal H$, there will be no a free field theory, and neccesarily the interactions are incorporated in the simplest model.

Furthermore, we may naturally ask
about the total charge of the full field $\psi$; once we have constructed a Lagrangian with the invariant form (\ref{physcon2}), the second quantization will show that the full field as particle carries no intrinsic angular momentum, and it will correspond to a {\it neutral} compound boson; such a spin-zero particle is constituted by two charged boson particles with opposite charge.

With these elements, the usual Lagrangian for the charged U(1) scalar field can be re-formulated on the hypercomplex space, considering the invariant form (\ref{square}), in order to have a $U(1)\times SO(1,1)$ global invariant expression
\begin{equation}
     {\cal L} (\psi, \overline{\psi}) = \int d^{n}x (\partial^{\mu}\psi \cdot \partial_{\mu} \overline{\psi} - m^{2} \psi\overline{\psi}),
\label{lag}
\end{equation}
in a $n$-dimensional background. In general $m^{2} \equiv m^{2}_{R} + ijm^{2}_{H}$, and the full Lagrangian is valued in the sub-set of Hermitian numbers, and it has the form $R+kR$, the generalization of a real number in the hypercomplex formalism. A possible interpretation is that one requires a pair of mass values for the pair of fields encoded in (\ref{physcon}),
$\{\psi_1, \psi_2\}\leftrightarrow \{ m_R, m_H\}$.

Hypercomplex Lagrangian can be formulated; for example holomorphic models in the conventional complex sense have been considered recently, and certain hidden gauge symmetries have been found \cite{vergara}; such complex gauge symmetries allow us to connect different real systems, and in the case of a holomorphic potential at Lagrangian level, those symmetries are manifiested through the usual Cauchy-Riemann conditions. Research along these lines are of our interest for the hypercomplex case, and it will be the subject of future explorations. In the present work, we shall focus on the natural Hermitian extension of the real valued Lagrangians, and thus a non-compact symmetry is automatically incorporated.

Rewriting the Lagrangian density as
\begin{equation}
     \partial^{\mu}\psi\cdot\partial_{\mu}\overline{\psi} - m^{2}\psi\overline{\psi} = \partial^{0}\psi\cdot\partial_{0}\overline{\psi} - \nabla\psi\cdot\nabla\overline{\psi} - m^{2}\psi\overline{\psi},
\end{equation}
we have the following expressions for the two canonically conjugated fields, and for the canonical Hamiltonian
\begin{eqnarray}
     \pi \!\! & \equiv & \!\! \frac{\partial L}{\partial \dot{\psi}} = \dot{\overline{\psi}}, \quad \overline{\pi} \equiv \frac{\partial L}{\partial\dot{\overline{\psi}}} = \dot{\psi}, \nonumber \\
     H \!\! & \equiv & \!\! \pi\dot{\psi} + \overline{\pi} \dot{\overline{\psi}} - {\cal L} = \pi\overline{\pi} + \nabla\psi\cdot\nabla\overline{\psi} + m^{2} \psi\overline{\psi};
\label{canon}
\end{eqnarray}
as well the Lagrangian, the Hamiltonian is a Hermitian field $\overline{H}=H$, and will have in general the form $H=H_1+kH_2$.

The symmetry under phase transformations leads to a $U(1)\times SO(1,1)$-Noether current, which has the same functional dependence on the fields $(\psi, \overline{\psi})$ that appears in the usual $U(1)$-Noether current,
\begin{equation}
     J_{\mu} \equiv \overline{\psi}\partial_{\mu}\psi - \partial_{\mu}\overline{\psi}\cdot\psi , \quad \partial_{\mu}J^{\mu} =0,
\label{noether}
\end{equation}
in particular the charge ${\cal Q}$ corresponding to this current will be given by
\begin{equation}
     {\cal Q} \equiv \int d^{3} {\bf x} (\overline{\psi}\dot{\psi} - \dot{\overline{\psi}}\psi) = \int d^{3} {\bf x} (\overline{\psi}\overline{\pi} -\pi\psi );
\label{charge}
\end{equation}
in terms of the real components of the fields, (see Eq. (\ref{ring})), the charge can be rewritten as
\begin{equation}
     Q = 2\int d^{3} {\bf x} [i(u\dot{y}-\dot{u}y+w\dot{v}-v\dot{w}) +j(u\dot{v}-v\dot{u}+y\dot{w}-\dot{y}w)],
\label{hypercharge}
\end{equation}
and hence, in the limit $v=w=0$, one recovers the U(1)-charge given by $i(u\dot{y}-\dot{u}y)$. Since ${\cal Q}$ has contributions coming from both complex units $(i,j)$, we can call it as the {\it hyper-charge}; under conjugation we have that $\bar{{\cal Q}}=-{\cal Q}$.

Furthermore, the energy-momentum tensor is given by
\begin{equation}
     T_{\mu\nu} = \frac{1}{2} \partial_{(\mu}\psi\cdot\partial_{\nu )}\overline{\psi} - \frac{1}{2} g_{\mu\nu} \partial^{\rho}\psi\cdot\partial_{\rho} \overline{\psi};
\label{emt}
\end{equation}
in particular, the total linear momentum is 
\begin{equation}
     {\bf P} = \frac{1}{2} \int d^{3} {\bf x} (\pi\nabla\psi + \overline{\pi}\nabla\overline{\psi}).
\label{momentum}
\end{equation}
The equation of motion read
\begin{equation}
   (\square + m^{2}) \psi =0; \qquad \square = \partial^{2} _{t} - \triangledown^{2}, \label{eqm1}
\end{equation}
corresponding equation holds for the complex conjugate field $\overline{\psi}$. The formal solution for the equation (\ref{eqm1}) is given by
\begin{equation}
\psi ({\bf x},t) = a e^{i(w_{1}t - {\bf k_{1}\cdot x})} e^{j(w_{2}t - {\bf k_{2}\cdot x})}+ b e^{-i(w_{1}t - {\bf k_{1}\cdot x})} e^{-j(w_{2}t - {\bf k_{2}\cdot x})}, \label{sol1}
\end{equation}
where $a,b$ are arbitrary coefficients, and $ w_{1,2} $, and $ {\bf k_{1,2}} $ are real parameters; the substitution of this expression into the equations of motion yields a generalized dispersion relation
\begin{equation}
{\bf k_{1}}^{2}- w_{1}^{2} + m^{2}_{R} = {\bf k_{2}} ^{2} - w_{2}^{2}, \quad 2w_{1}w_{2} - 2 {\bf k_{1}\cdot k_{2}} + am^{2}_{H} =0, \label{dispersion}
\end{equation}
from which one can obtain the usual dispersion relation for the $U(1)$ field theory in the limit $w_{2}=0= {\bf k_2}=m_H$.
Now, using the expression (\ref{rothyp}), the solution (\ref{sol1}) can be rewritten as
\begin{eqnarray}
\psi ({\bf x,t}) = a e^{i(w_{1}t -{\bf k_{1}\cdot x})} \left[ J^{+} e^{w_{2}t-{\bf k_{2}\cdot x}} + J^{-} e^{-w_{2}t + {\bf k_{2}\cdot x}}\right] \nonumber \\+
b e^{-i(w_{1}t -{\bf k_{1}\cdot x})} \left[ J^{-} e^{w_{2}t-{\bf k_{2}\cdot x}} + J^{+} e^{-w_{2}t + {\bf k_{2}\cdot x}}\right] ; \label{sol2}
\end{eqnarray}
where the two idempotent hyperbolic coefficients (\ref{nullbasis}) have appeared.
 However, the hyperbolic {\it phases} in (\ref{sol2}) are not bounded in general, and can not be used, as stand, as the base of the hypercomplex Fourier calculus. For example, if
\begin{equation}
{\bf k_{2}\cdot x} \geq 0, \label{constraint}
\end{equation}
then the modes $a J^{+}$, and $b J^{-} $ are convergent and $ a J^{-}$ and $b J^{+} $ are divergent, and conversely for the case with the sign changed in the constraint (\ref{constraint}). Hence, in the solution (\ref{sol2}) one has a combination of convergent and divergent modes, which are orthogonal in the sense of (\ref{ortho}), and only certain projections are bounded. For concreteness, we consider from now on, the constraint (\ref{constraint}), and the bounded projections $a J^{+}$, and $b J^{-} $, in order to construct the Laplace expansion of the field, understood now as an operator-valued distribution,
\begin{eqnarray}
\widehat{\psi} ({\bf x},t) \!\! & = & \!\! \int\limits^{+\infty}_{-\infty} N_{k_{1}} d^{3} {\bf k_{1}} \int\limits^{+\infty}_{0} N_{k_{2}} d^{3}  {\bf k_{2}}   \Big[J^{+} \hat{a} ({\bf k_{1}, k_{2}}) e^{i(w_{1}t- {\bf k_{1}\cdot x})}  e^{w_{2}t - {\bf k_{2}\cdot x}} \nonumber \\
\!\! & & \!\! +J^{-} \hat{b}^{\dagger} ({\bf k_{1}, k_{2}}) e^{-i(w_{1}t - {\bf k_{1}\cdot x})}  e^{w_{2}t -{\bf k_{2}\cdot x}} \Big] , 
\label{expansion}
\end{eqnarray}
where the two sets of creation and annihilation operators $ \{\widehat{a}, \widehat{a}^{\dagger}; \widehat{b}, \widehat{b}^{\dagger} \}$ are in general hypercomplex objects; the spectral parameters $ ({\bf k_{1}, k_{2}}) $ are real-valued, and $ (N_{k_{1}}, N_{k_{2}}) $ are normalization factors that will be fixed later; the substitution of the convergent solution (\ref{expansion}) into the equations of motion (\ref{eqm1}) leads essentially to the same dispersion relations 
(\ref{dispersion}), but enforcing the vanishing of the hybrid mass in the second relation:
\begin{eqnarray}
m_{H}=0, \quad \rightarrow \quad w_{1}w_{2} = {\bf k_{1}\cdot k_{2}}.
\label{dispersion2}
\end{eqnarray}
In the expansion (\ref{expansion}) we can identify the so called Poisson kernel,
\begin{equation}
\int\limits^{+\infty}_{-\infty} e^{ikx - \vert p x\vert} dx =  \frac{1}{ik+p} +\frac{1}{-ik+p} = \frac{2p}{p^{2}+ k^{2}};
\label{poisson}
\end{equation}
this integral will be used later in order to solve the expression (\ref{expansion}) to favor of the creation and annihilation operators.
Furthermore, for the conjugate momentum $ \widehat{\pi} \equiv \partial_{t} \widehat{\psi}^{\dagger} $, we have that
\begin{equation}
\widehat{\pi} ({\bf x},t) = \int\limits^{{+\infty}}_{-\infty} \overline{N}_{k_{1}} d^{3} {\bf k_{1}} \int\limits^{+\infty}_{0}    \overline{N}_{k_{2}} d^{3} {\bf k_{2}} \left[ (-iw_{1}+w_{2}) J^{-} \hat{a}^{\dagger} e^{-i(w_{1}t- {\bf k_{1}\cdot x})} + (iw_{1}+w_{2}) J^{+} \hat{b}e^{i(w_{1}t- {\bf k_{1}\cdot x})}\right] e^{w_{2}t- {\bf k_{2}\cdot x}};
\label{cm}
\end{equation}
and hence, the commutation relations at equal time for these field operators can be constructed in terms of the commutators for creation and annihilation operators
\begin{eqnarray}
\left[ \widehat{\psi} ({\bf x},t), \widehat{\pi} ({\bf x'},t) \right] \!\! & = & \!\! \int d_{1} \int d_{1'} \int d_{2}\int d_{2'} \Big\{ J^{+} (iw'_{1}+w'_{2})\cdot \left[ \hat{a} ({\bf k_{1},k_{2}}), \hat{b} ({\bf k'_{1},k'_{2}})\right] \nonumber \\
&& exp [ i(w_{1}+w'_{1})t -i ({\bf k_{1}\cdot x+k'_{1}\cdot x'}) ] 
+ J^{-} (-iw'_{1}+w'_{2}) \left[ \hat{b}^{\dagger} ({\bf k_{1},k_{2}}), \hat{a}^{\dagger} ({\bf k'_{1},k'_{2}})\right] \cdot \nonumber \\ 
&& exp \left[  -i(w_{1}+w'_{1})t +i ({\bf k_{1}\cdot x+k'_{1}\cdot x'}) \right] \Big\} exp\left[(w_{2} + w'_{2})t {\bf - k_{2}\cdot x - k'_{2}\cdot x'} \right] ,
\label{cc}
\end{eqnarray}
where the integration is denoted in compact form by $\int d_{1} = \int^{+\infty}_{-\infty} {N}_{k_{1}} d^{3}{\bf k_{1}} , \int d_{1'} = \int^{+\infty}_{-\infty} \overline{N}_{k'_{1}} d^{3}{\bf k'_{1}} $,  $\int d_{2} = \int^{+\infty}_{0} {N}_{k_{2}} d^{3}{\bf k_{2}}$, and $\int d_{2'} = \int^{+\infty}_{0} \overline{N}_{k'_{2}} d^{3}{\bf k'_{2}} $. Hence, assuming the following commutation rules
\begin{eqnarray}
      \left[ \widehat{a} ({\bf k_{1}}, {\bf k_{2}}), \widehat{b} ({\bf k'_{1}}, {\bf k'_{2}})\right] \!\! & = & \!\! \rho \delta ({\bf k_{1}} + {\bf k'_{1}}) \delta ({\bf k_{2}} +{\bf k'_{2}}), \nonumber \\
    \left[  \widehat{b}^{\dag} ({\bf k_{1}}, {\bf k_{2}}), \widehat{a}^{\dag} ({\bf k'_{1}}, {\bf k'_{2}})\right] \!\! & = & \!\! \overline{\rho} \delta ({\bf k_{1}} + {\bf k'_{1}}) \delta ({\bf k_{2}} +{\bf k'_{2}});
\label{cacgen}
\end{eqnarray}
where $\rho$ is an arbitrary element of the ring, $\rho \in {\cal H}$, which will be chosen conveniently later; hence
the commutator (\ref{cc}) reduces to
\begin{eqnarray}
\left[ \widehat{\psi} ({\bf x},t), \widehat{\pi} ({\bf x'},t) \right] \!\! & = & \!\! -\int \overline{N}_{k_{1}} d_{1} \int \overline{N}_{k_{2}} d_{2} \Big\{ J^{+} (iw_{1}+w_{2})\rho e^{ (-i{\bf k}_{1}-{\bf k_{2})}\cdot {\bf (x-x')} } \nonumber \\
\!\! & & \!\! +J^{-} (-iw_{1}+w_{2})\overline{\rho} e^{(i{\bf k}_{1}-{\bf k}_{2}) \cdot {\bf (x-x')}}\Big\} ,
\label{ccr}
\end{eqnarray}
where we have considered that the dispersion relations (\ref{dispersion}), and (\ref{dispersion2}), are invariant under the discrete transformation
\begin{equation}
({\bf k_{1}}, {\bf k_{2}}, w_{1}, w_{2}) \longrightarrow (-{\bf k}'_{1}, -{\bf k}'_{2}, -w'_{1}, -w'_{2}).
\label{sym}
\end{equation}
Furthermore, using the following integrals, where the second one corresponds to the new contribution of the hyperbolic modes,
\begin{equation}
\int^{+\infty}_{-\infty} \frac{d^{3}{\bf p}}{(2\pi)^{3}} e^{i {\bf p}\cdot ({\bf x}- {\bf x}')} = \delta ({\bf x-x'}), \quad \int^{+\infty}_{0} d^{3} {\bf p} e^{-{\bf p}\cdot ({\bf x}-{\bf x}')} = \frac{1}{{\bf x} - {\bf x}'},
\label{integrals}
\end{equation}
the expression (\ref{ccr}) reduces essentially to the product of the expressions (\ref{integrals}), 
\begin{eqnarray}
     \!\! & & \!\! \left[ \widehat{\psi} ({\bf x}, t), \widehat{\pi} ({\bf x'}, t)\right] = - (2\pi)^{3} |N_{K_{1}}|^{2} |N_{K_{2}}|^{2} \cdot 
     \left[J^{+} (iw_{1} + w_{2}) \rho + c.c. \right] \frac{\delta ({\bf x}-{\bf x}')}{{\bf x}-{\bf x}'}.
\label{commu}
\end{eqnarray}
This expression for the canonical commutator deserve some remarks; first, it is determined by the commutator $ [ \hat{a},\hat{b} ]=\rho$, which is fixed to zero in the usual approach for the quantization of the $U(1)$ Klein-Gordon theory; moreover, in that approach the canonical commutator is determined by the commutators $ [ \widehat{a}, \widehat{a}^{\dagger} ]  $, and $ [ \widehat{b}, \widehat{b}^{\dagger} ] $, which in fact do not appear in the computation of (\ref{commu}), since they turn out to be proportional to the null hyperbolic product $ J^+\cdot J^-=0 $; such commutators will appear in its due course for the case at hand. Moreover, since the fundamental field commutator $[\psi,\pi]$ must be nonvanishing in order to construct a non-trivial QFT, then it is crucial that the commutator $[ \hat{a},\hat{b} ]$ does not vanish; optionally the commutators
$ [ \widehat{a}, \widehat{a}^{\dagger} ]  $, and $ [ \widehat{b}, \widehat{b}^{\dagger} ]$ may be fixed to zero, as we shall see later by squeezing the vacuum and to generate squeezed boson states (see Eqs. (\ref{zeromix0})). Therefore, the Hilbert space will not be isomorphic to that of the 
quantum mechanical harmonic oscillators.

We realize that the Dirac delta distribution for the canonical commutator in the traditional $U(1)$ Klein-Gordon theory, has been substituted by its derivative, which can be understood as a {\it squeezing} effect on the usual delta function, due to the presence of the symmetry under 
hyperbolic rotations; it corresponds to a {\it tempered} distribution, which is more singular than the usual delta function. Such distributions describe the density operator for a mixture of quantum states that represent a coherent state (\cite{gerry}); the fact that they are more singular than the delta function, corresponds to a criterion for defining a coherent state as a nonclassical state.

Now we can find the explicit expressions for the creation and annihilation operators in terms of the canonical field operators; hence, from the Eq. (\ref{expansion}) we have that
\begin{eqnarray}
     \int\limits^{{+\infty}}_{-\infty} e^{(i{\bf P}_{1}-{\bf P}_{2})\cdot {\bf x}} \widehat{\psi} (x,t) d^{3}{\bf x} \!\! & = & \!\! N_{k_{1}} N_{k_{2}}\int\limits^{{+\infty}}_{-\infty} d^{3} {\bf k_{1}} \int\limits^{+\infty}_{0} d^{3} {\bf k_{2}}\nonumber  \\ 
          \!\! & & \!\! \Big\{ J^{+} e^{(iw_{1}+w_{2})t} \hat{a}({\bf k_{1}, k_{2}}) \big[\frac{1}{i{\bf k}_{1}-{\bf k}_{2}-(i{\bf P}_{1}+{\bf P}_{2})} + \frac{1}{-i{\bf k}_{1}-{\bf k}_{2}-(-i{\bf P}_{1}+{\bf P}_{2})} \big] \nonumber \\
     \!\! & + & \!\! J^{-} e^{-(iw_{1}-w_{2})t} \hat{b}^{\dag}({\bf k_{1}, k_{2}}) \big[\frac{1}{i{\bf k}_{1}-{\bf k}_{2}-(-i{\bf P}_{1}+{\bf P}_{2})} + \frac{1}{-i{\bf k}_{1}-{\bf k}_{2}-(i{\bf P}_{1}+{\bf P}_{2})} \big] \Big\}, \nonumber \\
      \label{trans1}
\end{eqnarray}
where the integration of the Poisson kernel (\ref{poisson}) has been used. To perform the $({\bf k_{1},k_{2}})$-integrals as contour integrals in the complex plane, we need to consider that there are two poles of the first order at $(i{\bf P}_{1}+{\bf P}_{2}, -i{\bf P}_{1}+{\bf P}_{2})$, as shown in the figure (\ref{polos}). The integration can be property analytically continued as follows,
\begin{equation}
     \int\limits^{{+\infty}}_{-\infty} d^{3}{\bf k}_{1} \int\limits^{{+\infty}}_{0} d^{3}{\bf k}_{2} =\int\limits^{{0}}_{-\infty} d^{3}{\bf k}_{1}\int\limits^{{+\infty}}_{0} d^{3}{\bf k}_{2} - \int\limits^{{0}}_{+\infty} d^{3}{\bf k}_{1} \int\limits^{{+\infty}}_{0} d^{3}{\bf k}_{2} = i\oint_{C_{+}} d {\bf Z} - i \oint_{C_{-}} d{\bf Z};
\label{bucle}
\end{equation}
where ${\bf Z} = -i{\bf k}_{1}+{\bf k}_{2}$, for the first term proportional to the operator $\hat{a}$ in (\ref{trans1}), and ${\bf Z} = i{\bf k}_{1}+{\bf k}_{2}$, for the second term. Explicitly we have that
\begin{equation}
     \int \frac{\hat{a}({\bf k}_{1},{\bf k}_{2})}{i{\bf k}_{1}-{\bf k}_{2}-(i{\bf P}_{1}+{\bf P}_{2})} = -i \big( \oint_{C_{+}} d{\bf Z} - \oint_{C_{-}} d{\bf Z}\big) \frac{\hat{a}({\bf k}_{1},{\bf k}_{2})}{{\bf Z}-{\bf Z}_{0}} = 2\pi \hat{a} ({\bf P}_{1},-{\bf P}_{2});
\label{integral1}
\end{equation}
where ${\bf Z}=-i{\bf k}_{1}+{\bf k}_{2}$, and ${\bf Z}_{0}=-i{\bf P}_{1}-{\bf P}_{2}$; note that the integration over $C_{-}$ vanishes, since the pole ${\bf Z}_{0}$ lies outside the region enclosed by $C_{-}$; similarly for the second term we have that,
\begin{equation}
     \int \frac{\hat{a}({\bf k}_{1},{\bf k}_{2})}{-i{\bf k}_{1}-{\bf k}_{2}-(-i{\bf P}_{1}+{\bf P}_{2})} = -i (\oint_{C_{+}} d{\bf Z} - \oint_{C_{-}} d{\bf Z}) \frac{\hat{a}({\bf k}_{1},{\bf k}_{2})}{{\bf Z}-{\bf Z}'_{0}} = 2\pi \hat{a} ({\bf P}_{1},-{\bf P}_{2});
\label{integral2}
\end{equation}
where ${\bf Z}=i{\bf k}_{1}+{\bf k}_{2}$, and ${\bf Z}'_{0}=i{\bf P}_{1}-{\bf P}_{2}$. Now the integration over $C_{+}$ vanishes due to the pole $Z'_{0}$ is enclosed by $C_{-}$.

Similarly along the same lines we have that
\begin{equation}
     \int \frac{\hat{b}^{\dag}({\bf k}_{1},{\bf k}_{2})}{i{\bf k}_{1}-{\bf k}_{2}-(-i{\bf P}_{1}+{\bf P}_{2})} = 2\pi \hat{b}^{\dag} (-{\bf P}_{1},-{\bf P}_{2}) = \int \frac{\hat{b}^{\dag}({\bf k}_{1},{\bf k}_{2})}{-i{\bf k}_{1}-{\bf k}_{2}-(i{\bf P}_{1}+{\bf P}_{2}))}.
\label{integral3}
\end{equation}
With these integrals the expression (\ref{trans1}) reduces to
\begin{equation}
     \int\limits^{+\infty}_{-\infty} d^{3}{\bf x} e^{(i{\bf P}_{1}-{\bf P}_{2})\cdot {\bf x}} \hat{\psi} ({\bf x},t) =  2\pi  N_{k_{1}} N_{k_{2}} \big[ J^{+} e^{(iw_{1}+w_{2})t} \hat{a} ({\bf P}_{1}, -{\bf P}_{2}) + J^{-} e^{(-iw_{1}+w_{2})t} \hat{b}^{\dag} (-{\bf P}_{1}, -{\bf P}_{2}) \big].
\label{integral4}     
\end{equation}     
A similar computation leads to
\begin{eqnarray}
     \int\limits^{+\infty}_{-\infty} d^{3}{\bf x} e^{(i{\bf P}_{1}-{\bf P}_{2})\cdot {\bf x}} \hat{\pi} ({\bf x},t) =- 2\pi  N_{k_{1}} N_{k_{2}} \Big[ (w_2-iw_1)J^{-} e^{(-iw_{1}+w_{2})t} \hat{a}^{\dag} (-{\bf P}_{1}, -{\bf P}_{2})\nonumber  \\+(w_2+iw_1) J^{+} e^{(iw_{1}+w_{2})t} \hat{b} ({\bf P}_{1}, -{\bf P}_{2}) \Big].
\label{integral5}     
\end{eqnarray}  

\begin{figure}[H]
  \begin{center}
    \includegraphics[width=.4\textwidth]{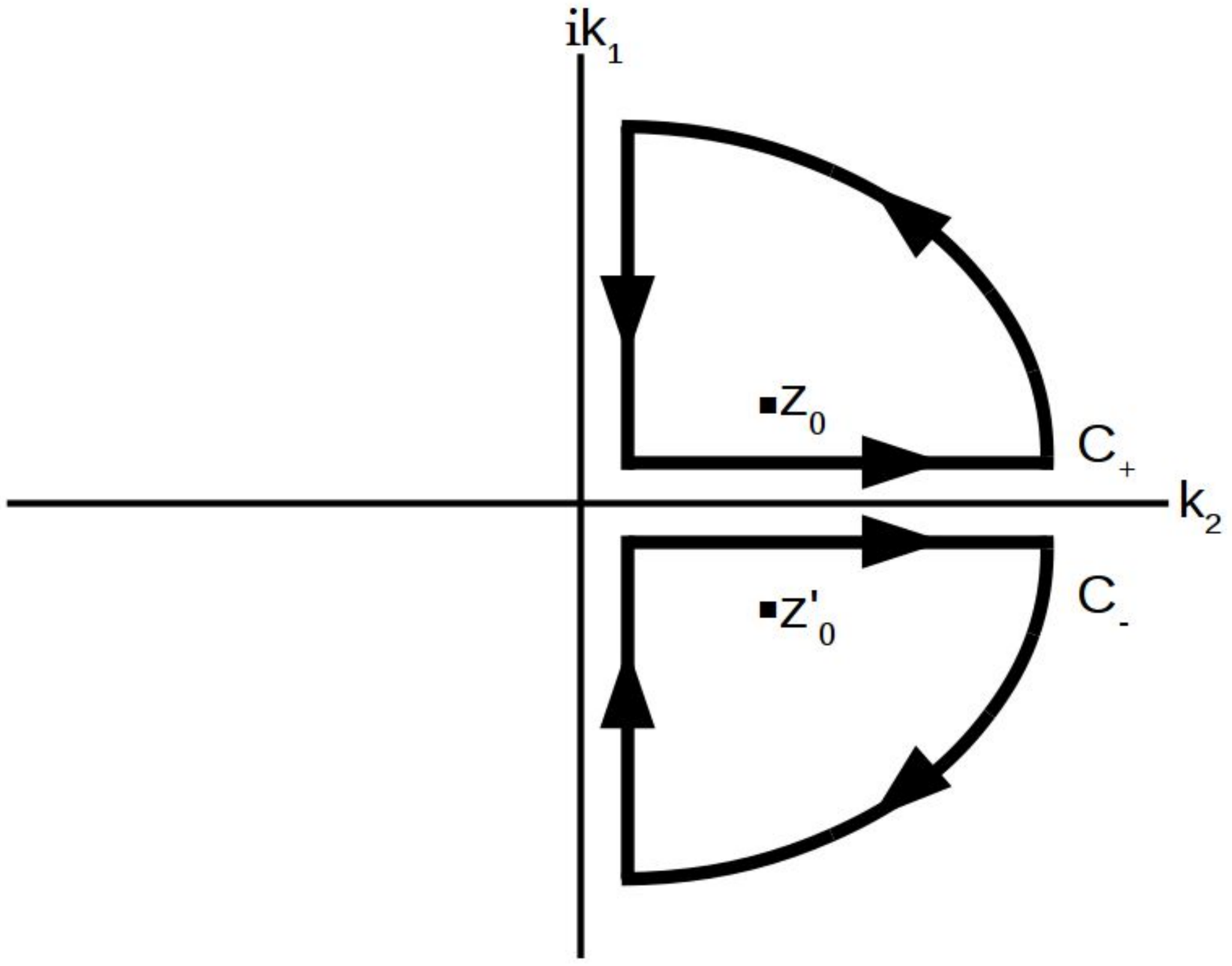}
  \caption{ There are two poles at ${\bf Z}_0=-i{\bf P_1-P_2}$, and ${\bf Z}^{'}_0=i{\bf P_1-P_2}$.}  
   \label{polos}
  \end{center}
\end{figure}

\section{Observables as quantum operators}
\label{quantum}

The expression $\pi\overline{\pi}$ is commutative as classical functional, and similarly for its quantum version in the case of the conventional $U(1)$ field theory, due basically to the use of the harmonic oscillators algebra. However, in the present scheme, such a product of operators has a nontrivial commutator; in order to promote the product $\pi\overline{\pi}$ as an operator-value distribution, it must be converted into the anti-commutator for the corresponding operators,
\begin{equation}
     \pi\overline{\pi} \rightarrow \frac{1}{2} \hat{\pi}\hat{\pi}^{\dag} + \frac{1}{2} \hat{\pi}^{\dag}\hat{\pi} \equiv  \frac{1}{2}\{\hat{\pi} , \hat{\pi}^{\dag} \},
\end{equation}
and then both the classical and quantum expressions are invariant under the corresponding interchange $(\pi\leftrightarrow \overline{\pi})$, and $(\hat{\pi}\leftrightarrow \hat{\pi}^{\dag})$. Now, using the expression (\ref{cm}) for obtaining the conjugate field operator to $\hat{\pi}$, the integration of the Poisson kernel (\ref{poisson}), and performing the resulting $({\bf k_{1}, k_{2}})$-integrations in a similar way to the contour integrations made in the Eqs. (\ref{bucle}) - (\ref{integral3}), one finds
\begin{eqnarray}
     \int d^{3}x \{\hat{\pi} ({\bf x},t), \hat{\pi}^{\dagger}({\bf x},t)\} \!\! & = & \!\! -4\pi |N_{k_{1}}|^{2} |N _{k_{2}}|^{2} \int\limits^{+\infty}_{-\infty} d^{3}{\bf k_{1}} \int\limits^{+\infty}_{0} d^{3}{\bf k_{2}}\Big[(w_{2}+iw_{1})^{2} J^{+} \{ \widehat{a}({\bf k_{1}, k_{2}}), \widehat{b}({\bf -k_{1}, -k_{2}}) \} \nonumber \\
     \!\! & & \!\! + h.c. \Big],
\label{H1}
\end{eqnarray}
where we have made use of the dispersion relations (\ref{dispersion}), and (\ref{dispersion2}); note that the addition of a Hermitian conjugate term generates an orthogonal term in the sense of the relation (\ref{ortho}).

Furthermore, along the same lines, one can prove that the additional terms in the Hamiltonian expression (\ref{canon}) will have exactly the same contribution displayed in the Eq. (\ref{H1}) in terms of creation and annihilation operators; therefore, the final Hamiltonian operator becomes
\begin{eqnarray}
     {\cal H} \!\! & = & \!\! \frac{1}{2} \int d^{3}{\bf x} \Big[ \{ \hat{\pi},\hat{\pi}^{\dag} \} + \{ \nabla\hat{ \psi} ,\nabla\hat{\psi}^{\dag} \} + m^{2} \{\hat{ \psi} ,\hat{\psi}^{\dag} \}\Big]  \nonumber \\
     \!\! & = & \!\! -4\pi |N_{k_{1}}|^{2} |N _{k_{2}}|^{2} \int\limits^{+\infty}_{-\infty} d^{3}{\bf k_{1}} \int\limits^{+\infty}_{0} d^{3}{\bf k_{2}}\Big[(w_{2}+iw_{1})^{2} J^{+} \{ \widehat{a}({\bf k_{1}, k_{2}}), \widehat{b}({\bf -k_{1}, -k_{2}}) \}     
    +h.c.\Big].\label{H2}
\end{eqnarray}
This form of the Hamiltonian as mixed pairing operators has appeared previuosly in the literature on quantum optics; the process of {\it nondegenerate parametric down-conversion}, a photon of the pump field is converted into two photons in modes $a$ and $b$, with a Hamiltonian of the form $\eta^*\hat{a}\hat{b}-\eta\hat{a}^{\dag}\hat{b}^{\dag}$, where $\eta$ defines the susceptibility of the medium \cite{gerry}. The quantum states generated by this process are entangled, and they have been studied in relation to experimental tests of quantum mechanics, and particularly in the quantum teleportation contexts; our Hamiltonian may be interpreted in the same sense, and we shall return to these issues in the concluding remarks.

From the classical expression (\ref{charge}), we can obtain the corresponding $\hat{Q}$-quantum operator,
\begin{equation}
     \hat{Q} = \frac{1}{2} \int d^{3}{\bf x} \Big(\{\widehat{\psi},\widehat{\pi} \} - \{\widehat{\psi}^{\dag}, \widehat{\pi}^{\dag} \} \Big),
\label{quantumQ}
\end{equation}
and in terms of the creation and annihilation operators we find that
\begin{eqnarray}
     \int d^{3}{\bf x} \{ \widehat{\psi},\widehat{\pi} \} \!\! & =  & \!\! -4\pi |N_{k_{1}}|^{2} |N_{k_{2}}|^{2} \cdot \int\limits^{+\infty}_{-\infty} d {\bf k}_{1} \int\limits^{+\infty}_{0} d{\bf k}_{2} \Big[ J^{+} (iw_{1}+w_{2}) \{ \hat{a} ({\bf k_{1}, k_{2}}), \hat{b}({\bf -k_{1}, -k_{2}})\} \nonumber \\
     \!\! & & \!\! + J^{-} (-iw_{1}+w_{2}) \{ \hat{a}^{\dag} ({\bf k_{1}, k_{2}}), \hat{b}^{\dag} ({\bf -k_{1}, -k_{2}})\} \Big];
\label{numberQ}
\end{eqnarray}
however, it is straightforward to prove that this expression is Hermitian,
\begin{equation}
     \int d^{3}{\bf x} \{\widehat{\psi},\widehat{\pi} \} = \int d^{3}{\bf x} \{\widehat{\psi}^{\dag}, \widehat{\pi}^{\dag} \},\quad \rightarrow \hat{Q}= 0;
\label{numberQ2}
\end{equation}
therefore, the global $\hat{Q}$-charge for all quantum states vanishes,and it is trivially conserved $[\hat{Q},\hat{H}]=0$.  Note that as well as the Hamiltonian operator, the $\hat{Q}$-charge is not a normal-ordered operator, since the ordering ambiguity is not present in the scheme at hand.

The conservation of the global $\hat{Q}$-charge implies that any quantum state will have the same number of $U(1)$-particles and of $SO(1,1)$-particles, and all quantum states will be {\it neutral}. In particular the vacuum state defined below, will contain nor $U(1)$-particles neither $SO(1,1)$-particles, and it will be annihilated trivially by the individual $\hat{Q_i}$-charge operators given by the operator (\ref{numberQ}) (see Eq. (\ref{individual})),
\begin{eqnarray}
   \hat{Q_i}\equiv \int d^{3}{\bf x} \{\widehat{\psi},\widehat{\pi} \}, \quad \hat{Q_i}|0> =0, \quad i=1,2;
\label{vacuumQ}
\end{eqnarray}
where the indices refers to the pair of charged fields in the expression (\ref{physcon}).

Similarly, the momentum operator can be obtained from the classical expression (\ref{momentum}),
\begin{equation}
     {\bf P} = \frac{1}{4} \int d^{3} {\bf x} [ \{\hat{\pi},\nabla\hat{\psi} \} + \{ \hat{\pi}^{\dag}, \nabla\hat{\psi}^{\dag} \} ],
\label{quantumP}
\end{equation}
where we have again a Hermitian expression, $  \int d^{3}{\bf x} \{ \hat{\pi}, \nabla\psi \}  =  \int d^{3}{\bf x} \{ \hat{\pi}^{\dag}, \nabla\hat{\psi}^{\dag} \}$, which explicitly reads
\begin{eqnarray}
     \int d^{3}{\bf x} \{ \hat{\pi}, \nabla\hat{\psi} \} \!\! & = & \!\! 4\pi |N_{K_{1}}|^{2} |N_{K_{2}}|^{2} \int\limits^{+\infty}_{-\infty} d{\bf k}_{1} \int\limits^{+\infty}_{0}d{\bf k}_{2} \Big[(w_{2}+iw_{1})(i{\bf k}_{1}+{\bf k}_{2}) J^{+} \{ \hat{a}({\bf k}_{1},{\bf k}_{2}), \hat{b}(-{\bf k}_{1}, -{\bf k}_{2}) \} + h.c.\Big].\nonumber\\
\label{quantumPP}
\end{eqnarray}
Additionally the angular momentum can be obtained from this expression for the linear momentum; however, the field carries no intrinsic angular momentum, as well as the individual fields, and then it will result into a spin-zero particle.

Now we are looking for a vacuum state that, besides satisfying the condition (\ref{vacuumQ}), must be translationally invariant, we mean annihilated by the operators $H$ and ${\bf P}$, and without invoking {\it normal ordering} of operators.

\section{The vacuum as a squeezed state: Bogoliubov transformations, and the elimination of normal ordering}
\label{vacuum}
Now, considering that we have two sets of annihilation and creation operators, $( \widehat{a},\widehat{a}^{\dag}; \widehat{b},\widehat{b}^{\dag})$, one may define the vacuum state $|0>$ in the usual way,
\begin{eqnarray}
     \hat{a} ({\bf k_{1}, k_{2}}) |0> = 0, \quad <0| \hat{a}^{\dag} ({\bf k_{1}, k_{2}}) =0;     \nonumber\\
       \hat{b} ({\bf k_{1}, k_{2}}) |0> = 0, \quad <0| \hat{b}^{\dag} ({\bf k_{1}, k_{2}}) =0;
\label{vac}
\end{eqnarray}
for all $({\bf k_{1}, k_{2}})$; however, this definition will lead to a trivial quantum field theory. We can see this fact as follows;
the expressions (\ref{cacgen}) imply that
\begin{equation}
     [\hat{a} ({\bf k}_{1}, {\bf k}_{2}), \hat{b} ({\bf k}'_{1}, {\bf k}'_{2})] =0, \qquad {\rm for} \qquad {\bf k}_{1} + {\bf k}'_{1}\neq 0, \qquad {\bf k}_{2} + {\bf k}'_{2} \neq 0;
     \label{vac1}
\end{equation}
and,
\begin{eqnarray}
     \!\! & & \!\! \int\limits^{+\infty}_{-\infty} d^{3} {\bf k}_{1} \int\limits^{+\infty}_{0} d{\bf k}_{2} [\hat{a}({\bf k}_{1}, {\bf k}_{2}) \hat{b} (-{\bf k}_{1}, -{\bf k}_{2}) - \hat{b} (-{\bf k}_{1}, -{\bf k}_{2}) \hat{a} ({\bf k}_{1}, {\bf k}_{2})] = \rho, \nonumber \\
     \!\! & & \!\! {\rm for} \quad  {\bf k}'_{1} = - {\bf k}_{1}, \qquad {\bf k}'_{2} = -{\bf k}_{2}; \label{vac2}
\end{eqnarray}
where we have integrated due to the divergence of the Dirac delta for vanishing argument. Therefore, the action of this integral operator on the vacuum state $|0>$ reads
\begin{equation}
     \int\limits^{+\infty}_{-\infty} d^{3} {\bf k}_{1} \int\limits^{+\infty}_{0} d{\bf k}_{2} [\hat{a}({\bf k}_{1}, {\bf k}_{2}) \hat{b} (-{\bf k}_{1}, -{\bf k}_{2}) |0> - \hat{b} (-{\bf k}_{1}, -{\bf k}_{2}) \hat{a} ({\bf k}_{1}, {\bf k}_{2}) |0> ] = \rho |0>, \label{vac3}
\end{equation}
the left hand side vanishes trivially according to the definition (\ref{vac}), and thus on the right hand side we have that $\rho =0$, which leads to trivial commutators (\ref{cacgen}) and (\ref{commu}), and therefore to a trivial QFT.

One may generalize the definition (\ref{vac}) in such a way that the vacuum state, is an annihilation operator coherent state,
\begin{equation}
     \hat{a}({\bf k}_{1}, {\bf k}_{2}) |0> = a|0>, \qquad \hat{b}({\bf k}_{1}, {\bf k}_{2}) |0> = b |0>, \label{coherent}
\end{equation}
where $a,b \in {\cal H}$, hence, the left hand side of Eq. (\ref{vac3}) reduces essentially to $(ab - ba)|0> =0$, since the ring ${\cal H}$ is commutative; again, on the right hand side in Eq. (\ref{vac3}) we will have that $\rho =0$, and again a trivial QFT.

The Eqs. (\ref{coherent}) can be obtained from Bogoliubov translations for coherent states
\begin{eqnarray}
     \hat{a} \!\! & \rightarrow & \!\! \breve{a}(a) \equiv \hat{a}+a, \qquad a \in {\cal H}, \nonumber \\
     \hat{b} \!\! & \rightarrow & \!\! \breve{b}(b) \equiv \hat{b}+b, \qquad b \in {\cal H},
     \label{bosontrans}
\end{eqnarray}
which preserve the commutator relations (\ref{cacgen}), and thus correspond to canonical transformations.
Specifically, since the canonical commutators (\ref{cacgen}) are preserved, then the Eq. (\ref{vac3}) is also valid for the new set of operators, with the substitutions $\hat{a}\rightarrow\breve{a}$, and $\hat{b}\rightarrow\breve{b}$. 
Therefore, the corresponding definitions of the new vacuum state given in Eq. (\ref{vac}), and (\ref{coherent}), in terms of the new operators $(\breve{a},\breve{b})$, also lead to a trivial quantum field theory with $\rho =0$. Hence, a Bogoliubov transformation does not allow us, unlike the known applications of such a transformation, to re-define a new vacuum without trivializing the quantum formulation.
 The reason is that in the known applications the commutator of operators that annihilate the vacuum state, vanishes trivially, as opposed to the case at hand; this fact will lead to looking for a novel definition for the vacuum state.

The appropriate definition of the vacuum state will require to invoke the concept of {\it squeezed coherent states}, which are of common use in quantum optics, and represent a generalization of the coherent states.
A squeezed state is an eigenstate of an operator that is a linear combination of creation and annihilation operators, which corresponds to a transformation that preserves the canonical commutation relations.\footnote{ Such an isomorphism of the commutation relation algebra appeared originally in the work of N. Bogoliubov on superfluidity \cite{bogo}.}

In the case at hand we propose the following linear transformation on the pair $(\hat{a},\hat{b})$ and their Hermitian conjugates,
\begin{eqnarray}
     \left( \begin{array}{c}
     \breve{a} \\
     \breve{a}^{\dag} \\
     \end{array} \right)
     = \left( \begin{array}{cc}
     \alpha_{1} & \alpha_{2} \\
     \overline{\alpha}_{2} & \overline{\alpha}_{1} \\
     \end{array} \right)
     \left( \begin{array} {c}
     \hat{a} \\
     \hat{a}^{\dag} \\
     \end{array} \right), \quad
\left( \begin{array}{c}
     \breve{b} \\
     \breve{b}^{\dag} \\
     \end{array} \right)
     = \left( \begin{array}{cc}
     \beta_{1} & \beta_{2} \\
     \overline{\beta}_{2} & \overline{\beta}_{1} \\
     \end{array} \right)
     \left( \begin{array} {c}
     \hat{b} \\
     \hat{b}^{\dag} \\
     \end{array} \right), \label{iso}
\end{eqnarray}
where the coefficients $(\alpha_{1}, \alpha_{2}, \beta_{1}, \beta_{2})$ are arbitrary elements of the ring ${\cal H}$; since these transformations are not mixing operators of different type, correspond in this sense to single-mode squeezing transformations. 

Additionally we assume non-trivial commutation relations,
\begin{eqnarray}
   \big [ \hat{a}({\bf k_{1},k_{2}}), \hat{a}^{\dag}(\bf{p_{1},p_{2}}) \big]\!\! & = & \!\! \varrho \delta (\bf{k_{1}\pm p_{1}})\delta (k_{2}\pm p_{2}), \label{aad} \\
        \big [\hat{b}({\bf k_{1},k_{2}}), \hat{b}^{\dag}(\bf{p_{1},p_{2}})  \big]\!\! & = & \!\! \zeta \delta (\bf{k_{1}\pm p_{1}})\delta (k_{2}\pm p_{2}), \label{bbd} \\
      \big [ \hat{a}({\bf k_{1},k_{2}}), \hat{b}^{\dag}({\bf{p_{1},p_{2}}}) \big]& = &\varepsilon  \delta (\bf{k_{1}\pm p_{1}})\delta (k_{2}\pm p_{2}), \label{abd}\\
       \big [ \hat{a}({\bf k_{1},k_{2}}), \hat{b}(\bf{p_{1},p_{2}})  \big]\!\! & = & \!\! \rho \delta (\bf{k_{1} + p_{1}})\delta (p_{2} + p_{2}), \label{ab}
\end{eqnarray}
where the coefficients $\varrho$, and $\zeta$ are Hermitian, and  $\varepsilon$, and, $\rho$ have the general form (\ref{ring}); the commutator (\ref{ab}) is simply the Eq. (\ref{cacgen}).

Hence, the preservation of this algebra of operators will restrict the coefficients of the transformation in the following form; the transformations (\ref{iso}) imply in relation to the commutatos (\ref{aad}), and (\ref{bbd}) that
\begin{equation}
     [\breve{a}, \breve{a}^{\dag}] = \big(|\alpha_{1}|^{2} - |\alpha_{2}|^{2}\big)[\hat{a},\hat{a}^{\dag}], \qquad [\breve{b},\breve{b}^{\dag}] = \big(|\beta_{1}|^{2} - |\beta_{2}|^{2}\big)[\hat{b},\hat{b}^{\dag}];
     \label{ABD}
\end{equation}
and thus the restrictions
\begin{equation}
     |\alpha_{1}|^{2} - |\alpha_{2}|^{2} =1, \qquad |\beta_{1}|^{2} - |\beta_{2}|^{2}=1,
     \label{determinant}
\end{equation}
are necessary for preserving the commutators (\ref{aad}), and (\ref{bbd}), and to establish thus part of the isomorphism; the rest of the commutators will be considered 
in short. Note that the above expressions fix incidentally the determinant of the transformation matrices (\ref{iso}), which turn out to be special. 
Therefore, we can obtain the inverse transformations, and to rewrite 
 the theory in terms of the new operators of creation and annihilation $(\breve{a}, \breve{a}^{\dag}, \breve{b}, \breve{b}^{\dag})$; 
\begin{eqnarray}
     \left( \begin{array}{c}
     \hat{a} \\
     \hat{a}^{\dag} \\
     \end{array} \right) 
     = \left( \begin{array}{cc}
     \overline{\alpha}_{1} & -\alpha_{2} \\
     -\overline{\alpha}_{2} & \alpha_{1} \\
     \end{array} \right)
     \left( \begin{array} {c}
     \breve{a} \\
     \breve{a}^{\dag} \\
     \end{array} \right), \quad
     \left( \begin{array}{c}
     \hat{b} \\
     \hat{b}^{\dag} \\
     \end{array} \right)
     = \left( \begin{array}{cc}
     \overline{\beta}_{1} & -\beta_{2} \\
     -\overline{\beta}_{2} & \beta_{1} \\
     \end{array} \right)
     \left( \begin{array} {c}
     \breve{b} \\
     \breve{b}^{\dag} \\
     \end{array} \right). \label{isoinverse}
\end{eqnarray}
The observables and expressions of interest contain basically the anti-commutator $J^{+}\{\hat{a},\hat{b}\}$, and simultaneously its Hermitian conjugate:
\begin{eqnarray}
J^{+}\{\hat{a},\hat{b}\}=J^{+} \overline{\alpha}_{1} \overline{\beta}_{1} \{\breve{a},\breve{b}\}-J^{+} \overline{\alpha}_{1}\beta_{2} \{\breve{a},\breve{b}^{\dagger}\}-J^{+}\alpha_{2}\overline{\beta}_{1}\{\breve{a}^{\dagger},\breve{b}\}+J^{+}\alpha_{2}\beta_{2}\{\breve{a}^{\dagger},\breve{b}^{\dagger}\},
\label{subs1}\\
J^{-}\{\hat{a}^{\dagger},\hat{b}^{\dagger}\}=J^{-} {\alpha}_{1} {\beta}_{1} \{\breve{a}^{\dagger},\breve{b}^{\dagger}\}-J^{-} {\alpha}_{1}\overline{\beta}_{2} \{\breve{a}^{\dagger},\breve{b}\}-J^{-}\overline{\alpha}_{2}{\beta}_{1}\{\breve{a},\breve{b}^{\dagger}\}+J^{-}\overline{\alpha}_{2}\overline{\beta}_{2}\{\breve{a},\breve{b}\},
\label{subs2}
\end{eqnarray}
now we can impose restrictions on the projections on the basis $(J^-,J^+)$ of the parameters $(\alpha, \beta)$ that eliminate the terms involving only creation operators, which 
will induce under Hermitian conjugation, second-order creation operators; considering the possible action
on the vacuum state (still undefined), such operators will generate excited states. A possible choice is (see Eq. (\ref{projJ})),
\begin{eqnarray}
J^{+}\beta_{2}=0, \quad (J^{-}\bar{\beta}_{2}=0); \quad J^{-}{\alpha}_{1}=0, \quad (J^{+}\bar{\alpha}_{1}=0); \quad |\alpha_1|^2=0, \quad |\beta_2|^2=0;
\label{projectionAB}
\end{eqnarray}
alternatively one may impose $J^{+}\alpha_{2}=0$, and $J^{-}\beta_{1}=0$. However, both choices exclude each other, due to the restrictions (\ref{determinant}), and thus once we have chosen the constraints (\ref{projectionAB}), automatically one must impose the complementary constraints 
\begin{eqnarray}
J^{+}\alpha_{2}\neq 0, \quad {\rm and}\quad  J^{-}\beta_{1}\neq 0.
\label{antiAB}
\end{eqnarray}
Now, the restrictions (\ref{projectionAB}) lead then to the vanishing of the corresponding terms
that involve purely annihilation operators, and additionally of the second terms in the Eqs. (\ref{subs1}), and  (\ref{subs2}),
\begin{eqnarray}
J^{+}\{\hat{a},\hat{b}\}=-J^{+}\alpha_{2}\overline{\beta}_{1}\{\breve{a}^{\dagger},\breve{b}\},
\label{quasivac}\\
J^{-}\{\hat{a}^{\dagger},\hat{b}^{\dagger}\}=-J^{-}\overline{\alpha}_{2}{\beta}_{1}\{\breve{a},\breve{b}^{\dagger}\}.
\label{quasivac2}
\end{eqnarray} 
hence, up to this point, we have reduced the expressions for observables to anti-commutators involving one-order annihilation and one-order creation operators; we shall return to these expression below (see Eqs. (\ref{quasi3}), and (\ref{quasi4})).
 In order to go further, we need to complete the analysis of the isomorphism between commutators 
 (\ref{abd}), and (\ref{ab}),
\begin{eqnarray}
[\breve{a},\breve{b}]=  \alpha_{1}\beta_{1}[\hat{a},\hat{b}]+\alpha_2\beta_{2}[\hat{a}^{\dagger},\hat{b}^{\dagger}]
 +\alpha_{1}\beta_{2}[\hat{a},\hat{b}^{\dagger}]+ \alpha_{2}\beta_{1}[\hat{a}^{\dagger},\hat{b}];
\label{abtransformed}
\end{eqnarray}
and
\begin{eqnarray}
[\breve{a},\breve{b}^{\dagger}]= {\alpha}_{1}\overline{\beta}_{2}[\hat{a},\hat{b}]+\alpha_2\bar{\beta}_{1}[\hat{a}^{\dagger},\hat{b}^{\dagger}]+{\alpha}_{1}\bar{\beta}_{1}[\hat{a},\hat{b}^{\dagger}]+\alpha_2\overline{\beta}_{2}[\hat{a}^{\dagger},\hat{b}];
\label{abdtransformed}
\end{eqnarray}
respectively. 
These commutators must be complemented with their Hermitian conjugate; note that as opposed to the Eq. (\ref{ABD}), the above expressions do not represent an one-to-one mapping between commutators, and we shall require a diagonalization of the following transformation matrix,
\begin{eqnarray}
     \left( \begin{array}{c}
     \big[\breve{a},\breve{b}\big] \\
     \big[\breve{a}^{\dag},\breve{b}^{\dag}\big]\\
     \big[\breve{a},\breve{b}^{\dag}\big]\\
     \big[\breve{a}^{\dag},\breve{b}\big]\\
     \end{array} \right)
     = \left( \begin{array}{cccc}
     \alpha_{1}\beta_{1} & \alpha_{2}\beta_{2} & \alpha_{1}\beta_{2} & \alpha_{2}\beta_{1} \\
     \overline{\alpha}_{2}\overline{\beta}_{2} & \overline{\alpha}_{1}\overline{\beta}_{1} & \overline{\alpha}_{2}\overline{\beta}_{1} & \overline{\alpha}_{1}\overline{\beta}_{2} \\
     \alpha_{1}\overline{\beta}_{2} & \alpha_{2}\overline{\beta}_{1} & \alpha_{1}\overline{\beta}_{1} & \alpha_{2}\overline{\beta}_{2} \\
     \overline{\alpha}_{2}\beta_{1} & \overline{\alpha}_{1}\beta_{2} & \overline{\alpha}_{2}\beta_{2} & \overline{\alpha}_{1}\beta_{1} \\
     \end{array} \right)
     \left( \begin{array} {c}
    \big[ \hat{a},\hat{b}\big]\\
    \big[ \hat{a}^{\dag},\hat{b}^{\dag}\big]\\
    \big[ \hat{a},\hat{b}^{\dag}\big]\\
    \big[ \hat{a}^{\dag},\hat{b}\big]\\
     \end{array} \right). \label{matrixiso}
\end{eqnarray}
The diagonalization will be simplified by incorporating the restrictions (\ref{projectionAB}), using the decomposition (\ref{directsum}) for hypercomplex numbers,
\begin{eqnarray}
     M \equiv \left( \begin{array}{cccc}
     J^{+}\alpha_{1}\beta_{1} & J^{-}\alpha_{2}\beta_{2} & 0 & \alpha_{2}\beta_{1}\\
     J^{+}\overline{\alpha}_{2}\overline{\beta}_{2} & J^{-}\overline{\alpha}_{1}\overline{\beta}_{1} & \overline{\alpha}_{2}\overline{\beta}_{1} & 0\\
     J^{+}\alpha_{1}\overline{\beta}_{2} & \alpha_{2}\overline{\beta}_{1} & J^{+}\alpha_{1}\overline{\beta}_{1} & J^{+}\alpha_{2}\overline{\beta}_{2}\\
     \overline{\alpha}_{2}\beta_{1} & J^{-}\overline{\alpha}_{1}\beta_{2} & J^{-}\overline{\alpha}_{2}\beta_{2} & J^{-}\overline{\alpha}_{1}\beta_{1}\\
     \end{array} \right). \label{matrixrest}
\end{eqnarray}
Furthermore, the determinant of this matrix  is given by $(|\alpha_2|^2|\beta_1|^2)^2=1$, where the iquality follows from the Eqs. (\ref{determinant}), and the constraints (\ref{projectionAB}).
Hence, a possible diagonalization of this matrix will lead to a realization of the isomorphism of the algebra of commutators in the following form
\begin{eqnarray}
     \left( \begin{array}{c}
     \big[\breve{a},\breve{b}\big] \\
     \big[\breve{a}^{\dag},\breve{b}^{\dag}\big]\\
    \big[ \breve{a},\breve{b}^{\dag}\big]\\
     \big[\breve{a}^{\dag},\breve{b}\big]\\
     \end{array} \right)
     \rightarrow \left( \begin{array}{cccc}
     \lambda_{1} &  &  &  \\
     & \lambda_{2} & & \\
     & & \lambda_{3} & \\
     & & & \lambda_{4} \\
     \end{array} \right)
     \left( \begin{array} {c}
     \big[\hat{a},\hat{b}\big]\\
     \big[\hat{a}^{\dag},\hat{b}^{\dag}\big]\\
     \big[\hat{a},\hat{b}^{\dag}\big]\\
     \big[\hat{a}^{\dag},\hat{b}\big]\\
     \end{array} \right), \label{diagonal}
\end{eqnarray}
where the $\lambda'$s correspond in general to elements of the ring, with $\lambda_2=-\overline{\lambda_1}$, and $\lambda_4=-\overline{\lambda_3}$, since
$ \big[\hat{a}^{\dag},\hat{b}^{\dag}\big]=-\big[\hat{a},\hat{b}\big]^{\dagger}$, and $\big[\hat{a},\hat{b}^{\dag}\big]=-\big[\hat{a}^{\dag},\hat{b}\big]^{\dagger}$. Since we are working on a ring, a spectral theory is not available to apply directly, and the diagonalization can not be made through a conventional procedure.

The characteristic polynomial can be determined from the matrix (\ref{matrixrest}), considering basically the orthogonality relation (\ref{ortho}),
\begin{eqnarray}
     |M -\lambda I| \!\! & = & \!\! \lambda^{4} - (\beta_{1}+\overline{\beta}_{1}) (J^{+}\alpha_{1} + J^{-}\bar{\alpha}_{1})\lambda^{3} + \Big[ |\beta_{1}|^{2} [(J^{+}\alpha_{1})^{2} + (J^{-}\overline{\alpha}_{1})^{2}]- |\alpha_{2}|^{2} [(\beta_{1})^{2} + (\overline{\beta}_{1})^{2} ]\Big] \lambda^{2} \nonumber \\
     \!\! & + & \!\! |\alpha_{2}|^{2} \Big[ (\beta_{1})^{2} [J^{+} \alpha_{1}\overline{\beta}_{1} + J^{-} \overline{\alpha}_{1}\overline{\beta}_{1}] + c.c. \Big] \lambda + (|\beta_{1}|^{2} |\alpha_{2}|^{2})^{2}; \label{charpol}
\end{eqnarray}
all coefficients of this quartic polynomial are Hermitian, and thus the finding of the roots will reduce to extract roots of Hermitian elements of the ring, which has been outlined previously in the section \ref{hf}.

For a quartic polynomial of the form \cite{irving}
\[
     ax^{4}+ bx^{3} + cx^{2} +dx + e = 0,
\]
the four roots can be obtained from the general formulas
\begin{eqnarray}
     (x_{1})^{\pm} \!\! & = & \!\! - \frac{b}{4a}- S \pm \frac{1}{2} \sqrt{-4S^{2}-2P+\frac{T}{S}}, \nonumber \\
     (x_{2})^{\pm} \!\! & = & \!\! - \frac{b}{4a}+S \pm \frac{1}{2} \sqrt{-4S^{2}-2P-\frac{T}{S}}, \label{4root}
\end{eqnarray}
where
\begin{eqnarray}
     P \!\! & = & \!\! \frac{8ac-3b^{2}}{8a^{2}}, \qquad T = \frac{b^{3}-4abc+8a^{2}d}{8a^{3}}, \nonumber \\
       \triangle_{0} \!\! & = & \!\! c^{2} -3bd+12ae, \qquad
       \triangle_{1}  = 2c^{3} -9bcd+27b^{2}e+27ad^{2}-72ace, \label{PDelta}
\end{eqnarray}
hence one can calculate,
\begin{eqnarray}
 Q = \sqrt[3]{\frac{\triangle_{1}+\sqrt{\triangle_{1}^{2}-4\triangle^{3}_{0}}}{2}},
 \label{Q}
 \end{eqnarray}
 and later
 \begin{eqnarray}
  S \!\! & = & \!\! \frac{1}{2} \sqrt{-\frac{2}{3}P + \frac{1}{3a}\Big(Q+\frac{\triangle_{0}}{Q}\Big)}.
  \label{S}
   \end{eqnarray}
Once we have identified the coefficients $(a,b,c,d,e)$ in the polynomial (\ref{charpol}), they must be substituted into the Eqs. (\ref{PDelta}), leading to very complicated expressions, which in their turn must be substituted into the Eqs. (\ref{Q}), and (\ref{S}); at the end, all expressions must be considered 
in the Eqs. (\ref{4root}) for obtaining the roots. Instead of performing long algebraic computations, we shall focus in proving that such roots exist, using essentially the extraction of roots of Hermitian numbers developed in section \ref{hf}. Hence, considering that $a=1$,  the polynomials in the Eqs. (\ref{PDelta}) for 
the Hermitian coefficients $(b,c,d,e)$ become also Hermitian, since they involve products, powers, and sums of Hermitian numbers. Similarly the polynomial  $\triangle_{1}^{2}-4\triangle^{3}_{0}$ and its square root
in Eq. (\ref{Q}) turn out to be Hermitian, according to the rules for powers and extracting of square roots in the section \ref{hf}. Since the Hermitian character is maintained under the cubic root, then $Q$ is Hermitian; furthermore, the expression $\frac{1}{Q}$ in the Eq. (\ref{S}) may  make no sense in a ring, however it is well defined for a Hermitian number according to the Eq. (\ref{inv}), and corresponds also to a Hermitian number. Under these considerations 
both $S$ and its inverse $\frac{1}{S}$ are Hermitian; therefore the roots (\ref{4root}) for the polynomial (\ref{charpol}) are Hermitian.

Put all together, including Eq. (\ref{ABD}), and (\ref{diagonal}), and considering the norms in Eqs. (\ref{projectionAB}), we have for the full isomorphism,
\begin{eqnarray}
     \left( \begin{array}{c}
     \big[ \breve{a},\breve{a}^{\dag}\big] \\
     \big[\breve{b},\breve{b}^{\dag}\big] \\
     \big[\breve{a},\breve{b}\big] \\
     \big[\breve{a}^{\dag},\breve{b}^{\dag} \big]\\
     \big[ \breve{a},\breve{b}^{\dag} \big] \\
     \big[\breve{a}^{\dag},\breve{b} \big] \\
     \end{array} \right)
     = \left( \begin{array}{cccccc}
     -|\alpha_{2}|^{2} &  &  & & &  \\
     & |\beta_{1}|^{2} & & & & \\
     & & \lambda_{1} & & & \\
     & & & -\lambda_{1} & & \\
     & & & & \lambda_{3} & \\
     & & & & & -\lambda_{3} \\
     \end{array} \right)
     \left( \begin{array} {c}
     \big[ \hat{a},\hat{a}^{\dag} \big] \\
     \big[ \hat{b},\hat{b}^{\dag} \big] \\
     \big[ \hat{a},\hat{b} \Big] \\
     \big[ \hat{a}^{\dag},\hat{b}^{\dag} \big] \\
     \big[ \hat{a},\hat{b}^{\dag} \big] \\
     \big[ \hat{a}^{\dag},\hat{b} \big] \\
     \end{array} \right), \label{fulliso}
\end{eqnarray}
where all eigenvalues are Hermitian.

We return now to the expressions (\ref{quasivac}), and (\ref{quasivac2}), in terms of which the observables are cons\-truc\-ted; in order to locate the annihilation
operators to the right hand side in both expressions we only require the vanishing of the following commutator
\begin{equation}
\big [ \breve{a}({\bf k_{1},k_{2}}), \breve{b}^{\dag}({\bf{p_{1},p_{2}}}) \big]  = 0, 
 \label{vanishabd} 
 \end{equation}
by fixing $\varepsilon=0$ in the Eq. (\ref{abd}); this simple constraint will allow us to avoid the ambiguities of {\it normal ordering}, since
\begin{eqnarray}
J^{+}\{\hat{a},\hat{b}\}=-J^{+}\alpha_{2}\overline{\beta}_{1}\{\breve{a}^{\dagger},\breve{b}\}=-2\alpha_{2}\overline{\beta}_{1}\breve{a}^{\dagger}(J^{+}\breve{b}),
\label{quasi3}\\
J^{-}\{\hat{a}^{\dagger},\hat{b}^{\dagger}\}=-J^{-}\overline{\alpha}_{2}{\beta}_{1}\{\breve{a},\breve{b}^{\dagger}\}=-2\overline{\alpha}_{2}{\beta}_{1}\breve{b}^{\dagger}(J^{-}\breve{a}).
\label{quasi4}
\end{eqnarray} 
Note that additionally we have grouped the projectors $(J^+, J^-)$ with the annihilation operators; the
Eqs. (\ref{quasi3}), and (\ref{quasi4}) suggest then the following definition of the new vacuum state,
\begin{eqnarray}
J^-\breve{a}|\breve{0}>=0, \quad J^+\breve{b}|\breve{0}>=0; \quad [J^-\breve{a},J^+\breve{b}]=0;
\label{newvac}
\end{eqnarray}
hence, only certain projections of the new annihilation operators are required for avoiding the trivialization of the quantum formulation; the vanishing of the commutator of operators that define the vacuum is a direct consequence of the orthogonality of the projectors $(J^+,J^-)$.

In terms of the new operators $(\breve{a},\breve{b})$ and the new vacuum, the projections on the basis $(J^+,J^-)$ of Eq. (\ref{vac3}) 
reduce to
\begin{eqnarray}
     \int\limits^{+\infty}_{-\infty} d^{3} {\bf k}_{1} \int\limits^{+\infty}_{0} d{\bf k}_{2} J^{+}\breve{b}({\bf -k_{1},-k_{2}}) \breve{a}({\bf k_{1},k_{2}})|\breve{0}> = -J^{+}{\rho}|\breve{0}>,\nonumber\\ 
      \int\limits^{+\infty}_{-\infty} d^{3} {\bf k}_{1} \int\limits^{+\infty}_{0} d{\bf k}_{2}J^{-}\breve{a}({\bf k_{1},k_{2}}) \breve{b}({\bf -k_{1},-k_{2}})|\breve{0}> = J^{-}{\rho}|\breve{0}> ; \label{vacsq}
\end{eqnarray}
since the projections $(J^{+}\rho ,J^{-}\rho )$ can be now different from zero, the new definition of vacuum will allow us to construct a non-trivial QFT.

At this point we emphazise that the commutator $[\hat{a},\hat{b}]$ must be always switched on, in order to have a nontrivial QFT; 
on the contrary, the commutator (\ref{vanishabd}) will be maintained always switched off, in order to avoid normal ordering; as we shall see, other consistency aspects of the present scheme will depend sensitively on this vanishing commutator. Additionally the single commutators $[\hat{a},\hat{a}^{\dag}]$, and $[\hat{b},\hat{b}^{\dag}]$ will be optionally switched on or switched off.

\section{Finite expectation values for observables and the vacuum structure}
\label{finite}

The vaccum state will be annihilated by the Hamiltonian (\ref{H2}) and by the momentum operator (\ref{quantumP}), described now in terms of the operators $(\breve{a},\breve{b},\breve{a}^\dagger, \breve{b}^\dagger)$,
\begin{eqnarray}
H  \!\! & =  & \!\! 8\pi |N_{k_{1}}|^2 |N_{k_{2}}|^2\Big[\alpha_{2}\overline{\beta}_{1}(\omega_{2}+i\omega_{1})^2\breve{a}^{\dagger}J^{+}\breve{b}+h.c.\Big], \quad H|\breve{0}>=0; \label{breveH}\\
{\bf P}  \!\! & =  & \!\! -8\pi |N_{k_{1}}|^2 |N_{k_{2}}|^2\Big[\alpha_{2}\overline{\beta}_{1}(\omega_{2}+i\omega_{1})(i{\bf k_{1}+k_{2}})\breve{a}^{\dagger}J^{+}\breve{b}+h.c.\Big], \quad {\bf P}|\breve{0}>=0; \label{breveP}
\end{eqnarray}
therefore the vacuum is translationally invariant in both senses, spatial and temporal. 

This Hamiltonian resembles the one that describes 
a beam splitter, with the form $k(\hat{a}\hat{b}^{\dag}+\hat{a}^{\dag}\hat{b})$, where $k$ is the coupling constant between the modes $a$ and $b$; the two output states of the beam splitter are uncorrelated. However, in the case at hand we shall obtain correlated states by squeezing the vacuum, and the possible applications in that sense are discussed in the concluding remarks. 

The density for the energy and momentum trivially vanish, explicitly we have
\begin{eqnarray}
     <\breve{0}| H |\breve{0}>=0, \quad <\breve{0}| {\bf P} |\breve{0}>=0.
     \label{vacE}
\end{eqnarray}
At this point it is mandatory to make a comparison with the U(1)-field theory, whose quantization is well known; the $U(1)$-Hamiltonian has the form $\hat{a}\hat{a}^{\dag} + \hat{b}^{\dag}\hat{b}$,  with nontrivial commutators $[\hat{a},\hat{a}^{\dag} ]$, and $[\hat{b}, \hat{b}^{\dag}]$. The strict use of the commutator 
$[\hat{a},\hat{a}^{\dag} ]=\delta$ in the first term will lead to an infinite contribution coming from the delta function. Such a divergence contains both {\it infrared} and {\it ultraviolet} divergences; the former comes from an infinite space, and it can be controlled by putting the system into a finite box, and by dividing by the total volume, one considers rather the energy density. However, the remaining integral is still divergent for large momenta, and a cut off
is necessary for controlling the UV divergences.
Normal ordering is required then for the first term in the $U(1)$-Hamiltonian for removing  the vacuum divergences.

In contrast, in the Hamiltonian (\ref{breveH}), crossing terms between creation and annihilation operators of the same type are absent; instead
crossing terms between creation and annihilation operators of different type are present, but with trivial commutator between them.
Therefore, any {\it ordering} of operators in the Hamiltonian (\ref{breveH}) does not have any effect.

Furthermore, we have shown in Section \ref{quantum} that the global charge for all compound particles vanishes (Eq. (\ref{numberQ2})); however the individual charge given by Eq.
(\ref{numberQ}) vanishes for the vacuum state, as expected,
\begin{eqnarray}
\hat{Q_i}|\breve{0}> =0, \quad i=1,2;
\label{individual}
\end{eqnarray}
according to the transformations (\ref{quasi3}), and (\ref{quasi4}), and the definition of the vacuum (\ref{newvac}); thus Eq. (\ref{vacuumQ}) makes sense now.

\section{Compound bosons, and number and energy eigen-states}
\label{fock}
 Furthermore, 
since the hyper-complex field (\ref{ring}) encodes two-complex scalar fields, there are no one-particle states, instead we have the two-particles states as the simplest particle states created by $ \breve{a}^{\dag}$ and $\breve{b}^{\dag}$,
\begin{eqnarray}
     \hat{a}^{\dag} ( {\bf k_{1}, k_{2}}) |0> = |{\bf k_{1}, k_{2}}>, \quad      <{\bf k_{1}, k_{2}}|= <0| \hat{a} ({\bf k_{1}, k_{2}});  \nonumber \\            
          \hat{b}^{\dag} ({\bf q_{1}, q_{2}}) |0> = |{\bf q_{1}, q_{2}}>, \quad <{\bf q_{1}, q_{2}}|= <0| \hat{b} ({\bf q_{1}, q_{2}});
          \label{2particle}
          \end{eqnarray}
these two-particle states are of different type, and we will refer them as ${\bf a}$-type and ${\bf b}$-type respectively. In order to simplify the quantum description, these two-particle states will be called {\it compound} particles; hence, the creation operators
create ${\bf a}$-type, and ${\bf b}$-type compound particles.

 Furthermore, the following four-particles states will contain one compound particle of each type,
\begin{eqnarray}
 \hat{a}^{\dag} ( {\bf k_{1}, k_{2}})  \hat{b}^{\dag} ({\bf q_{1}, q_{2}})|0> = |{\bf q_{1}, q_{2}};{\bf k_{1}, k_{2}}>,\nonumber \\
 \hat{b}^{\dag} ({\bf q_{1}, q_{2}}) \hat{a}^{\dag} ( {\bf k_{1}, k_{2}}) |0> =  |{\bf k_{1}, k_{2}};{\bf q_{1}, q_{2}}>,
 \label{4particle}
 \end{eqnarray}
these states are distinguishable to each  other under the interchange of compound particles of different types, due  to the nontrivial commutator (\ref{ab}) between the operators $ \hat{a}^{\dag}$, and $\hat{b}^{\dag}$.

Multi-(compound) particle states of the same type can be created by acting multiple times with the respective creation operator,
\begin{eqnarray}
 \hat{a}^{\dag} ( {\bf k_{1}, k_{2}}) \hat{a}^{\dag} ( {\bf k_{3}, k_{4}})...... \hat{a}^{\dag} ( {\bf k_{n}, k_{n+1}}) |0> = |{\bf k_{1}, k_{2}};.....;{\bf k_{n}, k_{n+1}}>,\nonumber\\
 \hat{b}^{\dag} ( {\bf q_{1}, q_{2}}) \hat{b}^{\dag} ( {\bf q_{3}, q_{4}})...... \hat{b}^{\dag} ( {\bf q_{n}, q_{n+1}}) |0> = |{\bf q_{1}, q_{2}};.....;{\bf q_{n}, q_{n+1}}>;
 \label{samespecie}
\end{eqnarray}
since the creation operators are acting $n$ times, these multi-particle states will have $n$ compound particles of the same type, and then $2n$ individual particles. Considering the  trivial commutator relations $[\hat{a}^{\dag}, \hat{a}^{\dag}]=0$, and $[\hat{b}^{\dag}, \hat{b}^{\dag}]=0$, we can see that each state in (\ref{samespecie}) is symmetric under the interchange 
of two compound particles of the same type; hence, a compound particle is a {\it boson}, and we have ${\bf a}$-type bosons and
${\bf b}$-type bosons.

Now we can excite simultaneously $n$ ${\bf a}$-type quanta and $m$ ${\bf b}$-type quanta, in an arbitrary order, generalizing the four-particle states (\ref{4particle}); for example,
\begin{eqnarray}
 \hat{a}^{\dag}_{1} \hat{b}^{\dag} _{1}... \hat{a}^{\dag}_{r} \hat{b}^{\dag} _{s}... \hat{b}^{\dag} _{m} \hat{a}^{\dag} _{n} |0> = 
 |1_a; 1_b; ....r_a;s_b;.....m_b;n_a>,\nonumber\\
 \hat{b}^{\dag}_{1} \hat{b}^{\dag} _{2}... \hat{b}^{\dag}_{r} \hat{a}^{\dag} _{s}... \hat{a}^{\dag} _{n} \hat{b}^{\dag} _{m}|0> = 
 |1_b; 2_b; ....r_b;s_a;.....n_a;m_b>,
 \label{hybrid}
\end{eqnarray}
etc; where $r_a=(k_r,k_{r+1})$, $s_b=(q_s, q_{s+1})$, and so on. However, unlike the pure-type states (\ref{samespecie}), the compound particles in the {\it hybrid} states (\ref{hybrid}) are not freely interchangeable, since two compound bosons of the same type susceptible to be interchanged, may be separated by a boson of the other type; hence, only compound particles of the same type and created by contiguous operators are interchangeable. Hybrid states are not symmetric under the interchange of particles, and we may have distinguishable hybrid states in which two bosons of the same type are interchanged.

The number operators are defined by
\begin{eqnarray}
     N_{a} \!\! & = & \!\! \int^{+\infty}_{-\infty} d^{3} {\bf k}_{1} \int^{+\infty}_{0} d^{3}{\bf k}_{2} \hat{a}^{\dag} ({\bf k}_{1},{\bf k}_{2}) J^- \hat{a} ({\bf k}_{1},{\bf k}_{2}), \qquad N_{a}|0>=0; \label{numbera} \\
     N_{b} \!\! & = & \!\! \int^{+\infty}_{-\infty} d^{3} {\bf k}_{1} \int^{+\infty}_{0} d^{3}{\bf k}_{2} \hat{b}^{\dag} ({\bf k}_{1},{\bf k}_{2}) J^+ \hat{b} ({\bf k}_{1},{\bf k}_{2}), \qquad   N_{b}|0>=0; \label{numberb}
\end{eqnarray}
where the insertions of the projectors $(J^{+},J^{-})$ are required in order to define  the old vacuum as that state that contains no particles, 
\begin{eqnarray}
J^- \hat{a}|0>=0, \qquad J^+ \hat{b}|0>=0, 
\label{oldvacuum}
\end{eqnarray}
in consistency with the Eqs. (\ref{newvac}).
Although the original vacuum state $|0>$ has no particles, the new vacuum state $|\breve{0}>$ will have, however,  an indefinite number of particles, whose statistics will be estudied in section \ref{bosonnum}.

The number operators satisfy the expected properties, considering the commutators (\ref{aad}), and (\ref{bbd}),
\begin{equation}
     \big[ N_{a}, \hat{a}^{\dagger} ({\bf q}_{1},{\bf q}_{2}) \big] = (J^-\varrho) \hat{a}^{\dagger} ({\bf q}_{1},{\bf q}_{2}), \qquad \big[ N_{b}, \hat{b}^{\dagger} ({\bf q}_{1},{\bf q}_{2}) \big] = (J^+\zeta )\hat{b}^{\dagger} ({\bf q}_{1},{\bf q}_{2}), \label{numberab}
\end{equation}
the relative sign in the arguments of the Dirac deltas in the commutators (\ref{aad}), and (\ref{bbd}), have been fixed to be 
$({\bf k}_1-{\bf p}_1)$ and $({\bf k}_2-{\bf p}_2)$.
Consequently the action on the excited states (\ref{samespecie}) is given by
\begin{eqnarray}
     N_{\bf{a}}| {\bf k}_{1},{\bf k}_{2}; \cdots ; {\bf k}_{n}, {\bf k}_{n+1}>_{\bf{a}} \!\! & = & \!\! (J^-\varrho ) n| {\bf k}_{1},{\bf k}_{2}; \cdots ; {\bf k}_{n},{\bf k}_{n+1}>_{\bf{a}}, \label{numberacting}\\
     N_{\bf{b}}| {\bf q}_{1},{\bf q}_{2}; \cdots ; {\bf q}_{n}, {\bf q}_{n+1}>_{\bf{b}} \!\! & = & \!\! (J^+\zeta ) n| {\bf q}_{1},{\bf q}_{2}; \cdots ; {\bf q}_{n},{\bf q}_{n+1}>_{\bf{b}}; \label{numberacting1}
\end{eqnarray}
therefore, $N_{a}$ and $N_{b}$ indeed count the number of (compound) particles of each type, and the pure-type states (\ref{samespecie}) are eigen-states of the number operators, which are traditionally termed as the Fock states; the sub-indices $\bf{a,b}$ in the eigenstates indicate the type of the modes excited, and consistently $N_{\bf{a}}|\cdot\cdot\cdot>_{\bf{b}}=0$, and 
 $N_{\bf{b}}|\cdot\cdot\cdot>_{\bf{a}}=0$, according to the vanishing commutator (\ref{vanishabd}) and the definition of the (old) vacuum
(\ref{oldvacuum}).
 Note however that, as opposed to the usual case, the eigen-states of the number operators do not correspond automatically to eigen-states for the Hamiltonian. Some squeezed states that we shall construct in the section \ref{disen} will be expanded in terms of Fock states; especifically
 we shall construct
 certain unitary operators that 
 generate the new vacuum $|\breve{0}>$ by squeezing the original one $|0>$.

Furthermore, the $J^-$-projected a-type eigenvalue equation (\ref{numberacting}) has its $J^+$-projected complement, as an eigenvalue equation, provided that the particles are fully massless. First we use the following commutator, 
which will determine the action of the $J^+$-projection of the Hamiltonian ( and other observables in terms of the old operators) on a-type single states,
\begin{eqnarray}
\int\limits^{+\infty}_{-\infty} d^{3}{\bf k_{1}} \int\limits^{+\infty}_{0} d^{3}{\bf k_{2}}\Big[(w_{2}+iw_{1})^{2} J^{+} \{ \widehat{a}({\bf k_{1}, k_{2}}), \widehat{b}({\bf -k_{1}, -k_{2}}) \}   
, \hat{a}^{\dag} ( {\bf q_{1}, q_{2}}) \hat{a}^{\dag} ( {\bf q_{3}, q_{4}})...... \hat{a}^{\dag} ( {\bf q_{n}, q_{n+1}})
\Big]=\nonumber \\
2J^+\varrho
\Big(
 \hat{a}^{\dag} ( {\bf q_{3}, q_{4}})...... \hat{a}^{\dag} ( {\bf q_{n}, q_{n+1}})
\hat{b}({\bf -q_{1}, -q_{2}})+
 \hat{a}^{\dag} ( {\bf q_{1}, q_{2}})...... \hat{a}^{\dag} ( {\bf q_{n}, q_{n+1}})
\hat{b}({\bf -q_{3}, -q_{4}})+\cdots +\nonumber\\
 \hat{a}^{\dag} ( {\bf q_{1}, q_{2}})...... \hat{a}^{\dag} ( {\bf q_{n-2}, q_{n-1}})
\hat{b}({\bf -q_{n}, -q_{n+1}})
\Big),\nonumber\\.
\label{eigen1}
\end{eqnarray}
where we have used basically the commutators $[\hat{a}, \hat{a}^{\dag}]=\varrho\delta$,  and
 $[\hat{a},\hat{b}^{\dag}]=0$; the later implies that 
 the annihilation $\hat{b}$ operator can be located to the right in a direct way,
and hence the action upon the original vacuum state, yields the vanishing of the right-hand side, due to the condition $J^+\hat{b}|0>=0$.
Therefore the action of the projection $J^{+}H$ of the original Hamiltonian operator (\ref{H2}), on the a-type multi-particle state reads
\begin{eqnarray}
\int\limits^{+\infty}_{-\infty} d^{3}{\bf k_{1}} \int\limits^{+\infty}_{0} d^{3}{\bf k_{2}}
\Big[(w_{2}+iw_{1})^{2} J^{+} \{ \widehat{a}({\bf k_{1}, k_{2}}), \widehat{b}({\bf -k_{1}, -k_{2}}) \} 
| {\bf q}_{1},{\bf q}_{2}; \cdots ; {\bf q}_{n}, {\bf q}_{n+1}>_{\bf{a}} =\nonumber\\
 \hat{a}^{\dag} ( {\bf q_{1}, q_{2}}) \hat{a}^{\dag} ( {\bf q_{3}, q_{4}})...... \hat{a}^{\dag} ( {\bf q_{n}, q_{n+1}})
\int\limits^{+\infty}_{-\infty} d^{3}{\bf k_{1}} \int\limits^{+\infty}_{0} d^{3}{\bf k_{2}}
(w_{2}+iw_{1})^{2} J^{+}  \hat{b}({\bf -k_{1}, -k_{2}})  \widehat{a}({\bf k_{1}, k_{2}})|0>,
\label{eigen2}
\nonumber\\
\end{eqnarray}
where we have omitted unessential constants, and on the right hand side, we have used again the vacuum condition $J^+\hat{b}|0>=0$; the integration on this side must be worked out. First, we rewrite the bilinear mixing operator as
\begin{eqnarray}
 \hat{b}({\bf -k_{1}, -k_{2}})  \widehat{a}({\bf k_{1}, k_{2}})=\lim_{\Delta\rightarrow 1}  \hat{b}\Big((\Delta-2){\bf k_{1}}, (\Delta-2){\bf k_{2}}\Big)  \widehat{a}({\bf k_{1}, k_{2}})\\
=\lim_{\Delta\rightarrow 1}\Big[\widehat{a}({\bf k_{1}, k_{2}})\hat{b}\Big((\Delta-2){\bf k_{1}}, (\Delta-2){\bf k_{2}}\Big)-\rho\delta\Big( (\Delta-1){\bf k_{1}}\Big)
\delta\Big( (\Delta-1){\bf k_{2}}\Big)
 \Big],
\label{eigen3}
\end{eqnarray}
where the commutator $[\hat{a},\hat{b}]=\rho\delta$ has been used. The substitution of the above expression into Eq. (\ref{eigen2}), and the use of the condition $J^+\hat{b}|0>=0$, 
lead to an eigenvalue equation for the multiparticle state,
\begin{eqnarray}
 (J^+) H| {\bf k}_{1},{\bf k}_{2}; \cdots ; {\bf k}_{n}, {\bf k}_{n+1}>_{a}      \!\! & = & \!\!    (J^+\rho)\lim_{\Delta\rightarrow 1} 
\int\limits^{+\infty}_{-\infty} d^{3}{\bf k_{1}} \int\limits^{+\infty}_{0} d^{3}{\bf k_{2}}
\Big[m^2_R-({\bf k}_{2}+i{\bf k}_{1})^{2}\Big] \nonumber\\
 \!\! & \cdot & \!\! \delta\Big( (\Delta-1){\bf k_{1}}\Big)
\delta\Big( (\Delta-1){\bf k_{2}}\Big)
| {\bf q}_{1},{\bf q}_{2}; \cdots ; {\bf q}_{n}, {\bf q}_{n+1}>_{\bf{a}} \nonumber\\
 \!\! & = & \!\!(J^+\rho)m^2_R \Big[ \lim_{\Delta\rightarrow 1}  \frac{1}{(\Delta-1)^2}\Big]| {\bf q}_{1},{\bf q}_{2}; \cdots ; {\bf q}_{n}, {\bf q}_{n+1}>_{\bf{a}},
\label{eigen4}
\end{eqnarray}
where additionally we have used the identity $(w_2+iw_1)^2=({\bf k}_2+i{\bf k}_1)^2-m^2_R$, obtained from the dispersion relations;
therefore, the eigenvalue diverges, and we need to impose the constraints
\begin{eqnarray}
 J^+\rho=0, \quad {\rm and/or} \quad m^2_R=0.
\label{eigen5}
\end{eqnarray}
In section \ref{secondmix}, we shall construct a squeezed boson state in terms of multiparticle states satisfying the Eq. 
(\ref{eigen4}), with eigenvalue zero, by choosing the second constraint; such a constraint will imply that the particles are fully massless.

However, there is no a $J^-$-projected complement for the b-type eigenvalue equation (\ref{numberacting1}), since the $J^-$-component of $H$  corresponds to creation operators of the form $\{\hat{a}^{\dag},\hat{b}^{\dag}\}$, and there are no expressions of the form (\ref{eigen1}), and (\ref{eigen2}).

We consider now the eigen-states constructed from hybrid states; we switch off the single commutators $[\hat{a},\hat{a}^{\dag}]=0$, and $[\hat{b}, \hat{b}^{\dag}]=0$ (see 
 Eqs. (\ref{zeromix0})), besides the vanishing commutator (\ref{vanishabd}), and we maintain the fundamental commutator $[\hat{a}, \hat{b}]=\rho \delta$; it is straightforward to show the following operator identity, and along the same lines, we can obtain the same  $J^+H$ eigen-states equation, which requires the same constraints (\ref{eigen5}),
\begin{eqnarray}
     [\{\hat{a},\hat{b}\}, (\hat{b}^{\dag}\hat{a}^{\dag})^{n}]=0
\quad \rightarrow \quad
 (J^+) H| ( {\bf k}_{1},{\bf k}_{2}; \cdots ; {\bf k}_{n}, {\bf k}_{n+1})_{a}; ( {\bf q}_{1},{\bf q}_{2}; \cdots ; {\bf q}_{n}, {\bf k}_{n+1} )_ {b}>
= 0;\quad n=1,2...;\nonumber\\
\label{stat1} 
\end{eqnarray}
In similarity to the previous case, there is no a $J^-$-projected version, due basically to the same reasons. This eigen-states will allow us to describe a mixture of a-bosons and b-bosons in section \ref{mixab}, which will look like a condensate.

\section{Squeezing operators}
\label{squeezed}
According to the Eqs. (\ref{coherent}), and (\ref{bosontrans}), translational transformations on the field operators lead to a trivial quantum field theory;
thus coherent states can not be constructed, since the bosonic displacements are turned off, and
the Bogoliubov transformations (\ref{iso}) will produce purely squeezed states.

In order to construct the {\it squeezed states}, we need first to look for the similarity transformations that generate the linear transformations (\ref{iso}) of the field operators
\begin{eqnarray}
     \breve{a} \!\! & = & \!\! \hat{S}_a (\alpha ,\beta ,\sigma ) \hat{a} \hat{S}_a(-\alpha ,-\beta,-\sigma), \label{similarity1} \\
     \breve{b} \!\! & = & \!\! \hat{S}_b (\alpha' ,\beta' ,\sigma' ) \hat{b} \hat{S}_b(-\alpha' ,-\beta',-\sigma'), \label{similarity2}
\end{eqnarray}
where the squeezing operators are given by
\begin{eqnarray}
      \hat{S}_a(\alpha,\beta,\sigma) \!\! & = & \!\! e^{\int dk(\alpha\hat{a}\hat{a} +\beta\hat{a}^{\dag}\hat{a}^{\dag} +\sigma\hat{a}\hat{a}^{\dag})}, \label{squeeze1} \\
      \hat{S}_b(\alpha',\beta',\sigma') \!\! & = & \!\! e^{\int dk(\alpha'\hat{b}\hat{b} +\beta'\hat{b}^{\dag}\hat{b}^{\dag} +\sigma'\hat{b}\hat{b}^{\dag})};
      \label{squeeze2}
\end{eqnarray}
where $\int d k=\int^{+\infty}_{-\infty} d^{3} {\bf k}_{1} \int^{+\infty}_{0} d^{3}{\bf k}_{2} $, and the coefficients $(\alpha,\alpha',\beta,\beta',\sigma,\sigma')$ must be determined by considering the restrictions  (\ref{determinant}), and  (\ref{projectionAB}) on the coefficients $(\alpha_{1},\alpha_{2},\beta_{1},\beta_{2})$, which can be parametrized as
\begin{eqnarray}
     \alpha_{1} =  J^{+}r_{1}e^{i\theta_{1}}, \qquad \alpha_{2} = je^{i\theta_{2}}, \qquad r_{1},\theta_{1}, \theta_{2} \in R, \label{para1}\\
     \beta_{1}  =  e^{i\theta_{3}}, \qquad \beta_{2} = J^{-}r_{4}e^{i\theta_{4}}, \qquad r_{4},\theta_{3}, \theta_{4} \in R, \label{para2}
\end{eqnarray}
where the Eqs. (\ref{projJ}), and (\ref{norm1}) have been taken into the account.

For determining the similarity transformations, we need the Baker-Campbell-Hausdorff (BCH) formula,
\begin{equation}
     e^{\hat{x}} \hat{y}e^{-\hat{x}} = \hat{y} + [\hat{x},\hat{y}] + \frac{1}{2!} [\hat{x},[\hat{x},\hat{y}]] + \frac{1}{3!}[\hat{x}, [\hat{x}, [\hat{x},\hat{y}]]] + \cdots;
     \label{BCH}
\end{equation}
for arbitrary operators $\hat{x},\hat{y}$. 

\subsection{$\hat{a}$-mode}
\label{amode}
We work out explicitly the case of the $\hat{a}$-mode squeezed operator (\ref{similarity1});
relevant commutators can be calculated by using the commutator (\ref{aad}),
\begin{eqnarray}
     \big[\hat{G}_{a},\hat{a}\big] \!\! & = & \!\! -\varrho\sigma\hat{a} - 2\varrho\beta\hat{a}^{\dag}, \label{BCH1}\\
     \big[\hat{G}_{a}, [\hat{G}_{a},\hat{a}]\big] \!\! & = & \!\! \varrho^{2}(\sigma^{2} - 4\alpha\beta )\hat{a}, \label{BCH2}\\
     \big[\hat{G}_{a}, [\hat{G}_{a}, [\hat{G}_{a},\hat{a}]]\big] \!\! & = & \!\! -\varrho^{3} (\sigma^{2}-4\alpha\beta ) (\sigma\hat{a}+2\beta\hat{a}^{\dag}), \label{BCH3}\\
     \big[ \hat{G}_{a}, [\hat{G}_{a}, [\hat{G}_{a}, [\hat{G}_{a},\hat{a}]]]\big] \!\! & = & \!\! \varrho^{5} (\sigma^{2}-4\alpha\beta )^{2}\hat{a}, \label{BCH4}
\end{eqnarray}
where the generator in the exponential (\ref{squeeze1}) has been identified with $\hat{G}_{a}=\int d k (\alpha\hat{a}\hat{a}+ \beta\hat{a}^{\dag}\hat{a}^{\dag}+ \sigma\hat{a}\hat{a}^{\dag})$. Furthermore,
by noting that all commutators from Eq. (\ref{BCH2}) and beyond, ad infinitum, are depending on the combination $\sigma^{2}-4\alpha\beta$, the convergence of the infinite series (\ref{BCH}) can be guaranteed by imposing
\begin{equation}
     \sigma^{2}-4\alpha\beta = 0, \label{baker1}
\end{equation}
and thus we can reduce the transformation (\ref{BCH}) to
\begin{equation}
     e^{\hat{G}_{a}}\hat{a}e^{-\hat{G}_{a}} = \hat{a} + [\hat{G}_{a},\hat{a}] = (1-\varrho\sigma)\hat{a} - 2\varrho\beta\hat{a}^{\dag}, \label{baker2}
\end{equation}
which must reproduce the transformation (\ref{iso}) with the parametrization (\ref{para1}), that we simplify now as follows,
\begin{equation}
     \alpha_{1} = 2J^{+}, \qquad \alpha_{2} =j, \qquad r_{1}=2, \quad \theta_{1}=0=\theta_{2}; \label{baker3}
\end{equation}
hence, a direct comparison between Eqs. (\ref{baker2}) and (\ref{baker3}) leads to
\begin{equation}
     \sigma\varrho =-j \qquad 2\beta\varrho =-j; \label{baker4}
\end{equation}
these constraints must be supplemented with the constraint (\ref{baker1}); fortunately one can solve this system of algebraic constraints in the ring ${\cal H}$. Since the constant $\varrho$ that defines the bosonic commutator $[\hat{a},\hat{a}^{\dag}]$ is Hermitian, then it has an inverse, according to Eq. (\ref{inv}),
\begin{equation}
     \varrho = \varrho_{1}+k \varrho_{2}; \qquad \varrho^{-1} = \frac{\varrho_{1}-k\varrho_{2}}{\varrho_{1}^{2}+\varrho_{2}^{2}}, \quad \varrho_{1}, \varrho_{2} \in R; \label{baker5}
\end{equation}
hence, from Eqs. (\ref{baker4}) we have 
\begin{equation}
     \sigma = -j\varrho^{-1} = \frac{i\varrho_{2}-j\varrho_{1}}{\varrho_{1}^{2}+\varrho_{2}^{2}}, \qquad \beta = -\frac{j}{2} \varrho^{-1} = \frac{1}{2} \frac{i\varrho_{2}-j\varrho_{1}}{\varrho_{1}^{2}+\varrho_{2}^{2}}, \quad \alpha=\beta; \label{baker6}
\end{equation}
where the last expression follow from the Eq. (\ref{baker1}), by considering that $\sigma=2\beta$; the generator $\hat{G}_{a}$ is determined fully in terms of the constant $\varrho$.

\subsection{$\hat{b}$-mode}
\label{bmode}

The case of the $b$-mode squeezed operator is somewhat different; first we simplify conveniently the parametrization (\ref{para2})
\begin{equation}
     \beta_{1} =1, \qquad \beta_{2} =2J^{-}; \qquad r_{4}=2, \ \theta_{3}=0=\theta_{4}; \label{haus}
\end{equation}
and thus the similarity transformation (\ref{similarity2}) must reproduce the linear transformation
\begin{equation}
     \breve{b} \equiv \hat{b} + 2J^{-}\hat{b}^{\dag}. \label{haus1}
\end{equation}
Furthermore, along the lines followed for the previous case, the algebraic constraint
\begin{equation}
     \sigma'^{2}- 4\alpha'\beta'=0, \label{haus2}
\end{equation}
ensures the convergence of the expansion (\ref{BCH}) for the case at hand, which will reduce to
\begin{equation}
     e^{\hat{G}_{b}} \hat{b} e^{-\hat{G}_{b}} = \hat{b}+ [\hat{G}_{b},\hat{b}] = (1-\zeta\sigma')\hat{b} - 2\zeta\beta'\hat{b}^{\dag},\quad \hat{G}_{b}=\int dk(\alpha'\hat{b}\hat{b} +\beta'\hat{b}^{\dag}\hat{b}^{\dag} +\sigma'\hat{b}\hat{b}^{\dag}),
 \label{haus3}
\end{equation}
where the Hermitian constant $\zeta$ determines the commutator $[\hat{b},\hat{b}^{\dag}]$ in Eq. (\ref{bbd}). Hence,
assuming that the constant $\zeta$ is different from zero, with the general form $\zeta=(\zeta_{1}+k \zeta_{2})$, a direct comparison between Eqs. (\ref{haus1}), and (\ref{haus3}) leads to
\begin{equation}
     \sigma'=0, \qquad \beta'=-J^{-}\zeta^{-1},  \quad \zeta^{-1}=\frac{\zeta_{1}-k \zeta_{2}} {\zeta_{1}^2+\zeta_2^2};    
     \label{haus4}
\end{equation}
additionally the vanishing of $\sigma'$ implies, according to the Eq. (\ref{haus2}), the vanishing of the product $\alpha'\cdot\beta'$, which allows us to make the identification
\begin{equation}
     \alpha'=-\bar{\beta}'=J^{+}\zeta^{-1}, \label{haus5}
\end{equation}
due to the orthogonality of the projectors $(J^{+}, J^{-})$.

\subsection{Unitarity}
\label{unitarity}
We consider now the unitarity of the squeezing operators;
taking into the account that the coefficients $(\alpha,\beta,\sigma)$ are anti-Hermitian, $\bar{\alpha}=-\alpha$, $\bar{\beta}=-\beta$, and $\bar{\sigma}=-\sigma$, due basically to the fact that they turn out to be linear combinations of purely imaginary terms, then the generator $\hat{G}_{a}$ is anti-Hermitian, since corresponds to a Hermitian operator multiplied by an anti-Hermitian factor; the squeezing operator $\hat{S}_{a}$ will be then unitary,
\begin{equation}
     \hat{G}_{a} = \alpha \int dk (\hat{a}\hat{a} + \hat{a}^{\dag}\hat{a}^{\dag} + 2\hat{a}\hat{a}^{\dag}) = -\hat{G}_{a}^{\dag},\quad \alpha=-\frac{j}{2}\varrho^{-1}, \qquad \hat{S}_{a}^{\dag} = \hat{S}_{a}^{-1}. \label{unitaryA}
\end{equation}
The unitarity of $\hat{S}_{b}$ is achieved in a slightly different way,
\begin{equation}
     \hat{G}_{b} = \int dk [\alpha'\hat{b}\hat{b} - \bar{\alpha}'\hat{b}^{\dag}\hat{b}^{\dag}] = -\hat{G}_{b}^{\dag}, \quad \alpha'=J^+\zeta^{-1}, \qquad \hat{S}_{b}^{\dag} = \hat{S}_{b}^{-1}; \label{unitaryB}
\end{equation}
where the $\sigma'$-term is absent. Furthermore, the generator $\hat{G}_{b}$ contains explicitly the idempotent projectors $(J^{+},J^{-})$, as opposed to the generator $\hat{G}_{a}$ that contains linear combinations of the imaginary units $(i,j)$; the later has besides a non-trivial $\sigma$-term contribution.

If one attempts to achieve certain balance between the description of a-bosons and b-bosons, enforcing for example the vanishing of the coefficient $\sigma$ in Eq. (\ref{baker6}), then it will lead to the vanishing of the constant $\varrho$, which rather will accentuate the differences.
In fact, the parameter $\sigma$ is neccesary for reproducing the projector $J^+$ as coefficient of the annihilation operator $\hat{a}$ in the Eq. (\ref{baker2}).
 Hence, there exist intrinsic differences in the physical properties of bosons of different type, which will be more evident once we shall see the squeezing effect on the vacuum state.

\subsection{Disentangling the operators: squeezing the vacuum}
\label{disen}

We consider first the case of the $b$-single squeezing operator. The fact that the generator $\hat{G}_{b}$ is constructed from the projectors $(J^{+},J^{-})$ leads to a trivialization of disentangling the operator $e^{\hat{G}_{b}}$,
\begin{equation}
     e^{\hat{G}_{b}} = e^{-\bar{\alpha}'\int dk \hat{b}^{\dag}\hat{b}^{\dag}} \cdot e^{\alpha'\int dk \hat{b}\hat{b}}; \label{disen1}
\end{equation}
according to the so-called Zassenhaus formula, the dual version of the BCH formula (\ref{BCH}),
\begin{equation}
     e^{\hat{A}+\hat{B}} = e^{\hat{A}} e^{\hat{B}} \prod^{\infty}_{i=2} e^{\hat{C}_{i}}, \label{zass}
\end{equation}
where $\hat{C}_{i}$ are polynomials of degree $i$ in the arbitrary operators $(\hat{A},\hat{B})$, and homogeneous in the commutator $[\hat{A},\hat{B}]$. The first polynomials are given by
\begin{eqnarray}
     \hat{C}_{2} & = & \frac{1}{2}[\hat{B},\hat{A}], \qquad \hat{C}_{3} = \frac{1}{3}[[\hat{B},\hat{A}],\hat{B}] + \frac{1}{6}[[\hat{B},\hat{A}], \hat{A]}, \nonumber \\
     \hat{C}_{4} & = & \frac{1}{8} \big( [[[\hat{B},\hat{A}],\hat{B}],\hat{B}] + [[[\hat{B},\hat{A}],\hat{A}],\hat{B}] \big) +\frac{1}{24} [[[\hat{B},\hat{A}],\hat{A}],\hat{A}]; \label{wilcox}
\end{eqnarray}
the higher order polynomials  are not displayed here, due to they are very long, and we shall take such expressions from the reference \cite{quesne}.

Furthermore, the vacuum state $|0>$ is an eigen-ket of the second exponential in Eq.\ (\ref{disen1}),
\begin{equation}
     e^{\alpha'\int dk\hat{b}\hat{b}} |0> = [1+\zeta^{-1} J^{+} \int dk \hat{b}\hat{b} + \cdots ] |0> = |0>,
     \label{disen2}
\end{equation}
where we have used the explicit form for the coefficient $\alpha'$ in terms of the projector $J^{+}$ given in Eq. (\ref{haus5}), and the fact that the vacuum state $|0>$ is annihilated by the operator $J^{+}\hat{b}$.

Thus, multiplying both sides of Eq. (\ref{similarity2}) by $J^{+}e^{\hat{G}_{b}}$, we arrive to the expression
\begin{equation}
     J^{+}\breve{b} [e^{\hat{G}_{b}} |0> ] = e^{\hat{G}_{b}} J^{+} \hat{b} |0> = 0; \label{disen3}
\end{equation}
where the resulting operator is acting upon the vacuum state $|0>$, which is annihilated by the operator $J^{+}\hat{b}$, and correspondingly the new vacuum state $|\breve{0}>$ can be identified with the state on the left hand side, which is annihilated by $J^{+}\breve{b}$; using the Eqs. (\ref{disen1}), and (\ref{disen2}) we can find an explicit expression for such a state,
\begin{equation}
     J^{+} e^{\hat{G}_{b}} |0> = J^{+} [1-\zeta^{-1} J^{-} \int dk \hat{b}^{\dag} \hat{b}^{\dag} + \cdots ] |0> = J^{+} |0>, \label{disen4}
\end{equation}
due to the orthogonality of the projectors; hence, the vacuum is an eigenket for the operator $(J^{+}) e^{\hat{G}_{b}}$, and there is no actually a b-mode squeezing effect on it. However for the a-mode the result will be different since a squeezed state of a-quanta will be generated by squeezing the vacuum $|0>$ by the action of $e^{\hat{G}_{a}}$; hence, similarly to the Eq. (\ref{disen3}), we have 
\begin{equation}
     J^{-}\breve{a} [e^{\hat{G}_{a}} |0> ] = e^{\hat{G}_{a}} J^{-} \hat{a} |0> = 0,
\label{disen5}
\end{equation}
and the new vacuum, annihilated by $J^-\breve{a}$ in the above equation and by  $J^+\breve{b}$ in Eq. (\ref{disen3}), corresponds to the vacuum annihilated by the observables in Eqs. (\ref{breveH}), and (\ref{breveP}), and defined originally in Eq.
(\ref{newvac}).

We attempt now the disentangling of the operator $e^{\hat{G}_{a}}$ with the identification
\begin{equation}
     \hat{A} \equiv \int dk \alpha (\hat{a}\hat{a}+\hat{a}^{\dag}\hat{a}), \qquad \hat{B} \equiv \int dk \alpha (\hat{a}^{\dag}\hat{a}^{\dag} + \hat{a}^{\dag}\hat{a}), \qquad [\hat{B},\hat{A}] = -2\alpha\varrho(\hat{A}+\hat{B})-2\alpha^2\varrho^2, \label{wilcox1}
\end{equation}
that yields a convergent but even entangled expression for the original operator; the polynomials in Eq. (\ref{wilcox}) read
\begin{eqnarray}
     \hat{C}_{2} & = & \frac{1}{2}[\hat{B},\hat{A}], \qquad \hat{C}_{3} = -\frac{j}{6}[\hat{B},\hat{A}], \qquad \hat{C}_{4} = \frac{1}{24} [\hat{B},\hat{A}], \nonumber \\
     \hat{C}_{5} & = & -\frac{j}{120} [\hat{B},\hat{A}], \qquad \hat{C}_{6} = \frac{1}{720} [\hat{B},\hat{A}]; \label{wilcox2}
\end{eqnarray}
and thus
\begin{equation}
     e^{\hat{G}_{a}} = e^{-j}e^{\hat{A}} e^{\hat{B}} e^{(\frac{1}{2}-\frac{j}{6}+\frac{1}{24}-\frac{j}{120}+\frac{1}{720}+\cdots ) [\hat{B},\hat{A}]}, \label{wilcox3}
\end{equation}
where the infinite series converges to
\begin{equation}
      \frac{1}{2}-\frac{j}{6}+\frac{1}{24}-\frac{j}{120}+\frac{1}{720}+\cdots = e^{-j} +j-1= \frac{1}{e} J^{+} +(e-2)J^{-}; \label{wilcox4}
\end{equation}
which can be proved using the expansion of the hyperbolic functions $\sinh$ and $\cosh$; the last equality follows from the identity (\ref{rothyp1}). Now, using the expression (\ref{unitaryA}) for $\alpha$, we arrive to the equation
\begin{equation}
     e^{\hat{A}+\hat{B}} = e^{-\frac{1}{2}[\frac{1}{e}J^{+}+(e-2)J^{-}]} e^{\hat{A}} e^{\hat{B}} e^{[\frac{1}{e}J^{+}+(2-e)J^{-}](\hat{A}+\hat{B})}, \label{entangled}
\end{equation} 
which relates the same exponential operator with different global factors; on the left hand side we have the original unitary operator, and on the right hand side we have a deformed version that is not unitary.  One can continue recursively to decompose again the exponential in the right hand side, and/or to use an $m$-order 
aproximant obtained for example by fractal decomposition \cite{fractal}. Moreover, such an approximate result will require a subsequent decomposition for the compound operators $\hat{A}$, and $\hat{B}$. 

If we renounce to the unitarity of the operator, then the disentangling is direct, by multiplying (from the right) both sides of the equality (\ref{entangled}) by the operator $ e^{-[\frac{1}{e}J^{+}+(2-e)J^{-}](\hat{A}+\hat{B})}$, and thus a non-unitary operator is obtained on the left hand side, since the global factor is not anti-Hermitian.

However, this is the correct way to achieve the disentangling. By returning to the original operator and considering
a redefinition of the generator by a global factor $g$ that belongs to the ring, we arrive to a scaling of the operators (\ref{wilcox1}),
\begin{equation}
     \hat{A}' = g\hat{A}, \qquad \hat{B}' = g\hat{B}, \qquad [\hat{B}',\hat{A}'] = -g^{2} \Big[2\alpha\varrho(\hat{A}+\hat{B}) +2\alpha^2\varrho^2\Big],
     \label{redefine}
\end{equation}
where the global factor must be restricted to be a pure hyperbolic number, due to unitarity,
\begin{equation}
     g=g_{1}+jg_{2}, \qquad g_{1}, g_{2} \in R. \label{redefine1}
\end{equation}
Therefore, along the same lines, we obtain the generalization of the Eq. (\ref{entangled})
\begin{eqnarray}
     e^{-(2j\alpha\varrho F_1+g_1)(\hat{A}+\hat{B}) } e^{-j(2j\alpha\varrho F_2+g_2)(\hat{A}+\hat{B})}= e ^{2(\alpha\varrho)^2(F_2+jF_1)}e^{-g\hat{B}} e^{-g\hat{A}};
      \label{redefine2}
\end{eqnarray}
with two real functions defined as,
\begin{eqnarray}
F_{1}(g_1,g_2)=\sinh g_{1}(\sinh g_{2}-\cosh g_{2})+g_1, \quad F_2(g_1,g_2)=
\cosh g_{1}(\cosh g_{2}-\sinh g_{2})+g_2-1;
\label{FS}
\end{eqnarray}
and additionally we have assumed that the product $\alpha\varrho$ has been maintained as a pure $j$-imaginary number (see Eq. (\ref{unitaryA})); note then that $j\alpha\varrho$ is real.
In this manner the unitarity on the left hand side of the equation (\ref{redefine2}) is preserved, provided that the hyperbolic phase vanishes,
\begin{equation}
     \gamma F_2+g_2=0, \quad \rightarrow \quad \cosh g_1= \frac{\gamma-(\gamma+1)g_2}{\gamma}e^{g_2}, \quad \gamma \equiv  2j\alpha\varrho;
\label{redefine3}
\end{equation} 
where $\gamma$ is real; finally the disentangled unitary operator can be written as
\begin{eqnarray}
     e^{-(\gamma F_1+g_1)(\hat{A}+\hat{B}) }=  e ^{\frac{1}{2}\gamma^2(F_2+jF_1)}
e^{-g\hat{B}} e^{-g\hat{A}}.
\label{redefine4}
\end{eqnarray} 
With this expression we have re-defined the generator modulo a real global factor
\begin{eqnarray}
\hat{G}_a \rightarrow -(\gamma F_1+g_1)\hat{G}_a,
\label{redefine5}
\end{eqnarray}
and we can redo the computations of section \ref{amode} with this new scaled generator,
obtaining basically that same Eqs. (\ref{baker4}), and \ref{baker6}), with a scaling by the same global factor for the coefficients $\sigma$, $\beta$, and $\alpha$; the Eq. (\ref{baker1}) remains as it stands, and implies that $\alpha=\beta=\sigma/2$. Thus, we have the generalization for the constraint (\ref{baker4}),
\begin{eqnarray}
 (\gamma F_1+g_1)\gamma=1, \quad \rightarrow \quad e^{g_2}=\frac{\gamma^2\sinh g_1}{\gamma(\gamma+1)g_1-1},
\label{redefine6}
\end{eqnarray}
this constraint together with the constraint (\ref{redefine3}) can be solved for the parameter $\gamma$ in terms of the  combination $(g_1-g_2)$, which will be taken as the effective squeezing parameter,
\begin{eqnarray}
\gamma(g_{0})=\frac{-g_0e^{g_0}\pm\sqrt{(g_0e^{g_0})^2-4e^{g_0}(g_0e^{g_0}-e^{g_0}+1)}}{2(g_0e^{g_0}-e^{g_0}+1)}, \quad g_2-g_1\equiv g_0.
\label{solconst}
\end{eqnarray}
For concreteness, we shall focus below on the positive root in the above equation, 
the case of the negative root will be studied elsewhere.

Note now that the vacuum is an eigenket of the first exponential operator in Eq. (\ref{redefine4})
\begin{equation}
     J^{-} e^{-g\hat{A}} |0> = J^{-} |0>, \label{redefine7}
\end{equation}
since $J^{-}\hat{a}|0>=0$. Thus, we need to achieve the disentangling of the second exponential operator $e^{-g\hat{B}}$ with the identification
\begin{equation}
     \hat{B}_{1} = -g\alpha \int dk \hat{a}^{\dag}\hat{a}^{\dag}, \qquad \hat{B}_{2} = -g\alpha\int dk \hat{a}^{\dag}\hat{a}; \label{redefine8}
\end{equation}
along the same lines followed previously we obtain a disentangled expression
\begin{equation}
     e^{-g\hat{B}} = e^{\hat{B}_{2}} \cdot e^{F_{0}\int dk \hat{a}^{\dag}\hat{a}^{\dag}}, \qquad F_{0} (\varrho^{-1}, \gamma gj) \equiv \frac{\varrho^{-1}}{2} [(1+\gamma gj) + (1-\gamma gj) e^{\gamma gj}]; \label{redefine9}
\end{equation}
where we have used basically the quadratic commutator $\int dk[\hat{a}^{\dag}\hat{a}, \hat{a}^{\dag}\hat{a}^{\dag}] = 2\varrho \hat{a}^{\dag}\hat{a}^{\dag}$.

Although the vacuum is an eigenket of the exponential operator $(J^{-})e^{\hat{B}_{2}}$, such an operator can not act directly since the operators in Eq. (\ref{redefine8}) do not commute. Hence, the only possibility (without invoking normal ordering of exponential operators) is to consider the explicit expansion of both operators, and the direct action on the vacuum using the identity
\begin{equation}
     J^{-} \int dk (\hat{a}^{\dag}\hat{a})^{M} (\hat{a}^{\dag})^{N} |0> = J^{-}\int dk (N\varrho )^{M} (\hat{a}^{\dag})^{N} |0>, \qquad M,N: 0,1,2,3\ldots ; \label{redefine10}
\end{equation}
which can be shown by using iteratively the commutator $[\hat{a},\hat{a}^{\dag}] =\varrho\delta_{Dirac}$, and the fact that $J^{-}\hat{a}|0>=0$; note that for $M=1$ the above expression reduces exactly to the number operator described in (\ref{numberacting}).
Therefore, the final result is
\begin{equation}
     (J^{-})e^{-g\hat{B}} |0> = J^{-} e^{F_{0}(\varrho^{-1}, -\gamma gj) \int dk \hat{a}^{\dag}\hat{a}^{\dag}} \cdot |0>; \label{redefine11}
\end{equation}
putting all together, the final squeezing effect on the vacuum will read
\begin{equation}
     (J^{-})e^{-(\gamma F_1+g_1)\hat{G}_a}\cdot |0> = 
\sqrt{e}(J^{-})e^{-\frac{\gamma gj}{2}} e^{F_{0}(\varrho^{-1}, -\gamma gj) \int dk \hat{a}^{\dag}\hat{a}^{\dag}} \cdot |0>;
\label{redefine12}
\end{equation}
which can be interpreted as a statistical mixture of $a$-quanta, with an indefinite number of particles, and described as an expansion of Fock-number eigenstates (see Eq. (\ref{numberacting})); note that this quantum state contains only terms with even compound boson numbers, which follows from the bilinear nature of the squeezing operator in the creation operator. Note also that the purely hyperbolic function $F_{0}$ has suffered a change of sign in the argument $\gamma gj$, respect to the original form in Eq. (\ref{redefine9}). Using the property (\ref{absorb}) we can determine the $J^-$-projections for this function and for the global exponential in Eq. (\ref{redefine12}),
\begin{eqnarray}
(J^-)F_{0}(\varrho, \pm \gamma gj)\!\!& = & \!\!(J^-)f_{\pm}\equiv (J^-)\frac{\varrho^{-1}}{2}\Big[ 1\pm\gamma(g_2-g_1) +\Big( 1 \mp\gamma(g_2-g_1)\Big)e^{\pm\gamma(g_2-g_1)} \Big], 
\label{fpm}\\
(J^+)e^{-\frac{\gamma gj}{2}}\!\!&=&\!\!(J^+)e^{-\frac{\gamma}{2}(g_2-g_1)};\label{globalexp}
\end{eqnarray}
$f_{-}$ appears in Eq. (\ref{redefine12}), and $f_{+}$ will appear below in an alternative state constructed by using {\it normal ordering}; thus, a direct comparison will be possible.
Therefore, the final expressions depend again on the same effective combination $(g_2-g_1)$ that appears in the Eq. (\ref{solconst}), and then 
\[
\gamma(g_2-g_1)= \gamma(g_{0})g_0;
\]
and hence
the real squeezing functions $f_{\pm}$ can be expressed fully in terms of the only squeezing parameter $g_{0}$. 
We choose the positive root for $\gamma(g_{0})$ in the Eq. (\ref{solconst}) for a detailed analysis; 
by fixing the constant $\varrho=1$, these functions are illustrated in the figure \ref{ff}, in the interval $g_0 \in (0,+\infty)$.
\begin{figure}[H]
  \begin{center}
    \includegraphics[width=.4\textwidth]{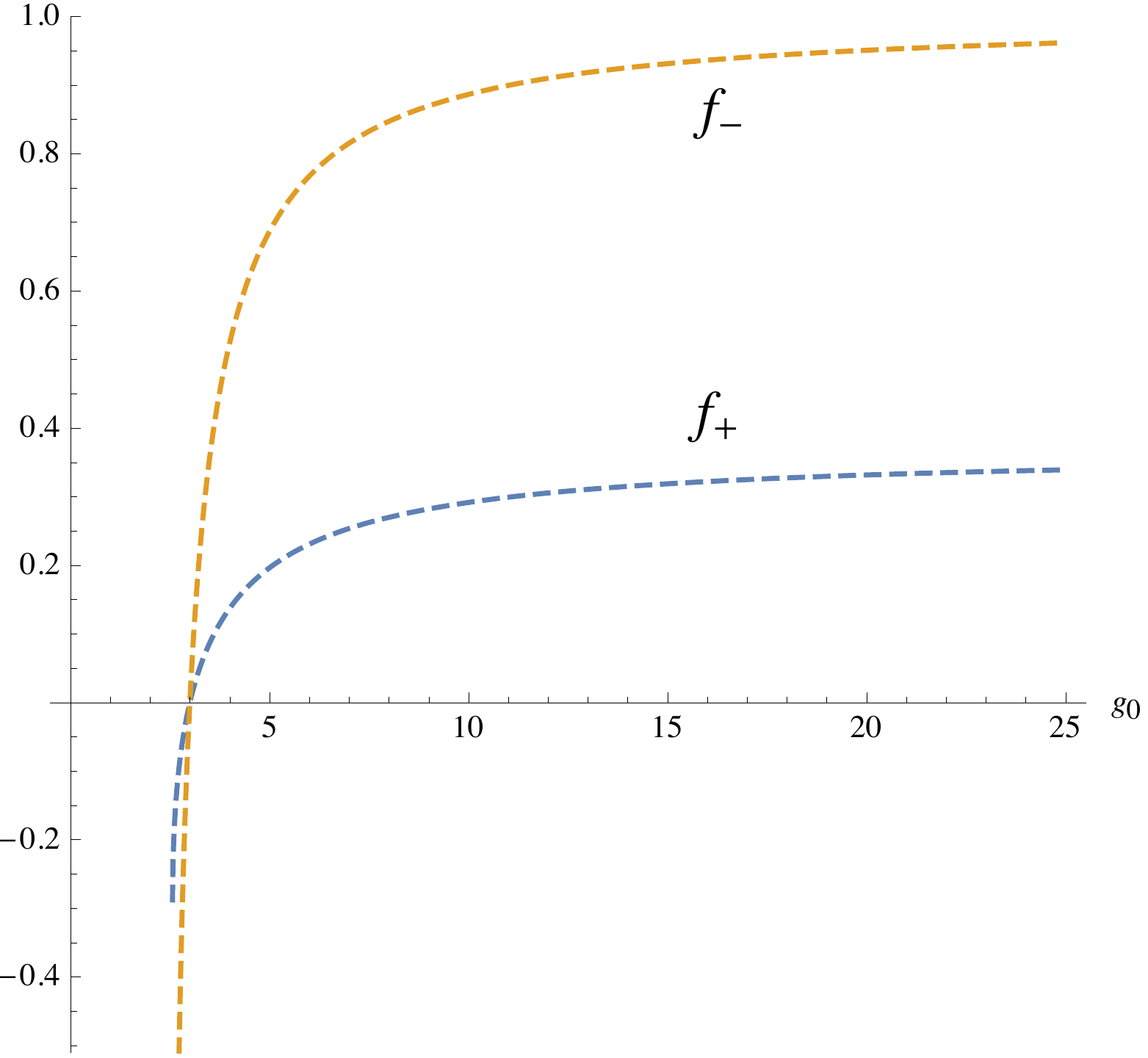}
  \caption{ The functions have a common root localized at $g_0\approx 2.9921$; to the right of the root, both functions are positive definite, with the asymptotic values
$\lim_{_{g_0\rightarrow +\infty}} f_{-}=1$, and $\lim_{_{g_0\rightarrow +\infty}} f_{+}=\frac{1}{e}$; additionally the y-axis is a common asymptote for both functions, $\lim_{_{g_0\rightarrow 0}} f_{\pm}=-\infty$.
} 
   \label{ff}
  \end{center}
\end{figure}
In the figure \ref{ff} we have omitted by simplicity the profile of the functions $f_{\pm}$ in the interval $g_0 \in (0,-\infty)$, which is disconnected from the profile in the interval $g_0 \in (0,+\infty)$ due to the presence of the asymptote at $g_0=0$; we shall analyze the complementary region in subsequent works. Similarly the case for the negative root in Eq. (\ref{solconst}) will be analyzed later.

\subsection{Boson-number distributions: with and without normal ordering}
\label{bosonnum}
We determine now the probability of finding the (compound) boson number $n$ for the state (\ref{redefine12}),
\begin{eqnarray}
P(n; g_0)\equiv |<n|\sqrt{e}e^{-\frac{\gamma (g_2-g_1)}{2}} e^{f_{-} \int dk \hat{a}^{\dag}\hat{a}^{\dag}} \cdot |0>|^2= e^{(1-\gamma (g_2-g_1))}\frac{(f_{-})^n}{(\frac{n}{2}!)^2}, \quad  n=0,2,4,6,....;
\label{redefine121}
\end{eqnarray}
although $f_{-}$ is not positive definite for $g_0 \in R^+$ , the probability 
(\ref{redefine121}) corresponds to even powers, and thus is positive definite in that interval; $P(n,g_0)$ is shown in figure \ref{nonormal} as a function of $g_0$ for the first values of $n$; the curves for lower boson numbers are difficult to display, since they are strongly {\it compressed} on the 'x'-axis. Since the asymptotic behavior of $f_-$, and its powers, is bounded as $g_0\rightarrow +\infty$, as opposed to the limit $g_0\rightarrow 0$, we shall focus the analysis in the region $[ {\rm root}, +\infty)$.
\begin{figure}[H]
  \begin{center}
    \includegraphics[width=.4\textwidth]{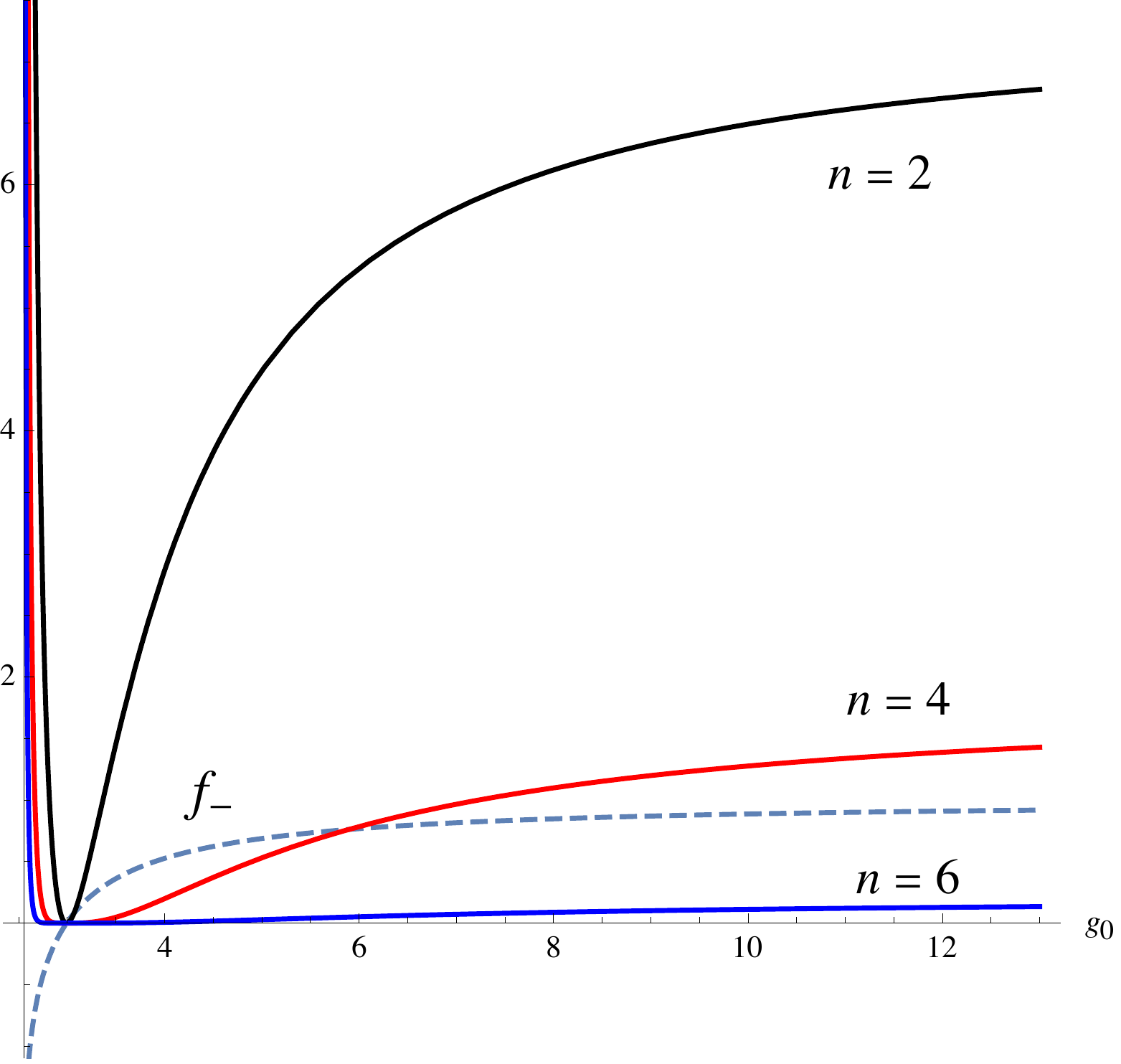}
  \caption{The curves are increasing monotonically from the root to their asymptotic values $P(n; g_0\rightarrow +\infty)= \frac{e^2}{(\frac{n}{2}!)^2}$; hence each curve has a horizontal asymptote.} 
   \label{nonormal}
  \end{center}
\end{figure}
Although in this work we are not invoking the {\it normal ordering} (NO) of operators, it will be of interest to make a comparison with a squeezed state obtained from the assumption of such
an ordering,
\begin{eqnarray}
     (J^{-}):e^{-(\gamma F_1+g_1)\hat{G}_a}:|0>& = &\sqrt{e}(J^{-})e^{-\frac{\gamma gj}{2}} e^{F_{0}(\varrho^{-1}, +\gamma gj) \int dk \hat{a}^{\dag}\hat{a}^{\dag}} \cdot |0>\nonumber\\ 
&=&
\sqrt{e}(J^-)e^{-\frac{\gamma (g_2-g_1)}{2}} e^{f_{+} \int dk \hat{a}^{\dag}\hat{a}^{\dag}}
\cdot |0>; \label{redefine13}
\end{eqnarray}
thus, as opossed to the Eq. (\ref{redefine12}), the change of sign in the argument $\gamma gj$ is absent, and thus this state is described by $f_+$. Furthermore, for this case $P^{NO}(n;g_0)$ is given by the expression (\ref{redefine121})
with the substitution $f_{-} \rightarrow f_{+}$, and is illustrated in figure \ref{NO}.
\begin{figure}[H]
  \begin{center}
    \includegraphics[width=.4\textwidth]{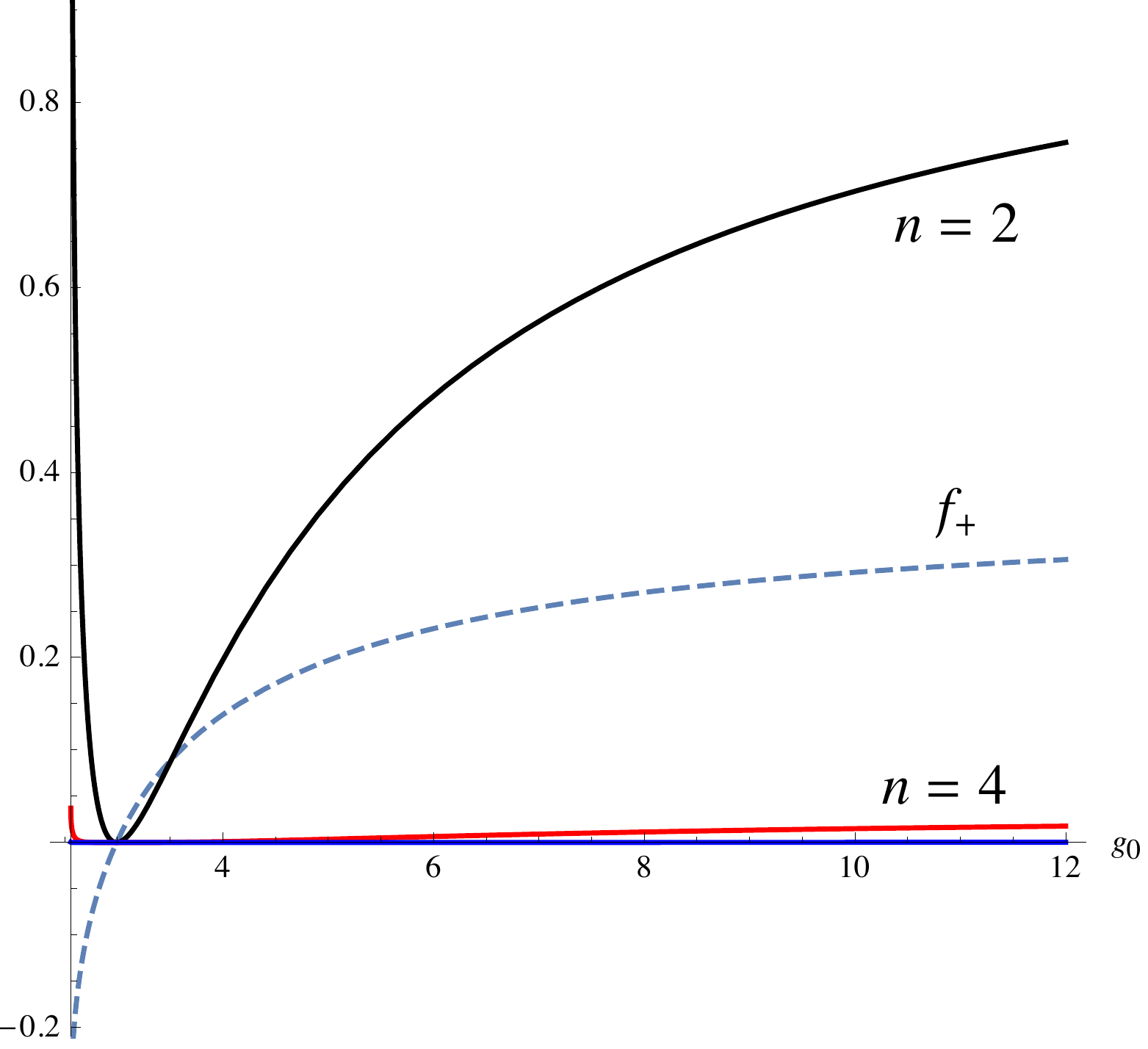}
  \caption{ The asympotic behavior is $P^{NO}(n; g_0\rightarrow +\infty)= \frac{e^2}{(\frac{n}{2}!)^2}\frac{1}{e^n}$; hence, each $n$ value defines a horizontal asymptote.} 
   \label{NO}
  \end{center}
\end{figure}
Therefore both approaches, with and without normal ordering, give
finite measurements for squeezed boson-number statistics, and a direct comparison is possible;
for example in the asymptotic region, we have that 
\[P^{NO}(n,g_0\rightarrow +\infty)=\frac{P(n,g_0\rightarrow +\infty)}{e^n},\] 
and considering that $\frac{1}{e^2}\approx 1.3 \times 10^{-1}$, and that $n$ is even, then for a given $n$  there exist $\frac{n}{2}$ orders of magnitude in favor of $P$. As a complementary result, in the figure \ref{pn} we display  
$P^{NO}$, and $P$ as functions of $n$, and for a parameter $g_0$ fixed.
\begin{figure}[H]
  \begin{center}
    \includegraphics[width=.4\textwidth]{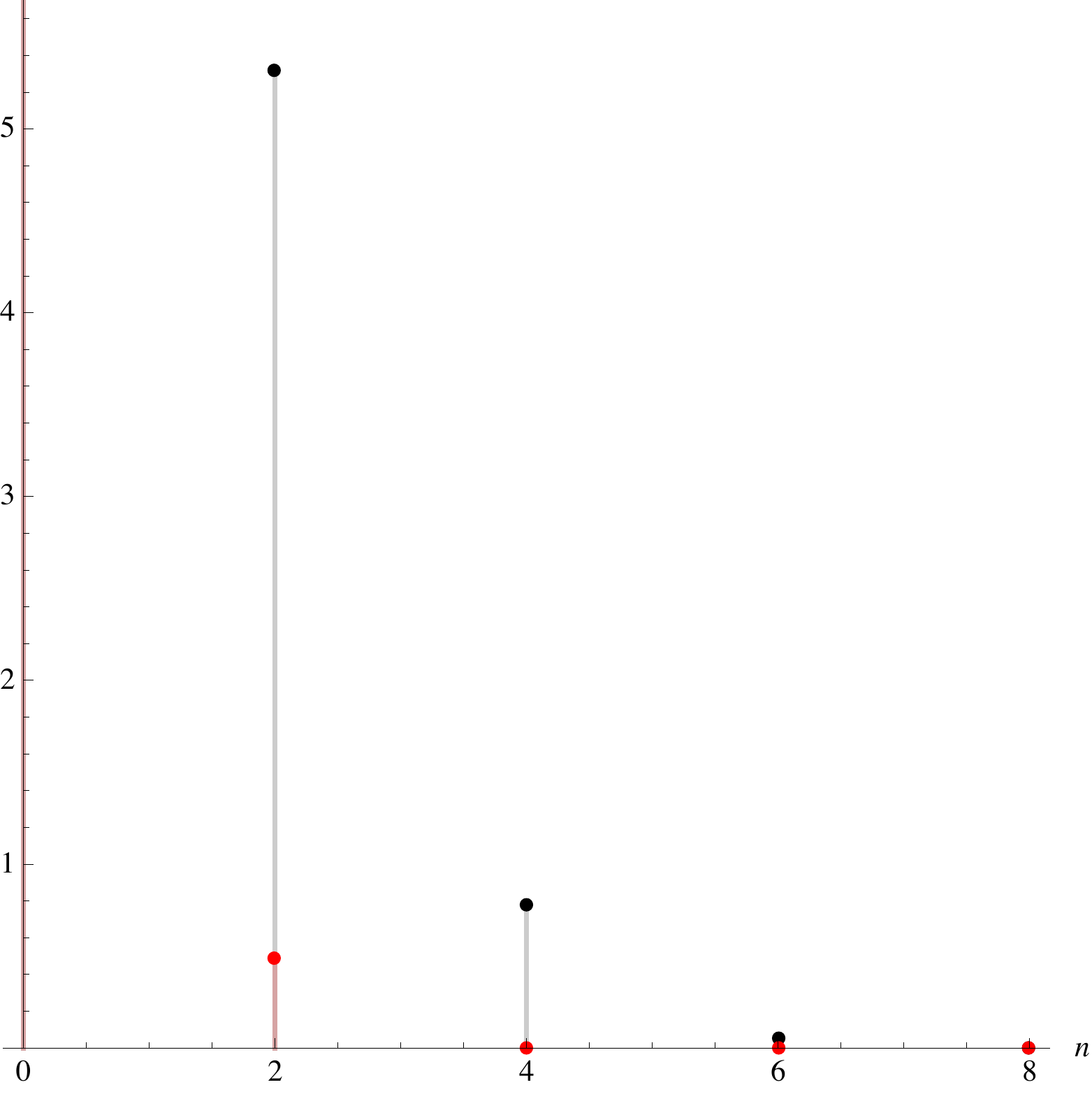}
  \caption{For a fixed value of the parameter, say $g_0=6$, $P(n;g_0=6)$ is shown in black points for the discrete values of $n$; $P^{NO}(n;g_0=6)$ is shown in red. It is evident that for $n=2$ there exists a difference around one order of magnitude in favor of $P(n=2,g_0=6)$, in similarity to the asymptotic regime. Similar results can be found for higher boson-numbers.} 
   \label{pn}
  \end{center}
\end{figure}
The profiles shown in figure \ref{pn} correspond to deformed Poisson distributions, retaining certain global features of conventional particle number statistics; for example in quantum optics  the probability of measuring $2m$ photons in a squeezed state is given by  $P(2m)=\frac{(\tanh r)^{2m}}{\cosh(r)}\frac{(2m)!}{2^{2m}(m!)^2}$, where $r$ is the squeezing parameter \cite{gerry}. The similarities between this case and the cases at hand are as follows: 

i) the probabilities vanish for all odd particle numbers.

ii) the probabilities go to zero as $n\rightarrow \infty$.

iii)  the maximum values for the probabilities are localized at the minimum value for the particle number $n=2$; from the maximun values, the distributions decrease monotonically to zero.

iv) therefore, all these distributions resemble that of the thermal radiation.

These squeezed states must be compared with the trivial state generated with the operator $e^{\hat{G}_b}$ in Eq. (\ref{disen4}); furthermore,
 other state of a-quanta will be constructed in section \ref{secondmix} by using a multi-mode squeezing operator, which will show physically different properties.

\section{Multimode squeezed states}
\label{multi}
The squeezing effects exist not only for the single modes developed previously, but for correlated states constructed by mixing the two modes as well.
Hence, in the single modes, the squeezing operators (\ref{squeeze1}), and (\ref{squeeze2}),  depend only on the a-type or b-type field operators, and they are able to generate the single transformations (\ref{similarity1}) and (\ref{similarity2}); at the end, a state of single type is generated (see Eq. (\ref{redefine12})). However, the commutator $[\hat{a},\hat{b}]$ is not trivial, and this fact will allow us to propose generators that mix the two modes, and that, restricted by unitarity, will generate the transformations (\ref{similarity1}) and (\ref{similarity2}). At the end, correlated states will induce a state of a-quanta and b-quanta.

\subsection{$\hat{G}^{0}_{ab}$}
\label{mixab}
We consider the following generator
\begin{equation}
     \hat{G}^{0}_{ab} \equiv \int dk [\gamma_{1}\hat{a}\hat{b}^{\dag} + \gamma_{2}\hat{b}\hat{a}^{\dag} + \gamma_{3}\hat{a}\hat{b} + \gamma_{4}\hat{b}^{\dag}\hat{a}^{\dag}], \label{zeromix}
\end{equation}
where we have bilinear terms that mix the field operators; since the generator must be anti-Hermitian, such mixed terms correspond to pairs with elements that are Hermitian conjugates to each other, namely, $(\hat{a}\hat{b}^{\dag}, b\hat{a}^{\dag})$, and $(\hat{a}\hat{b}, \hat{b}^{\dag}\hat{a}^{\dag})$. The anti-Hermitian character $(\hat{G}^{0}_{ab})^{\dag}=-\hat{G}^{0}_{ab}$, implies that $\gamma_{1}=-\bar{\gamma}_{2}$, and $\gamma_{3}=-\bar{\gamma}_{4}$.

In order to simplify the computation we can impose the vanishing of the commutators
\begin{eqnarray}
     [\hat{a},\hat{a}^{\dag}] \!\! & = & \!\! 0, \quad \varrho =0, \nonumber \\
     \big[\hat{b},\hat{b}^{\dag}\big] \!\! & = & \!\! 0, \quad \zeta=0, \label{zeromix0}
\end{eqnarray}
in addition to the vanishing commutator (\ref{vanishabd}), which implies that the normal ordering is unnecessary; hence, our fundamental commutator $[\hat{a},\hat{b}]=\rho\delta$ remains as the only non-vanishing commutator, and we fix $\rho$ as a real parameter; this algebra contrasts with the traditional harmonic oscillator algebra defined by $[\hat{a},\hat{a}^{\dag}]\neq0$, $[\hat{b}, \hat{b}^{\dag}]\neq0$, and $[\hat{a},\hat{b}]=0$. In the next sub-section, the commutators 
(\ref{zeromix0}) will be turned on again, and a second multimode operator will be constructed; similarly in the Appendix,  more general cases without the constrictions (\ref{zeromix0}) are considered.

Hence by using the remaining commutator, we have that
\begin{equation}
     [\hat{G}^{0}_{ab},\hat{a}]= -\gamma_{3}\rho\hat{a}+ \bar{\gamma}_{1}\rho\hat{a}^{\dag}, \qquad [\hat{G}^{0}_{ab},\hat{a}^{\dag}] = \gamma_{1}\rho\hat{a} - \bar{\gamma}_{3}\rho\hat{a}^{\dag}. \label{zeromix1}
\end{equation}
Therefore, along the same lines followed in section (\ref{amode}), the restriction
\begin{equation}
     \gamma_{1}\bar{\gamma}_{1} - \gamma_{3}\bar{\gamma}_{3} =0, \quad \rightarrow \quad \gamma_{1}=-\bar{\gamma}_{3}; \label{zeromix2}
\end{equation}
is the constraint analogue of Eq. (\ref{baker1}), which ensures the convergence of the corresponding BCH expansion. The similarity transformation will read,
\begin{equation}
     e^{\hat{G}^{0}_{ab}} \hat{a} e^{-\hat{G}^{0}_{ab}} = \hat{a} + [\hat{G}^{0}_{ab},\hat{a}] = (1+\rho\bar{\gamma}_{1})\hat{a} + \rho\bar{\gamma}_{1}\hat{a}^{\dag}, \label{zeromix3}
\end{equation}
which must reproduce the transformation (\ref{iso}) with the parametrization (\ref{baker3}), and hence
\begin{equation}
     \gamma_{1}\rho = -j, \qquad \hat{G}^{0}_{ab} = \frac{j}{\rho} \int dk (\hat{a}\hat{b} + \hat{b}^{\dag}\hat{a}^{\dag} - \hat{a}\hat{b}^{\dag} - \hat{b}\hat{a}^{\dag}). \label{zeromix4}
\end{equation}
Now, the attempt of disentangling with the identification of compound operators $\hat{A}\equiv \frac{j}{\rho}\int dk (\hat{b}^{\dag}\hat{a}^{\dag} - \hat{b}\hat{a}^{\dag})$, and $\hat{B}\equiv \frac{j}{\rho}\int dk (\hat{a}\hat{b} - \hat{a}\hat{b}^{\dag})$, which are (anti)-Hermitian conjugate to each other, will lead to an expression similar to Eq. (\ref{entangled}), due basically to the commutator reduces to the same generator, $[\hat{A},\hat{B}]=j(\hat{A}+\hat{B})-1$ (see Eq. (\ref{wilcox1})). Therefore, the technique of the scaling by a hyperbolic factor of the form (\ref{redefine1}) will work for the case at hand, by using essentially the map $-2\alpha\varrho\rightarrow j$ in the Eqs. (\ref{redefine2})-(\ref{redefine5}); the disentangled unitary operator reads
\begin{equation}
     e^{\frac{1}{\gamma}(\hat{A}+\hat{B})} = e^{\frac{-j}{2}(g-\frac{1}{\gamma})} e^{g\hat{A}} e^{g\hat{B}}; \label{zeromix5}
\end{equation}
where $\gamma \equiv j\rho\gamma_{1}$, which is real according to the original expression in (\ref{zeromix4});
additionally the hyperbolic factor $g=g_{1}+jg_{2}$, is determined fully in terms of the parameter $\gamma$:
\begin{equation}
     \tanh g_{1}=\gamma, \qquad g_{2}=\ln\cosh g_{1}. \label{zeromix6}
\end{equation}
Now, the vacuum is an eigenket for the first exponential,
\begin{eqnarray}
     J^{+} e^{g\hat{B}} |0> = J^{+} |0>, \quad  J^{-} e^{g\hat{B}} |0> =e^{g_{2}-g_{1}} J^{-} |0>.  \label{zeromix7}
\end{eqnarray}
 Furthermore, the vacuum condition (\ref{newvac}) are not sufficient for determining in general the action of higher-order operators, and more conditions must be added for the bilinear operators in the above equation. Thus, we have imposed the bilinear condition $\hat{a}\hat{b}^{\dag} |0>=0$, on the one-particle states.
 
Thus, we need only to disentangle the second exponential $e^{g\hat{A}}$  with the identification
\begin{equation}
     \hat{A}' \equiv g\frac{j}{\rho}\int d
     k \hat{b}^{\dag}\hat{a}^{\dag}, \quad \hat{B}' \equiv -g\frac{j}{\rho}\int dk \hat{b}\hat{a}^{\dag}, \label{zeromix8}
\end{equation}
and the polynomials (\ref{wilcox}) are proportional to the same operator $\hat{B}'$;
\begin{eqnarray}
     \hat{C}_{2} & = & \frac{1}{2}\frac{g^{2}}{\rho}\int dk \hat{b}\hat{a}^{\dag}, \quad \hat{C}_{3} = -\frac{j}{6}\frac{g^{3}}{\rho}\int dk \hat{b}\hat{a}^{\dag}, \quad \hat{C}_{4} = \frac{1}{24}\frac{g^{4}}{\rho}\int dk \hat{b}\hat{a}^{\dag}, \nonumber \\
     \hat{C}_{5} & = & -\frac{j}{120}\frac{g^{5}}{\rho}\int dk \hat{b}\hat{a}^{\dag}, \quad \hat{C}_{6} = \frac{1}{720}\frac{g^{6}}{\rho}\int dk \hat{b}\hat{a}^{\dag}, \label{zeromix9}
\end{eqnarray}
and hence
\begin{eqnarray}
e^{g\hat{A}}=e^{\hat{A}'+\hat{B}'} & = & e^{\hat{A}'} e^{\frac{1}{\rho}[J^{-}e^{g}+J^{+}e^{-g}-1]\hat{b}\hat{a}^{\dag}}, \label{zeromix10} \\
     (J^{\pm}) e^{g\hat{A}}|0> & = & J^{(\pm)} e^{\hat{A}'} |0> = J^{(\pm)} e^{\frac{g_{2}\pm g_{1}}{\rho}\int dk \hat{b}^{\dag}\hat{a}^{\dag}} |0>, \label{zeromix11}
\end{eqnarray}
where we have imposed an additional bilinear condition on the vacuum $\hat{b}\hat{a}^{\dag} |0>=0$, and we have used the properties
(\ref{absorb}) of the projectors. Finally we can put all together and determine the action of the full operator (\ref{zeromix5}) upon the vacuum
\begin{eqnarray}
 (J^{\pm}) e^{\frac{1}{\gamma}(\hat{A}+\hat{B})} |0> = J^{(\pm)}e^{-\frac{1}{2}g_{1}} e^{ \pm\frac{1}{2}(\frac{1}{\gamma}-g_{2}) }
 e^{\frac{g_{2}\pm g_{1}}{\rho}\int dk \hat{b}^{\dag}\hat{a}^{\dag}} |0>. \label{zeromix12}
   \end{eqnarray}
We consider now in particular the $J^+$-projected state;
 all quantum state of a-bosons and b-bosons in the mixture,  accommodates at the same zero-energy level, according to the Eq.  (\ref{stat1}), which is achieved with the vanishing mass constraint $m^2_R=0$; thus, all bosons are massless.
 Such an eigenvalue equation implies that the full $J^+$-mixture is also an eigen-state for the original Hamiltonian and with the same eigenvalue zero,
\begin{equation}
     (J^{+}H) e^{\frac{1}{\gamma}(\hat{A}+\hat{B})} |0> = 0 ;
\label{stat3}
\end{equation}
therefore we have obtained a condensate state of massless  a-bosons and b-bosons.
In contrast, the $J^-$-mixture in (\ref{zeromix12}) is not an eigenstate for $H$, since the $J^-$-component of $H$ corresponds to creation operators of the form $\{\hat{a}^{\dag},\hat{b}^{\dag}\}$.

Boson-number distributions of the condensate (\ref{stat3}) will be developed elsewhere,  due to the large volume of this manuscript.

\subsection{$\hat{G}^{2}_{ba}$}
\label{secondmix}
For this case we switch on again the commutators (\ref{zeromix0}) with $\varrho\neq 0$, and $\zeta\neq 0$; the multimode character will appear in a novel way, through an entangling of the constants that define the nontrivial commutators.

In analogy with the generator (\ref{multi2}), we can consider the second mixed generator,
\begin{equation}
     \hat{G}^{2}_{ba} \equiv \int dk [\gamma_{1}\hat{a}\hat{b}^{\dag} + \gamma_{2}\hat{b}\hat{a}^{\dag} +\gamma_{3}\hat{a}\hat{b} + \gamma_{4}\hat{b}^{\dag}\hat{a}^{\dag} + \gamma_{5}\hat{a}\hat{a}^{\dag} + \gamma_{6}\hat{a}\hat{a} + \gamma_{7}\hat{a}^{\dag}\hat{a}^{\dag}],
     \label{mixed1}
\end{equation}
where the first four mixed terms are corrected now by purely $(\hat{a},\hat{a}^{\dag})$-operators. Along the same lines, we can arrive to the equation
\begin{equation}
     \hat{G}^{2}_{ba} = \int dk [-2\zeta^{-1}\rho (\bar{\gamma}_{7}\hat{a}\hat{b}^{\dag}-\gamma_{7}\hat{b}\hat{a}^{\dag}) + \zeta^{-1}\rho\gamma_{5} (\hat{a}\hat{b}+\hat{b}^{\dag}\hat{a}^{\dag}) + \gamma_{5}\hat{a}\hat{a}^{\dag} - \bar{\gamma}_{7}\hat{a}\hat{a} + \gamma_{7}\hat{a}^{\dag}\hat{a}^{\dag}],
     \label{mixed2}
\end{equation}
which is analogous to the Eq. (\ref{multi6}) that describes the generator $\hat{G}^{1}_{ab}$ in terms of the two coefficients $(\gamma_{5},\gamma_{7})$. Furthermore, the convergence of the corresponding BCH series can be achieved by imposing the same constraint in Eq. (\ref{multi10}), $4\gamma_{7}\bar{\gamma}_{7} - \gamma_{5}\bar{\gamma}_{5}=0$, and thus the transformation on the field operator $\hat{b}$ will reduce to
\begin{equation}
     e^{\hat{G}^{2}_{ba}} \hat{b} e^{-\hat{G}^{2}_{ba}} = \hat{b} + [\hat{G}^{2}_{ba},\hat{b}] = (1+\zeta^{-1}\rho^{2}\gamma_{5})\hat{b} - 2\zeta^{-1}\rho^{2}\bar{\gamma}_{7}\hat{b}^{\dag}. \label{mixed3}
\end{equation}
therefore, a direct comparison with the Eq. (\ref{haus1}) leads to
\begin{equation}
     \gamma_{5}=0, \qquad \bar{\gamma}_{7} = -\frac{\zeta}{\rho^{2}} J^{-}, \label{mixed4}
\end{equation}
which also solve the afore mentioned constraint in a slightly different way to that considered in the Eq. (\ref{multi10}). At this point we have considered that $\rho$ is Hermitian, and thus has an inverse; in particular its $J^+$ and $J^-$ projections are non-vanishing.

The corresponding squeezing operators can be disentangled straightforwardly as follows,
\begin{eqnarray}
     e^{\hat{G}^{2}_{ba}} & = & exp \Big[ -J^{+}\int dk (\frac{\zeta}{\rho^{2}}\hat{a}^{\dag}\hat{a}^{\dag} + \frac{2}{\rho}\hat{a}^{\dag}\hat{b}) + J^{-}\int dk (\frac{2}{\rho}\hat{b}^{\dag}\hat{a}+ \frac{\zeta}{\rho^{2}}\hat{a}\hat{a}) \Big] \nonumber \\
     & = & exp \Big[ -J^{+} \int dk (\frac{\zeta}{\rho^{2}}\hat{a}^{\dag}\hat{a}^{\dag} + \frac{2}{\rho}\hat{a}^{\dag}\hat{b}) \Big] exp \Big[J^{-}  \int dk (\frac{2}{\rho}\hat{b}^{\dag}\hat{a} + \frac{\zeta}{\rho^{2}}\hat{a}\hat{a}) \Big] \label{mixed55} \\
     & = & e^{-J^{+}\int dk \frac{\zeta}{\rho^{2}}\hat{a}^{\dag}\hat{a}^{\dag}} \cdot e^{-J^{+}\int dk \frac{2}{\rho}\hat{a}^{\dag}\hat{b}} \cdot e^{J^{-}\int dk \frac{2}{\rho}\hat{b}^{\dag}\hat{a}} \cdot e^{J^{-}\int dk \frac{\zeta}{\rho^{2}}\hat{a}\hat{a}}; \label{mixed56}
\end{eqnarray}
the Eq. (\ref{mixed55}) follows from the orthogonality of the projectors $(J^{+},J^{-})$, and then the Eq. (\ref{mixed56}) follows from the vanishing commutator (\ref{vanishabd}), which implies that $[\hat{b}^{\dag}\hat{a}, \hat{a}\hat{a}]=0= [\hat{a}^{\dag}\hat{b}, \hat{a}^{\dag}\hat{a}^{\dag}]$. Furthermore, considering that the vacuum is annihilated by $J^{-}\hat{a}$, and $J^{+}\hat{b}$, the vacuum will be an eigenket for the last three exponential operators in the Eq. (\ref{mixed56}), and thus the full squeezing effect on the vacuum is given by
\begin{equation}
     e^{\hat{G}^{2}_{ba}} |0> = e^{-\frac{\zeta}{\rho^{2}}J^{+}\int dk \hat{a}^{\dag}\hat{a}^{\dag}} |0>, \label{mixed7}
\end{equation}
which corresponds seemingly to a single-type squeezed state, instead of a multi-mode state;
each component in the mixture can correspond to a $J^+H$ eigen-state provided that the requirements established in Eqs. (\ref{eigen4}), and (\ref{eigen5}), are satisfied, and then the full mixture is also an eigen-state with the same zero eigen-value; hence, by imposing the vanishing mass restriction, we obtain a quantum state compound of massless squeezed bosons. This state must be compared directly with the previous single state constructed in Eq. (\ref{redefine12}) in terms of eigen-states of the number operator; first, the squeezing effect upon the vacuum in the Eq. (\ref{mixed7}) is, by construction, on the $J^{+}$-direction, and in Eq. (\ref{redefine12}) it is projected on the $J^{-}$-direction. In Eq. (\ref{mixed7}) the operator $\hat{G}^{2}_{ba}$ generates a transformation on the operator $\hat{b}$; this fact is encoded in the presence of the constant $\zeta$, which defines the commutator $[\hat{b},\hat{b}^{\dag}]$; if this constant vanishes, then  the squeezing effect in the Eq. (\ref{mixed7}), depending on the $a$-type field, will be trivial. On the other hand, $\hat{G}_{a}$ generates a transformation on the operator $\hat{a}$, and the 
final state (\ref{redefine12}) is independent on the constant $\zeta$, and hence it is a fully single state. Finally the eigen-states of the number operator in Eq. (\ref{redefine12}) are massive, since the constraints (\ref{eigen5}) used in this section are not involved.

The squeezed states obtained in previous sections required a re-definition of the squeezing operators by a global factor of the form $g=g_1+j g_2$, which defined at the end the corresponding squeezing parameters. The disentanglement in the expression (\ref{mixed56}) is achieved without such a procedure, hence there is no a squeezing parameter. However, the quotient $\zeta/(\rho)^2$ in the state (\ref{mixed7}) may play the role of  an effective squeezing parameter, and a boson-number statistics is possible.

\section{Concluding remarks}

Squeezed states appear in the so called {\it quantum Lagrangians}, which describe problems in\-vol\-ving {\it quantum friction} or fluctuation-disipation phenomena \cite{bishop};
laser quantum theory and photon detection correspond to the traditional applications. The preliminary results that we have obtained by using our approach suggest that a QFT on a ring represents a convenient scheme for studying those phenomena, and will be the subject of future communications.

Two-mode squeezed vacuum, such as the state constructed in the section \ref{multi}, is used as the entangled resource in a quantum communication protocol, in which a quantum state is transferred between two locations. Diverse quantum teleportation experiments have been reported \cite{tele,tele1,tele2,tele3}; considerable efforts are employed for refining the protocols, and teleport increasingly complex quantum states; our model that  naturally accommodates correlated states may be useful in this context.

Once the interactions are turned on, new divergences occur, and the standard QFT formalism requires  further subtractions beyond normal ordering subtractions; the renormalization group scheme introduces then a regulator. It is of our interest the incorporation of (higher order) interactions in the present scheme, and to study the structure at high energies of QFT constructed on a ring.

\section{Appendix}

\subsection{$\hat{G}^{1}_{ab}$}
We consider first the following generator,
\begin{equation}
     \hat{G}^{1}_{ab} \equiv \int dk [ \gamma_{1}\hat{a}\hat{b}^{\dag} + \gamma_{2}\hat{b}\hat{a}^{\dag} + \gamma_{3}\hat{a}\hat{b} + \gamma_{4}\hat{b}^{\dag}\hat{a}^{\dag} + \gamma_{5}\hat{b}\hat{b}^{\dag} + \gamma_{6}\hat{b}\hat{b} + \gamma_{7}\hat{b}^{\dag}\hat{b}^{\dag}], \quad \gamma_{i} \in {\cal H}, \label{multi1}
\end{equation}
where the hypercomplex coefficients $\gamma_{i}$ must be determined.
This generator is constructed as follows; the first four terms correspond to all bilinear terms mixing  the field operators; since the generator must be anti-Hermitian, such mixed terms correspond to pairs with elements that are Hermitian conjugates to each other, $(\hat{a}\hat{b}^{\dag}, \hat{b}\hat{a}^{\dag})$, and $(\hat{a}\hat{b}, \hat{b}^{\dag}\hat{a}^{\dag})$. The commutators of these bilinear terms with the operator $\hat{a}$ will yield unwanted terms, which can be canceled out with those coming from the commutators constructed with the purely b-terms in the generator, which correspond to the last three terms in the Eq. (\ref{multi1}). Hence, using the commutators (\ref{aad})-(\ref{ab}), and (\ref{vanishabd}), we arrive at the expression
\begin{equation}
     \big[ \hat{G}^{1}_{ab}, \hat{a} \big] = -\gamma_{3}\rho\hat{a} - \gamma_{2}\rho\hat{a}^{\dag} - (\gamma_{2}\varrho+2\gamma_{6}\rho )\hat{b} - (\gamma_{4}\varrho+\gamma_{5}\rho)\hat{b}^{\dag}, \label{multi2}
\end{equation}
and the similarity transformation (\ref{similarity1}) can be generated through the exponential operator $e^{\hat{G}^{1}_{ab}}$, provided that the unwanted b-terms in the above equation vanish,
\begin{equation}
     \gamma_{2} = -2\varrho^{-1}\rho\gamma_{6}, \qquad \gamma_{4} = -\varrho^{-1}\rho\gamma_{5}; \label{multi3}
\end{equation}
where $\varrho^{-1}$ is given in Eq. (\ref{baker5}). Similarly, the vanishing of the unwanted b-terms in the commutator $[\hat{G}^{1}_{ab}, \hat{a}^{\dag}]$, and the anti-Hermiticity of the generator, lead to more constraints between the $\gamma$-coefficients,
\begin{eqnarray}
     \big[ \hat{G}^{1}_{ab},\hat{a}^{\dag}\big] & = & \gamma_{1}\bar{\rho}\hat{a} + \gamma_{4}\bar{\rho}\hat{a}^{\dag}; \qquad \gamma_{1} = -2\varrho^{-1}\bar{\rho}\gamma_{7}, \qquad \gamma_{3} = -\varrho^{-1}\bar{\rho}\gamma_{5}; \label{multi4} \\
     (\hat{G}^{1}_{ab})^{\dag} & = & -\hat{G}^{1}_{ab}; \qquad \gamma_{6} = -\bar{\gamma}_{7}, \qquad \bar{\gamma}_{5} = -\gamma_{5}; \label{multi5}
\end{eqnarray}
putting all together, the generator can be rewritten only in terms of two $\gamma$-coefficients, say $(\gamma_{5},\gamma_{7})$,
\begin{equation}
     \hat{G}^{1}_{ab} =\int dk [  -2\varrho^{-1}\bar{\rho}\gamma_{7}\hat{a}\hat{b}^{\dag} + 2\varrho^{-1}\rho\bar{\gamma}_{7}\hat{b}\hat{a}^{\dag} - \varrho^{-1}\bar{\rho}\gamma_{5}\hat{a}\hat{b} - \varrho^{-1}\rho\gamma_{5}\hat{b}^{\dag}\hat{a}^{\dag} + \gamma_{5}bb^{\dag} - \bar{\gamma}_{7}\hat{b}\hat{b} + \gamma_{7}\hat{b}^{\dag}\hat{b}^{\dag}]. \label{multi6}
\end{equation}
Relevant commutators can be determined then in terms of the coefficients $(\gamma_{5},\gamma_{7})$,
\begin{eqnarray}
     \big[ \hat{G}^{1}_{ab}, \hat{a}\big] \!\!& = & \!\! \varrho^{-1}\rho [\gamma_{5}\bar{\rho}\hat{a}-2\bar{\gamma}_{7}\rho\hat{a}^{\dag}] ,\quad [\hat{G}^{1}_{ab},\hat{a}^{\dag}]= -\varrho^{-1}\bar{\rho} [2\gamma_{7}\bar{\rho}\hat{a} + \gamma_{5}\rho\hat{a}^{\dag}], \label{multi7} \\
     \big[ \hat{G}^{1}_{ab}, [\hat{G}^{1}_{ab},\hat{a}] \big] & = & (\varrho^{-1}\rho\bar{\rho})^{2} (4\gamma_{7}\bar{\gamma}_{7} - \gamma_{5} \bar{\gamma}_{5}) \hat{a}, \label{multi8} \\
     \big[ \hat{G}^{1}_{ab}, [\hat{G}^{1}_{ab},[\hat{G}^{1}_{ab},\hat{a}]]\big] & = & (\varrho^{-1}\rho)(\varrho^{-1}\rho\bar{\rho})^{2} (4\gamma_{7}\bar{\gamma}_{7} - \gamma_{5}\bar{\gamma}_{5})[\gamma_{5}\bar{\rho}\hat{a}-2\bar{\gamma}_{7}\rho\hat{a}^{\dag}];
\label{multi85}\\
\Big[\hat{G}^{1}_{ab}, \big[ \hat{G}^{1}_{ab}, [\hat{G}^{1}_{ab},[\hat{G}^{1}_{ab},\hat{a}]]\big]\Big] & = & (\varrho^{-1}\rho\bar{\rho})^{4} (4\gamma_{7}\bar{\gamma}_{7} - \gamma_{5}\bar{\gamma}_{5})^{2} \hat{a}; \label{multi9}
\end{eqnarray}
these expressions are analogous to the Eqs. (\ref{BCH1})-(\ref{BCH4}), and similarly the convergence of the expansion (\ref{BCH}) can be guaranteed by imposing the constraint
\begin{eqnarray}
     4\gamma_{7}\bar{\gamma}_{7} - \gamma_{5}\bar{\gamma}_{5} = 0, \quad \rightarrow \quad \gamma_{5}=2\gamma_{7};\label{multi10}
\end{eqnarray}
where we have indicated a simple solution. Therefore, the similarily transformation on the operator $\hat{a}$ reduces to
\begin{equation}
     e^{\hat{G}^{1}_{ab}} \hat{a} e^{-\hat{G}^{1}_{ab}} = \hat{a} + [\hat{G}^{1}_{ab},\hat{a}] = (1+\varrho^{-1}\gamma_{5}|\rho |^{2})\hat{a} + \varrho^{-1} \gamma_{5}\rho^{2}\hat{a}^{\dag}, \label{multi11}
\end{equation}
this expression is the analogous to that of Eq. (\ref{baker2}), and must reproduce the transformation (\ref{iso}); we use the parametrization (\ref{baker3}), for simplicity and for a direct comparison.
Under these considerations we arrive to the following expressions
\begin{equation}
     \gamma_{5} = j\frac{\varrho}{\rho^{2}}, \qquad \rho =\bar{\rho}; \label{multi12}
\end{equation}
where $\rho$ is fixed now to be Hermitian, and thus the cocient $\frac{\varrho}{\rho^{2}}$ is Hermitian (see Eqs. (\ref{aad})-(\ref{ab})). In terms of these constants
the generator takes the following final form,
\begin{equation}
     \hat{G}^{1}_{ab} = - \frac{j}{\rho} \int dk \Big[\hat{a}\hat{b}^{\dag} + \hat{b}\hat{a}^{\dag} + \hat{a}\hat{b} + \hat{b}^{\dag}\hat{a}^{\dag} - \frac{\varrho}{\rho} (\hat{b}\hat{b}^{\dag} + \frac{1}{2} \hat{b}\hat{b} + \frac{1}{2} \hat{b}^{\dag}\hat{b}^{\dag})\Big]. \label{multi13}
\end{equation}
\subsubsection{The disentanglement}
\label{thedisen}
In order to attempt the disentangling, there exist several possibilities for grouping the seven operators occurring in $\hat{G}^{1}_{ab}$; the following choice yields a first direct disentangling due to the presence of a vanishing commutator
\begin{equation}
     \hat{A} \equiv \frac{-j}{\rho}\int dk (\hat{a}\hat{b}^{\dag}+ \hat{b}\hat{a}^{\dag}+\hat{a}\hat{b}+\hat{b}^{\dag}\hat{a}^{\dag}); \qquad \hat{B} = \frac{j\varrho}{\rho^{2}}\int dk (\hat{b}\hat{b}^{\dag}+\frac{1}{2}\hat{b}\hat{b}+\frac{1}{2}\hat{b}^{\dag}\hat{b}^{\dag}), \qquad [\hat{A},\hat{B}]=0; \label{entan1}
\end{equation}
and thus
\begin{equation}
     e^{\hat{G}^{1}_{ab}} = e^{\frac{-j}{\rho}\int dk (\hat{a}\hat{b}^{\dag}+\hat{b}\hat{a}^{\dag}+\hat{a}\hat{b}+\hat{b}^{\dag}\hat{a}^{\dag})} \cdot e^{\frac{j\varrho}{\rho^{2}}\int dk (\hat{b}\hat{b}^{\dag}+\frac{1}{2}\hat{b}\hat{b}+\frac{1}{2}\hat{b}^{\dag}\hat{b}^{\dag})}; \label{entan2}
\end{equation}
and subsequently one must attempt the disentangling for each separated exponential.

The disentangling of the operator $e^{\hat{B}}$ will be made by considering again its re-definition by a global factor of the form (\ref{redefine1}),
\begin{eqnarray}
     g\hat{B} & \equiv & g\hat{B}_{1} + g\hat{B}_{2} + jg\frac{\varrho\zeta}{\rho^{2}}; \nonumber \\
     \hat{B}_{1} & \equiv & j\frac{\varrho}{2\rho^{2}} \int dk (\hat{b}\hat{b} + \hat{b}^{\dag}\hat{b}), \qquad \hat{B}_{2} \equiv j\frac{\varrho}{2\rho^{2}} \int dk (\hat{b}^{\dag}\hat{b}^{\dag} + \hat{b}^{\dag}\hat{b}), \nonumber \\
    \big[ \hat{B}_{1},\hat{B}_{2} \big] & = & -j \frac{\varrho\zeta}{\rho^{2}} (\hat{B}_{1}+\hat{B}_{2}) - \frac{1}{2} \Big( \frac{\varrho\zeta}{\rho^{2}} \Big)^{2}; \label{entan3}
\end{eqnarray}
in similarity with the case developed in detail in Section (\ref{disen}), we can arrive to the following entangled expression
\begin{equation}
     e^{g(\hat{B}_{1}+\hat{B}_{2})} = e^{\frac{-j}{2}[F_{1}+jF_{2}-g\frac{\varrho\zeta}{\rho^{2}}]} e^{g\hat{B}_{1}} e^{g\hat{B}_{2}} \cdot e^{\frac{-\rho^{2}}{\varrho\zeta}[F_{1}+jF_{2}-g\frac{\varrho\zeta}{\rho^{2}}](\hat{B}_{1}+\hat{B}_{2})}, \label{entan4}
\end{equation}
where the real functions read
\begin{equation}
     F_{1} = \sinh \big(g_{1}\frac{\varrho\zeta}{\rho^{2}}\big) \big[\cosh \big(g_{2}\frac{\varrho\zeta}{\rho^{2}}\big) + \sinh \big(g_{2}\frac{\varrho\zeta}{\rho^{2}}\big)\big], \qquad F_{2} = \cosh \big(g_{1}\frac{\varrho\zeta}{\rho^{2}}\big) \big[\cosh \big(g_{2}\frac{\varrho\zeta}{\rho^{2}}\big) + \sinh (g_{2}\frac{\varrho\zeta}{\rho^{2}})-1\big]. \label{entan5}
\end{equation}
Furthermore, under the condition of the vanishing hyperbolic phase for the compound operator in Eq. (\ref{entan4}),
\begin{equation}
     F_{2} = 0, \qquad \rightarrow \quad g_{2} = - \frac{\rho^{2}}{\varrho\zeta} \ln\cosh (g_{1}\frac{\varrho\zeta}{\rho^{2}}), \label{entan6}
\end{equation}
we arrive to a disentangled unitary operator,
\begin{equation}
     e^{-\frac{\rho^{2}}{\varrho\zeta}\tanh (g_{1}\frac{\varrho\zeta}{\rho^{2}}) [\hat{B}_{1}+\hat{B}_{2}]} = e^{\frac{1}{2}\ln\cosh (g_{1}\frac{\varrho\zeta}{\rho^{2}})} e^{\frac{j}{2}[\tanh (g_{1}\frac{\varrho\zeta}{\rho^{2}}) - g_{1}\frac{\varrho\zeta}{\rho^{2}}]} \cdot e^{-(g_{1}+jg_{2})\hat{B}_{2}} e^{-(g_{1}+jg_{2})\hat{B}_{1}}, \label{entan7}
\end{equation}
where there is only one free parameter $g_{1}$, according to the Eq. (\ref{entan6}), and we have assumed by simplicity that the combination $\frac{\varrho\zeta}{\rho^{2}}$ is real valued.

Furthermore, the vacuum is an eigenket for the first exponential operator,
\begin{equation}
     (J^{+}) e^{-g\hat{B}_{1}} |0> = (J^{+}) |0>, \label{entan8}
\end{equation}
due to the definition of vacuum $J^{+} \hat{b} |0>=0 $; hence we need the disentangling of the second exponential, which can be achieved by establishing a correspondence with the disentangling performed in Eqs. (\ref{redefine8})-(\ref{redefine9}), namely,
\begin{eqnarray}
     \hat{a} & \leftrightarrow & \hat{b}, \nonumber \\
     -g\alpha & \leftrightarrow & -gj\frac{\varrho}{2\rho^{2}}, \nonumber \\
     F_{0}(\varrho^{-1}, 2g\alpha\varrho) & \leftrightarrow & \tilde{F}_{0}(\zeta^{-1}, j\frac{g\varrho}{\rho^{2}}\zeta); \label{entan9}
\end{eqnarray}
where
\begin{equation}
     \tilde{F}_{0}(\zeta^{-1}, j\frac{g\varrho}{\rho^{2}}\zeta) = \frac{\zeta^{-1}}{2}\big[ (1+j\frac{g\varrho}{\rho^{2}}\zeta) + (1-j\frac{g\varrho}{\rho^{2}}\zeta) e^{j\frac{g\varrho}{\rho^{2}}\zeta}\big] , \label{entan10}
\end{equation}
and thus the disentangled version for the operator $e^{-g\hat{B}_{2}}$ can be mapped under the correspondence, directly from the Eq. (\ref{redefine9}),
\begin{eqnarray}
     e^{-g\hat{B}_{2}} & \equiv & \exp [-gj\frac{\varrho}{2\rho^{2}} \int dk \hat{b}^{\dag}\hat{b} - gj\frac{\varrho}{2\rho^{2}} \int dk \hat{b}^{\dag}\hat{b}^{\dag}] \nonumber \\
     & = & e^{-gj\frac{\varrho}{2\rho^{2}} \int dk \hat{b}^{\dag}\hat{b}} \cdot e^{\tilde{F}_{0}\int dk \hat{b}^{\dag}\hat{b}^{\dag}}, \label{entan11}
\end{eqnarray}
where the quadratic commutator involved is $\int dk [\hat{b}^{\dag}\hat{b}, \hat{b}^{\dag}\hat{b}^{\dag}] =2\zeta\int dk \hat{b}^{\dag}\hat{b}^{\dag}$.

Now we need to calculate the action of this operator on the vacuum, and the analogous of the Eq. (\ref{redefine10}) is required,
\begin{equation}
     J^{+} \int dk (\hat{b}^{\dag}\hat{b})^{M} (\hat{b}^{\dag})^{N} |0> = J^{+}\int dk (N\zeta)^{M} (\hat{b}^{\dag})^{N} |0>, \qquad M,N: 0,1,2,3,\ldots ;  \label{entan12}
\end{equation}
and consequently the analogous of Eq. (\ref{redefine11}) reads
\begin{equation}
     J^{+} e^{-g\hat{B}_{2}}\cdot |0> = J^{+} \cdot e^{\tilde{F}_{0}(\zeta^{-1}, -j\frac{g\varrho}{\rho^{2}}\zeta)\int dk \hat{b}^{\dag}\hat{b}^{\dag}} \cdot |0>; \label{entan13}
\end{equation}
this equation allows us to determine the action of the operator (\ref{entan7}) upon the vacuum; however, this action is only partial, since we have disentangled only a part of the complete
operator $\hat{G}^{1}_{ab}$ in (\ref{entan2}).

\subsection{$\hat{G}^{3}_{ba}$}
We consider now the third case that consists of a combination of the two previous cases, 
\begin{equation}
     \hat{G}^{3}_{ab} \equiv \int dk [\gamma_{1}\hat{a}\hat{b}^{\dag} + \gamma_{2}\hat{b}\hat{a}^{\dag} +\gamma_{3}\hat{a}\hat{b} + \gamma_{4}\hat{b}^{\dag}\hat{a}^{\dag} + \gamma_{5}\hat{a}\hat{a}^{\dag} + \gamma_{6}\hat{a}\hat{a} + \gamma_{7}\hat{a}^{\dag}\hat{a}^{\dag} + \gamma'_{5}\hat{b}\hat{b}^{\dag} + \gamma'_{6}\hat{b}\hat{b} + \gamma'_{7}\hat{b}^{\dag}\hat{b}^{\dag}];
     \label{mixx1}
\end{equation}
in this expression the mixed terms are deformed by the simultaneous addition of purely $\hat{a}$- and $\hat{b}$-terms, and hence it will generate similarity transformations on the operators $\hat{a}$, and $\hat{b}$. We consider  the $\hat{b}$-case, and the $\hat{a}$-case will be developed elsewhere.

Following the same procedure, and omitting the details, the generator takes the following form in terms of four coefficients $(\gamma_{5},\gamma'_{5}, \gamma_{7},\gamma'_{7})$;
\begin{eqnarray}
     \hat{G}^{3}_{ab}  & = & \int dk [-2\zeta^{-1}\rho (\bar{\gamma}_{7}\hat{a}\hat{b}^{\dag}-\gamma_{7}\hat{b}\hat{a}^{\dag}) + \zeta^{-1}\rho\gamma_{5} (\hat{a}\hat{b}+\hat{b}^{\dag}\hat{a}^{\dag}) + \gamma_{5}\hat{a}a^{\dag} - \bar{\gamma}_{7}\hat{a}\hat{a} + \gamma_{7}\hat{a}^{\dag}\hat{a}^{\dag} + \gamma'_{5}\hat{b}\hat{b}^{\dag} \nonumber \\
     & & - \bar{\gamma}'_{7}\hat{b}\hat{b} + \gamma'_{7}\hat{b}^{\dag}\hat{b}^{\dag}],  \label{mixx2} \\
     & & 4\Big(\bar{\gamma}_{7} + \frac{\zeta^{2}}{\rho^{2}}\gamma'_{7}\Big)\Big(\gamma_{7} +\frac{\zeta^{2}}{\rho^{2}}\bar{\gamma}'_{7}\Big) - \Big(\gamma_{5} -\frac{\zeta^{2}}{\rho^{2}}\gamma'_{5}\Big)\Big(\bar{\gamma}_{5} -\frac{\zeta^{2}}{\rho^{2}}\bar{\gamma}'_{5}\Big) = 0, \label{mixx3}
\end{eqnarray}
with $\gamma_{5}=-\bar{\gamma}_{5}$, and $\gamma'_{5}=-\bar{\gamma}'_{5}$; the constraint (\ref{mixx3}) is imposed for the convergence of the corresponding BCH series, and the transformation on the operator $\hat{b}$ is reduced basically to the Eq. (\ref{mixed3}), and then 
\begin{equation}
     \gamma_{5} - \frac{\zeta^{2}}{\rho^{2}}\gamma'_{5}=0, \qquad \bar{\gamma}_{7}+ \frac{\zeta^{2}}{\rho^{2}}\gamma'_{7} = -\frac{\zeta}{\rho^{2}}J^{-}, \label{mixx4}
\end{equation}
which reduce to the Eq. (\ref{mixed4}) by fixing $\gamma'_{5}=0=\gamma'_{7}$, as expected; these adjusted version of the Eqs. (\ref{mixed4}) will allow us to construct multi-mode states.

{\bf Acknowledgements:}
This work was supported by the Sistema Nacional de Investigadores (M\'exico), and the Vicerrectoria de Investigaci\'on y Estudios de Posgrado
(BUAP). A.J.C. Ju\'arez-Dom\'{\i}nguez acknowledges the financial support by CONACyT (Mexico) under Grant No. CB-2014-01/240781.
 The numerical analysis and graphics have been made using Mathematica.

\end{document}